\documentclass[aps,twocolumn,prx,showpacs,notitlepage,longbibliography]{revtex4-2}
\usepackage{CJKutf8}
\usepackage{graphicx}
\usepackage{xcolor}
\usepackage{enumerate}
\usepackage{amsmath, amssymb}
\usepackage{float}
\usepackage[pdftex,colorlinks=true,linkcolor=blue,citecolor=blue]{hyperref}

\usepackage[linesnumbered]{algorithm2e}
\RestyleAlgo{ruled}

\usepackage{soul}
\setstcolor{red}

\DeclareMathOperator*{\argmax}{arg\,max}

\newcommand\undermat[2]{
  \makebox[0pt][l]{$\smash{\underbrace{\phantom{%
    \begin{matrix}#2\end{matrix}}}_{\text{$#1$}}}$}#2}

\let\emptyset\varnothing

\usepackage{lineno}

\begin{document}
\begin{CJK*}{UTF8}{bkai}
\title{Approximate maximum likelihood decoding with $K$ minimum weight matchings}

\author{Mao Lin (林茂)}
\affiliation{Amazon Braket, Seattle, WA, 98170, USA}

\date{\today}

\begin{abstract}
    The minimum weight matching (MWM) and maximum likelihood decoding (MLD) are two widely used and distinct decoding strategies for quantum error correction. For a given syndrome, the MWM decoder finds the most probable physical error corresponding to the MWM of the decoding graph, whereas MLD aims to find the most probable logical error. Although MLD is the optimal error correction strategy, it is typically more computationally expensive compared to the MWM decoder. In this work, we introduce an algorithm that approximates MLD with $K$ MWMs from the decoding graph. Taking the surface code subject to graphlike errors as an example, we show that it is possible to efficiently find the first $K$ MWMs by systematically modifying the original decoding graph followed by finding the MWMs of the modified graphs. For the case where the $X$ and $Z$ errors are correlated, despite the MWM of the decoding hypergraph cannot be found efficiently, we present a heuristic approach to approximate the MLD by finding the $K$ MWMs in the $X$ and $Z$ subgraphs. We benchmark the efficacy of our algorithm for the surface code subject to graphlike errors, the surface-square Gottesman-Kitaev-Preskill (GKP) code and surface-hexagonal GKP code subject to the Gaussian random displacement errors, showing that the fidelity approaches that of the exact MLD (for the first two cases) or the tensor-network decoder (for the last case) as $K$ increases.
\end{abstract}

\maketitle
\end{CJK*}

\tableofcontents

\section{Introduction}
An efficient and accurate decoding algorithm is indispensable for building a fault-tolerant quantum computer \cite{demarti2023decoding}. In order to detect and correct  errors from the environmental {noise} and imperfect controls, the classical decoder is responsible to process the error syndromes generated by the quantum device and infer a recovery operation to restore the quantum information. The decoder not only has to be reliable, but also needs to be fast enough to avoid accumulating more errors and be scalable to larger quantum error correction (QEC) codes \cite{battistel2023real, delfosse2023choose}.

The performance and efficiency requirements impose a trade-off for the design and implementations of the decoding strategies.
For example, various efficient decoding algorithms have been proposed for the surface code, including the {minimum-weight-perfect-matching (MWPM) decoder} \cite{fowler2013minimum}, the union-find (UF) decoder \cite{delfosse2021almost, delfosse2022toward, delfosse2021union}, the renormalization group decoder \cite{duclos2010fast, duclos2010renormalization} and others \cite{demarti2023decoding}. 
Recently, two variants of the MWPM decoder, which finds the minimum-weight-matching (MWM) on a decoding graph, have been shown to be more efficient than the conventional MWPM decoder \cite{wu2023fusion, higgott2023sparse, footnote1}.
The {MWPM} and UF decoders have also been applied to color codes \cite{delfosse2014decoding, kubica2023efficient, delfosse2021almost, sahay2022decoder, zhang2021quantum} and low-density-parity-check (LDPC) codes \cite{delfosse2021union, delfosse2022toward}.
However, these decoders typically exhibit worse QEC performance compared to the optimal but more computationally-demanding decoder, such as the maximum-likelihood decoder (MLD), for both surface codes \cite{bravyi2014efficient} and color codes \cite{katzgraber2009error, chubb2021general}. 
Indeed, as recently demonstrated with surface codes realized on a superconducting device, MLD is shown to yield lower logical errors, despite being several orders of magnitude slower than the MWM decoder \cite{google2023suppressing}.
Similar result was also obtained for distance-3 heavy-hexagon code \cite{sundaresan2022matching}.
Besides, UF decoder and its variants have also been used in QEC experiments, such as the recent experiment based on reconfigurable atom arrays \cite{bluvstein2024logical, cain2024correlated}, or even implemented on chips \cite{barber2023real}.
It is now widely accepted that decoder design should be customized for different quantum computers  based on the QEC code used, device topology and throughput requirements \cite{battistel2023real}.
For example, each QEC round typically takes $1\ \mu s$ for superconducting-qubit experiments \cite{van2019electronic, jeffrey2014fast, sundaresan2022matching, krinner2022realizing, google2023suppressing, jones2024improved} which implies that the decoder needs to be fast enough to process the syndrome data in this timescale; on the other hand, the QEC-cycle time is usually on the order of milliseconds for trapped-ion systems \cite{ryan2021realization, egan2021fault}, opening up the possibility of using more accurate decoders. 
Given that more diverse qubit technologies, such as bosonic qubits \cite{gottesman2001encoding, sivak2022real, campagne2020quantum, de2022error}, spin qubits \cite{abobeih2022fault, o2016silicon, takeda2022quantum, barthel2010fast, vandersypen2017interfacing}, neutral atoms \cite{henriet2020quantum, bluvstein2024logical}, are {maturing}, it calls for more effort to find new decoders.

There are several approaches to devise better decoding strategies. One approach is to improve either the accuracy or the efficiency of currently available decoders.
For example, there have been several attempts to improve the accuracy of the {MWPM and MWM decoders} \cite{fowler2013optimal, delfosse2014adecoding, criger2018multi, yuan2022modified, iolius2023performance, higgott2023improved, delfosse2023splitting, shutty2024efficient}. 
These modified  decoders typically involve certain heuristic methods to modify the weights in the graphs to reduce the logical error rate of the specific QEC codes studied.
New decoding strategies can also be developed through finding the connections between different families of decoders. 
Recently, it was discovered that the UF decoder can be viewed as an approximate implementation of the blossom algorithm used for the {MWPM} decoder \cite{wu2022interpretation}.
This interesting connection later inspired the authors to {discover the more efficient MWM decoder} \cite{wu2023fusion}. 
The connection between the MWM decoder and MLD has previously been noted in studies of the surface code threshold using a statistical-mechanical approach \cite{dennis2002topological}.

In this work, we introduce an algorithm to approximate MLD {by deterministically enumerating the first $K$ MWMs in} a given decoding graph.
One of the most well known applications of the MWM decoder is for {graphlike error models} where each error can produce at most two nontrivial stabilizer measurement outcomes. Graphlike error models have been used to simulate, for example, the surface code subject to either the $X$-type or $Z$-type Pauli errors.
For a given syndrome, the idea is to construct a weighted graph, whose edges and vertices correspond to the qubits and stabilizers of the surface code respectively, highlight a set of vertices based on the syndrome, followed by finding a set of edges that matches the highlighted vertices in pairs with minimum weight. 
This is the MWM of the decoding graph, which corresponds to the most probable \emph{physical} error, the set of qubits that most likely experience errors according to the syndrome.
Although {the} MWM decoder is an efficient decoding strategy, its decoding accuracy is typically worse than MLD, which explores more instances of physical errors and determines the most probable \emph{logical} error.
Since there is more than one way to match the highlighted vertices in a decoding graph, one can show that the physical errors, which are consistent with the syndrome, are one-to-one corresponding to the matchings in the decoding graph.
Because distinct matchings could have the same weight, a matching with weight larger than or equal to $(K-1)$ distinct matchings in the graph will be referred to as the $K$-th MWM hereafter (the first MWM will be simply referred to as the MWM).
We further show that the probability of a physical error is inversely proportional to the exponential of the weight of the corresponding matching. 
With that, we propose a decoding strategy which identifies $K$ MWMs for a given decoding graph, and determines the most probable logical errors based on the weights of the matchings.
We refer to this strategy as the $K$-MWM decoder, where K is the number of MWMs used from the decoding graph.

For a given decoding graph, although its MWM can be identified efficiently using either the blossom algorithm \cite{fowler2013minimum}, or its variants \cite{higgott2023sparse,wu2023fusion}, it is not necessarily straightforward to find the second and subsequent MWMs. 
One may find a new matching by replacing certain edges in the MWM while maintaining the highlighted vertices matched pairwise, but the resultant matching is not guaranteed to have lower weight compared to other unexplored matchings. 
In order to address this issue, we take inspirations from the Chegireddy-Hamacher's algorithm \cite{chegireddy1987algorithms}, which finds the first $K$ MWPMs in a complete weighted graph.
In particular, the $K$-MWM decoder finds all the candidates of the second MWM by finding the MWMs of a series of reduced graphs, which are constructed by systematically removing the edges in the MWM of the original decoding graph.
The second MWM is then the matching with the minimum {weight} among the candidates.
To generalize this process to find subsequent MWMs, we construct a \emph{decoding tree}, where the root of the tree is the MWM of the graph, the nodes of the tree are the explored matchings, and the $K$-th MWM corresponds to the leaf node of the tree with {the} minimum weight. 
To efficiently find the $(K+1)$-th MWM, the key observation is that we only need to find the child nodes for the $K$-th MWM such that the new and existing leaf nodes are the candidates for the $(K+1)$-th MWM.

The algorithm can also be extended to handle cases where $X$ and $Z$ errors are correlated and the decoding graph is replaced by a decoding hypergraph. For that, we could start by finding the first $K$ MWMs of the $X$ decoding graph, followed by modifying the weights of the $Z$ decoding graph, which then leads to $K$ distinct $Z$ decoding graphs. One could proceed to find $K$ MWMs for each $Z$ decoding graph and approximate the MLD with the resultant $K^2$ MWMs.
A more practically efficient approach, which is adopted in this work, is to use a decoding tree to keep track of the explored MWMs, and only find the child nodes for the $K$-th MWM. 
It is important to emphasize that neither approach can guarantee to find \emph{the} MWM of the decoding hypergraph, which generally requires runtime {exponential in the size} of the hypergraph. This is in sharp contrast to the case with a decoding graph, for which the MWM can be identified efficiently.
Nevertheless, our numerical {simulations} demonstrate that the fidelity of the $K$-MWM decoder increases with respect to $K$, and approaches that of the tensor-network decoder, another method to approximate the MLD \cite{bravyi2014efficient}.

The $K$-MWM decoder can also be applied to QEC codes with continuous variables, such as the Gottesman-Kitaev-Preskill (GKP) codes. In Ref.~\cite{lin2023closest}, it has been shown that the MWM of the decoding graph corresponds to the closest point in the symplectic dual GKP lattice. We generalize such correspondence to that between the MWMs in the decoding graph and the lattice coset representatives, providing a lattice perspective for the $K$-MWM decoder. 
The source code and data used in this work is available through the package LatticeAlgorithms.jl \cite{latticealgorithms}, where we implement the $K$-MWM decoder for the concatenated-GKP codes, and apply the decoder to the corresponding qubit stabilizer code as a special case.

The remainder of the paper is organized as follows. 
In Sec.~\ref{sec: K-MWM decoding for graphlike errors} and \ref{sec: K-MWM decoding for correlated errors}, we describe the $K$-MWM decoder for graphlike and correlated errors respectively, using the surface code as an example. 
In Sec.~\ref{sec: $K$-MWM decoding for GKP codes}, we describe the application of the $K$-MWM decoder to the GKP code, highlighting the interpretation from a lattice perspective. 
In Sec.~\ref{sec:Finding the $K$-th matching for the decoding graph}, we show the main result of the paper, an efficient algorithm to find $K$ MWMs given a decoding graph. In particular, we explain how to find the second MWM in Sec.~\ref{sec:Finding the second matching for the syndrome graph}, and generalize it to $K$ MWMs in Sec.~\ref{sec:Finding K matchings for the decoding graph}. We prove the correctness of the algorithm in Sec.~\ref{sec: Proof for finding the second matching of a decoding graph}. {Graph} related terminologies are introduced in Sec.~\ref{sec: The MWM decoder} and Sec.~\ref{sec:Finding the $K$-th matching for the decoding graph} can be read without other prior sections. 
In Sec.~\ref{sec:Numerical results}, we present the numerical results of applying the $K$-MWM decoder to surface-square, surface-hexagonal GKP codes, and qubit surface code subject to the $Z$-error.
We conclude and discuss some future directions in Sec.~\ref{sec:Discussion}.
In App.~\ref{sec: Preliminary} and App.~\ref{sec:Maximum likelihood decoder for concatenated GKP codes: A lattice perspective}, we expand the discussion in Sec.~\ref{sec: $K$-MWM decoding for GKP codes} and provide a more complete description of the $K$-MWM decoder from a lattice perspective. In App.~\ref{sec: Minimum weight cycle of a graph}, we provide some details of finding {the} minimum weight cycle in a decoding graph, an important part of the $K$-MWM decoder.

\section{$K$-MWM decoding for graphlike errors}
\label{sec: K-MWM decoding for graphlike errors}

In this section, we describe the general idea of $K$-MWM decoding for graphlike errors, using the surface code subject to the $Z$-type error as an example. For a given syndrome, we show that the physical errors are one-to-one corresponding to the matchings in the decoding graph. To determine the most probable logical error, we identify the first $K$ MWMs and their corresponding logical errors, followed by summing up the probability of the physical errors that correspond to the same logical error. The details of how to find subsequent MWMs of a decoding graph will be presented in Sec.~\ref{sec:Finding the $K$-th matching for the decoding graph}. 

\subsection{The MWM decoder}
\label{sec: The MWM decoder}

\begin{figure}
\centering
\includegraphics[width=\linewidth]{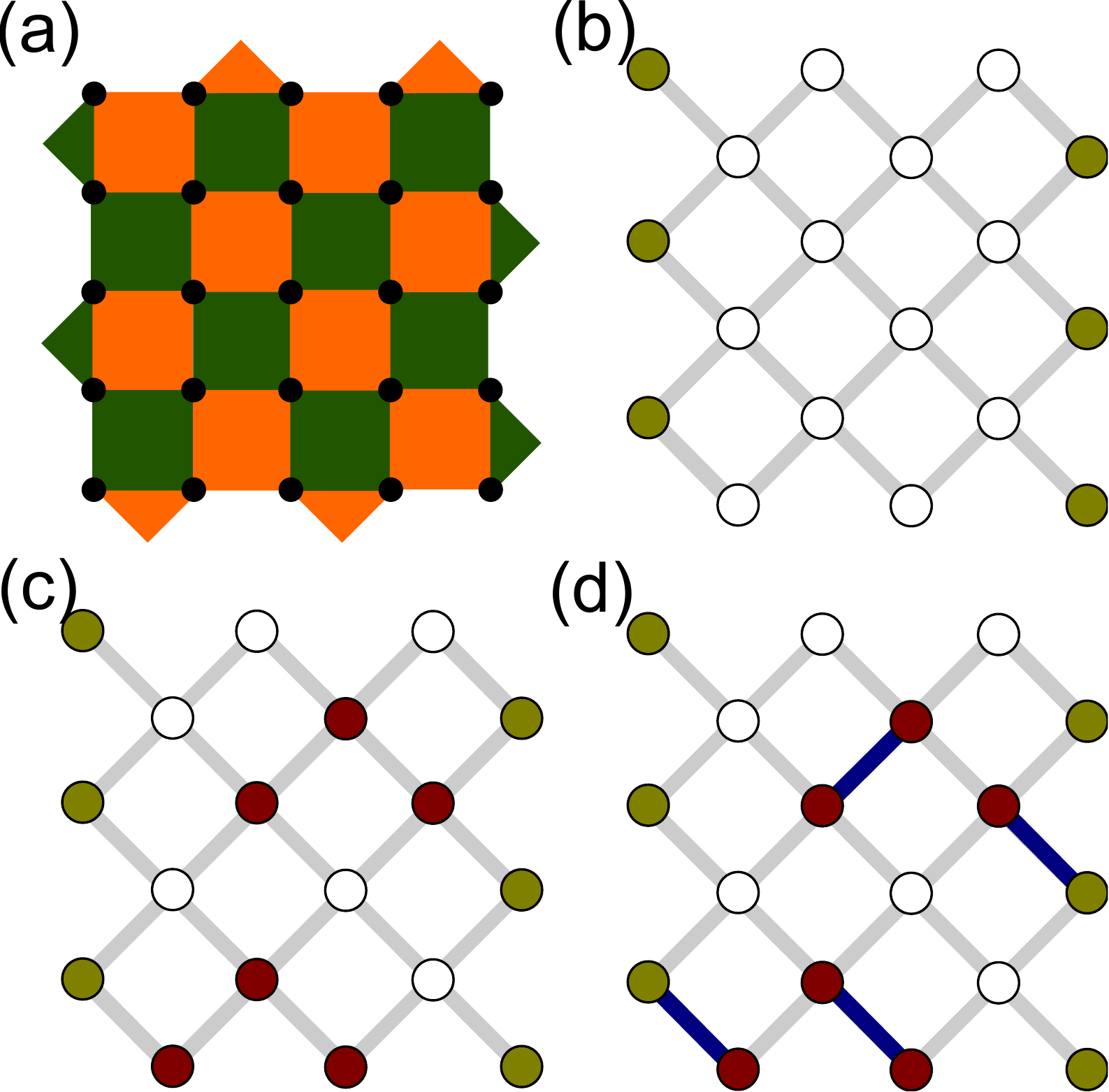}
 \caption{The MWM decoder for the surface code subject to graphlike errors. 
 (a) The $d=5$ surface code where the black dots represent the data qubits and the $X$-type and $Z$-type stabilizers are shown in orange and green respectively.
 (b) The model graph for the $Z$-type errors where the white vertices and edges correspond to the $X$-type stabilizers and data qubits of the code respectively. The {green} circles represent virtual vertices that are all connected via edges with weight zero which are not shown for clarity reason.
 (c) The decoding graph where the red vertices represent nontrivial stabilizer measurement outcomes, and the weights of the edges are given in Eq.~\eqref{eq:weight_e_k}. 
 (d) An example of MWM which matches the highlighted vertices pairwise with lowest possible weight.
}
 \label{fig:MWM}
\end{figure}

We start by reviewing the MWM decoder {for \emph{graphlike} errors (also referred to as stringlike errors in the literature) where each error can produce
at most two nontrivial stabilizer measurement outcomes}. For convenience, we illustrate the $d = 5$ surface code in Fig.~\ref{fig:MWM}(a) where the black dots represent the data qubits and the $X$-type and $Z$-type stabilizers are shown in orange and green respectively.

The first step of the MWM decoder is to construct the \emph{model graph} \cite{wu2023fusion} as shown in Fig.~\ref{fig:MWM}(b). Because the surface code is a Calderbank-Shor-Steane (CSS) code and we assume it is subject to a graphlike error, the model graph consists of two disjoint subgraphs for the $X$-type and $Z$-type errors respectively.
For clarity, we have only shown the model graph for the $Z$-type errors in Fig.~\ref{fig:MWM}(b) and the model graph for the $X$-type errors can be constructed and analyzed similarly.
In the model graph, each white vertex has either two or four incident edges, depending on the Hamming weight of the corresponding stabilizer, and two distinct white vertices will share an edge if the corresponding stabilizers share a data qubit. 
For edges that are only incident to one (white) vertex, corresponding to data qubits that are only in one $X$-type stabilizer, we connect each of them to a \emph{virtual vertex}. The virtual vertices are all connected via edges with weight zero, hence not shown in Fig.~\ref{fig:MWM}(b) for clarity reason \cite{noh2020fault}.
{It is worth {emphasizing} that we have assumed there is no duplicated edge in the model graph throughout this work.}
The process of constructing the model graph can be done prior to correcting the errors and be used repeatedly for different input syndromes.

For a given syndrome, the MWM decoder aims to find the most probable physical error. The probability of a physical error $\boldsymbol{\eta}\in\mathbb{F}_2^N$ is given by
\begin{align}
\label{eq:prob_physical_error}
    P(\boldsymbol{\eta}) = \prod_{i=1}^N\left(\frac{\epsilon_i}{1-\epsilon_i}\right)^{\eta_i}\prod_{i=1}^N(1-\epsilon_i),
\end{align}
where $N$ is the number of data qubits, and $\epsilon_i$ is the probability of an $Z$-type error in the $i$-qubit. We will omit the second product in Eq.~\eqref{eq:prob_physical_error} since it is independent of $\boldsymbol{\eta}$. Equivalently, the MWM decoder aims to minimize
\begin{align}
\label{eq:prob_physical_error_log}
    -\log P(\boldsymbol{\eta}) = \sum_{i=1}^N\eta_i\log\left(\frac{1-\epsilon_i}{\epsilon_i}\right).
\end{align}
For that, a \emph{decoding graph} \cite{wu2023fusion} is constructed on top of the model graph as shown in Fig.~\ref{fig:MWM}(c). More concretely, let $\boldsymbol{s}\in\mathbb{F}_2^{N_X}$ denote the syndrome measurement outcomes for the $N_X$ stabilizer generators of $X$-type, which correspond to the (non-virtual) vertices in the model graph $(\mathcal{V}, \mathcal{E})$. We highlight those vertices with nontrivial stabilizer measurement outcomes ($s_a=1$) and define the set of highlighted vertices as
\begin{align}
    \mathcal{V}_h = \left\{v_a\in\mathcal{V} ~|~ 1\leq a\leq N_X, s_a=1\right\}.
\end{align}
Further, we assign to the edge $e_i\in\mathcal{E}$ the following weight
\begin{align}
\label{eq:weight_e_k}
    w(e_i) = \log\left(\frac{1-\epsilon_i}{\epsilon_i}\right),
\end{align}
and denote the set of weights as $\mathcal{W}$. The resulting decoding graph is $G = (\mathcal{V}, \mathcal{E}, \mathcal{W}, \mathcal{V}_h)$, corresponding to the set of vertices, edges, weights and highlighted vertices respectively. 

After the decoding graph is set up, a \emph{matching} $\mathcal{M}$ is defined to be a subset of edges that satisfies the following relation
\begin{align}
\label{eq:def_matching}
    \mathcal{V}_h = \oplus_{e\in\mathcal{M}}e,
\end{align}
where $\oplus$ denotes symmetric difference of two sets, i.e., 
\begin{align}
    \label{eq:def_symmetric_diff}
    e_1\oplus e_2 \equiv \left\{v ~|~ v\in e_1 \text{ or } v\in e_2 \text{ but not both}\right\}.
\end{align}
From Eq.~\eqref{eq:def_matching}, it suggests that if $v_a\in\mathcal{V}_h$, then there {is an} odd number of edges in $\mathcal{M}$ that are incident to $v_a$; otherwise, the number of incident edges is even.

For a matching satisfying Eq.~\eqref{eq:def_matching}, its weight is defined as 
\begin{align}
\label{eq:weight_highlighted_matching_0}
    w(\mathcal{M}) \equiv \sum_{e\in \mathcal{M}}w(e).
\end{align}
Further, we can define a vector $\boldsymbol{\eta}^\mathcal{M}\in\mathbb{F}_2^N$ as
\begin{align}
\label{eq:def_physical_error_from_matching}
    \eta^\mathcal{M}_j = \begin{cases}
0 & \text{if } e_j\notin \mathcal{M} \\
1 & \text{if } e_j\in \mathcal{M}
\end{cases}.
\end{align}
Combining Eq.~\eqref{eq:prob_physical_error_log} with Eq.~\eqref{eq:weight_highlighted_matching_0}-\eqref{eq:def_physical_error_from_matching}, we arrive at
\begin{align}
\label{eq:-logP=w(M)}
    -\log P(\boldsymbol{\eta}^\mathcal{M}) &= w(\mathcal{M}),
\end{align}
which indicates that the probability of the error $\boldsymbol{\eta}^\mathcal{M}$ is inversely proportional to the exponential of the weight of the matching.

As will be shown in Sec.~\ref{sec: Approximate the MLD with $K$ MWMs}, the physical errors that are consistent with the syndrome are one-to-one corresponding to the matchings in the decoding graph. As a result, minimizing $-\log P(\boldsymbol{\eta})$ is equivalent to finding a matching that matches the highlighted vertices pairwise with the lowest possible weight, or an MWM. We show one example of an MWM in Fig.~\ref{fig:MWM}(d). After the most probable physical error is found, we identify its corresponding logical error, followed by applying certain operations to attempt to correct the logical error, which concludes the MWM decoder.

\subsection{The MLD}
\label{sec: The MLD}

Because the actual error is a priori unknown, there is always a chance that we may misidentify the logical error that happens in the QEC code. In contrast to the MWM decoding, the MLD aims to find the most probable logical error, which is the optimal strategy to minimize the probability of getting a logical error after the attempted error correction \cite{bravyi2014efficient}. For that, we first identify a candidate error $\boldsymbol{\eta}$ for the given syndrome and consider the following coset probability
\begin{align}
\label{eq:coset_prob}
    P([\boldsymbol{\eta}]) \equiv \sum_{\boldsymbol{g}^Z\in\mathcal{G}^Z}P(\boldsymbol{g}^Z\ominus\boldsymbol{\eta}),
\end{align}
where $\boldsymbol{g}^Z\in\mathbb{F}_2^N$ is the binary vector representations of the $Z$-type stabilizer group elements and $\boldsymbol{g}^Z\ominus\boldsymbol{\eta}$ indicate element-wise binary subtraction (addition) over $\mathbb{F}_2$. Here 
\begin{align}
\label{def:coset}
    [\boldsymbol{\eta}]\equiv\left\{\boldsymbol{\eta}\ominus\boldsymbol{g}^Z ~|~ \boldsymbol{g}^Z\in\mathcal{G}^Z\right\}
\end{align}
denotes the set of physical errors that differ from $\boldsymbol{\eta}$ only by a stabilizer. The MLD aims to solve 
\begin{align}
\label{eq:MLD}
    \argmax_{\boldsymbol{l}}P_{\boldsymbol{l}} \equiv \argmax_{\boldsymbol{l}}P([\boldsymbol{\eta}\ominus\boldsymbol{l}]),
\end{align}
where $\boldsymbol{l}\in\mathbb{F}_2^N$ are distinct logical errors. For the surface code subject to the $Z$-type error, the MLD aims to find the maximum among $P([\boldsymbol{\eta}])$ and $P([\boldsymbol{\eta}\ominus\boldsymbol{l}^Z])$, the coset probability for logical identity and $Z$ operators. 

Because the size of $\mathcal{G}^Z$ typically scales exponentially with the number of qubits, it is generally difficult to evaluate the coset probability exactly. Nevertheless, it turns out that the MWM decoding can be viewed as the ``zeroth order'' approximation of the MLD. To see that, suppose $\epsilon_i\rightarrow0$ for all the qubits, then we only need to {be concerned with} the largest summand in $P([\boldsymbol{\eta}\ominus\boldsymbol{l}])$, i.e.,
\begin{align}
    P([\boldsymbol{\eta}\ominus\boldsymbol{l}]) \approx\max_{\boldsymbol{g}^Z\in\mathcal{G}^Z}P(\boldsymbol{\eta}\ominus\boldsymbol{g}^Z\ominus\boldsymbol{l}).
\end{align}
With that, the MLD reduces to 
\begin{align}
\label{eq:MWM_qubit}
    \argmax_{\boldsymbol{l}}P_{\boldsymbol{l}}\approx&\text{argmax}_{\boldsymbol{l}} \max_{\boldsymbol{g}^Z\in\mathcal{G}^Z}P(\boldsymbol{\eta}\ominus\boldsymbol{g}^Z\ominus\boldsymbol{l})\nonumber\\
    =&\text{argmax}_{\tilde{\boldsymbol{l}}\in\tilde{\mathcal{G}}^Z} P(\boldsymbol{\eta}\ominus\tilde{\boldsymbol{l}})\\
    =&\text{argmin}_{\tilde{\boldsymbol{l}}\in\tilde{\mathcal{G}}^Z} \sum_{i=1}^N(\eta_i\ominus \tilde{l}_i)\log\left(\frac{1-\epsilon_i}{\epsilon_i}\right),    \nonumber
\end{align}
where $\tilde{\mathcal{G}}^Z$ is the $Z$-type normalizer group of the code. We recognize that the summation in Eq.~\eqref{eq:MWM_qubit} takes the same form as Eq.~\eqref{eq:prob_physical_error_log}, which shows that the most probable logical error can indeed be approximated by finding the most probable physical error. The essence of the $K$-MWM decoding is to improve this approximation systematically by including more MWMs in the decoding graph.

\subsection{Approximate the MLD with $K$ MWMs}
\label{sec: Approximate the MLD with $K$ MWMs}

The correctness of the MWM decoding relies on the one-to-one correspondence between the physical errors and the matchings in the decoding graph, which we now elaborate. For a physical error $\boldsymbol{\eta}$ that is consistent with the syndrome $\boldsymbol{s}$, it satisfies the condition
\begin{align}
\label{eq:anticommutation_relation_physical_error}
    H\boldsymbol{\eta} = \boldsymbol{s},
\end{align}
where $H\in\mathbb{F}_2^{N_X\times N}$ and its $a$-th row corresponds to the $a$-th $X$-type stabilizer generator $\boldsymbol{g}^{X,a}$.
We note that the vector multiplication in Eq.~\eqref{eq:anticommutation_relation_physical_error} is carried out over $\mathbb{F}_2$, which represents the anti-commutation relations between the physical error and the $X$-type stabilizers. The matching corresponding to $\boldsymbol{\eta}$ can be defined as
\begin{align}
\label{eq:def_matching_from_physical_error}
    \mathcal{M} = \left\{e_i\in\mathcal{E} ~|~ 1\leq i\leq N \text{ and } \eta_i = 1\right\},
\end{align}
which contains an edge iff the corresponding component in $\boldsymbol{\eta}$ is not zero. Eq.~\eqref{eq:def_matching_from_physical_error} can be regarded as a converse of Eq.~\eqref{eq:def_physical_error_from_matching}. 
To prove that the set of edges defined in Eq.~\eqref{eq:def_matching_from_physical_error} is indeed a matching that satisfies the condition in Eq.~\eqref{eq:def_matching}, we consider a vertex $v_a\in\mathcal{V}_h$, or an $X$-type stabilizer $\boldsymbol{g}^{X,a}$ with $s_a=1$. The condition in Eq.~\eqref{eq:anticommutation_relation_physical_error} suggests that there {is an} odd number of components for which both $\boldsymbol{\eta}$ and $\boldsymbol{g}^{X,a}$ take values of unity which, combined with Eq.~\eqref{eq:def_matching_from_physical_error}, indicates there {is an} odd number of edges in $\mathcal{M}$ that are incident to $v_a$. Hence $v_a\in\oplus_{e\in\mathcal{M}}e$ or $\mathcal{V}_h\subset\oplus_{e\in\mathcal{M}}e$. With a similar argument, we have $v_a\notin\mathcal{V}_h$ implies $v_a\notin\oplus_{e\in\mathcal{M}}e$, which concludes the proof. 

Conversely, for a matching satisfying Eq.~\eqref{eq:def_matching}, we consider a vector $\boldsymbol{\eta}$ as defined in Eq.~\eqref{eq:def_physical_error_from_matching}. Using a similar argument as the one above, we can show that $\boldsymbol{\eta}$ satisfies the condition in Eq.~\eqref{eq:anticommutation_relation_physical_error}, and hence is a physical error consistent with the syndrome. 

With the correspondence at hand, the $K$-MWM decoding works by finding $K$ distinct matchings $\mathcal{M}_1, \mathcal{M}_2, .., \mathcal{M}_K$ such that 
\begin{align}
\label{eq:sequence_M_0}
    w(\mathcal{M}_1) \leq w(\mathcal{M}_2) \leq \cdots \leq w(\mathcal{M}_K) \leq w(\mathcal{M})
\end{align}
for $\forall \mathcal{M}\neq \mathcal{M}_1, \cdots, \mathcal{M}_K$, where the weight of a matching is defined in Eq.~\eqref{eq:weight_highlighted_matching_0}. For each matching, we determine the corresponding physical error $\boldsymbol{\eta}$ via Eq.~\eqref{eq:def_physical_error_from_matching} followed by determining the corresponding logical error $\boldsymbol{l}$. From Eq.~\eqref{eq:-logP=w(M)}, the contribution of the physical error to the logical error probability $P_{\boldsymbol{l}}$ reads $P(\boldsymbol{\eta})=\exp(-w(\mathcal{M}))$ such that the $K$-MWM decoding approximates the logical error probability as
\begin{align}
\label{eq:approx_P_l_MWMs}
    P_{\boldsymbol{l}}\approx \sum_{\mathcal{M}\in\Xi_{\boldsymbol{l}}} \exp\left[-w(\mathcal{M})\right].
\end{align}
Here $\Xi_{\boldsymbol{l}}$ denotes the set of matchings that result in the logical error $\boldsymbol{l}$. 
From Eq.~\eqref{eq:approx_P_l_MWMs}, we observe that including subsequent MWMs can improve decoding fidelity when there exist matchings belonging to different logical classes than the first MWM, and with sufficiently low weights. This behavior is illustrated with a simple example in App.~\ref{sec: An example of decoding with $K$ MWMs}.

In summary, we have described the general idea for $K$-MWM decoding for graphlike errors. To complete the algorithm, in Sec.~\ref{sec:Finding the $K$-th matching for the decoding graph}, we will describe how to find the $K$ MWMs for a given decoding graph that satisfies Eq.~\eqref{eq:sequence_M_0}.

\section{$K$-MWM decoding for correlated errors}
\label{sec: K-MWM decoding for correlated errors}

In this section, we switch gears to consider the case where the $X$ and $Z$ errors are correlated and the decoding graph is replaced by a decoding hypergraph. Although the notion of a matching (a subset of hyperedges) can be similarly defined, there is no known algorithm that can find the MWM of the decoding hypergraph efficiently, let alone its first $K$ MWMs. Instead, the $K$-MWM decoder works by finding MWMs in the $X$ and $Z$ decoding graphs, followed by combining the matchings to approximate the logical error probability. 

\subsection{The MLD for correlated errors}
\label{sec: The MLD for correlated errors}
We start by reviewing the MLD for correlated errors. Suppose there are $N$ data qubits, let $\epsilon_i^{X,Y,Z}$ denote the probability for the $X$-type, $Y$-type and $Z$-type error of the $i$-th qubit respectively. We will use $\boldsymbol{\eta}^X, \boldsymbol{\eta}^Z\in\mathbb{F}_2^{N}$ to denote the $X$-errors and $Z$-errors for a given syndrome $\boldsymbol{s}\in\mathbb{F}_2^{N'}$ where $N'$ is the number of $X$ and $Z$ stabilizer generators combined. Let $\boldsymbol{\eta}\equiv[\boldsymbol{\eta}^X; \boldsymbol{\eta}^Z]\in\mathbb{F}_2^{2N}$ denotes a physical error, its probability reads
\begin{align}
\label{eq:P_e}
    P(\boldsymbol{\eta}) &\equiv P(\boldsymbol{\eta}^X, \boldsymbol{\eta}^Z)\\
    &=\prod_{i=1}^N\left(\frac{\epsilon_i^X}{{\epsilon^I_i}}\right)^{\eta_i^X}\left(\frac{\epsilon_i^Z}{{\epsilon^I_i}}\right)^{\eta_i^Z}\left(\frac{\epsilon_i^Y{\epsilon^I_i}}{{\epsilon^X_i\epsilon^Z_i}}\right)^{\eta_i^X\eta_i^Z},\nonumber
\end{align}
where ${\epsilon^I_i=1-(\epsilon_i^X+\epsilon_i^Y+\epsilon_i^Z)}$ and we have omitted a factor that is independent with $\boldsymbol{\eta}$. For the case of correlated errors, the MLD aims to solve exactly the same equation as Eq.~\eqref{eq:MLD}, but with the coset probability replaced by
\begin{align}
\label{eq:P_l_correlated}
    P([\boldsymbol{\eta}\ominus\boldsymbol{l}]) = \sum_{\boldsymbol{g}^X\in\mathcal{G}^X}\sum_{\boldsymbol{g}^Z\in\mathcal{G}^Z}P(\boldsymbol{\eta}^X\ominus\boldsymbol{g}^X\ominus\boldsymbol{l}^X, \boldsymbol{\eta}^Z\ominus\boldsymbol{g}^Z\ominus\boldsymbol{l}^Z),
\end{align}
where $\boldsymbol{l}\equiv[\boldsymbol{l}^X; \boldsymbol{l}^Z]\in\mathbb{F}_2^{2N}$ denotes the logical error, and the summation is carried out over all the $X$-type and $Z$-type stabilizers in $\mathcal{G}^{X,Z}$. In the case where $\epsilon_i^{X,Y,Z}$ are small, the most probable logical error can be approximated by the most probable physical error, which is equivalent to minimizing
\begin{align}
\label{eq:-logP_correlated_errors}
    &-\log P(\boldsymbol{\eta}) \\
    = &\sum_{i=1}^N\eta_i^Z\log\left(\frac{{\epsilon^I_i}}{\epsilon_i^Z}\right)+\eta_i^X\left[\log\left(\frac{{\epsilon^I_i}}{\epsilon_i^X}\right)+\eta_i^Z\log\left(\frac{{\epsilon^X_i\epsilon^Z_i}}{\epsilon_i^Y{\epsilon^I_i}}\right)\right].\nonumber
\end{align}
Because of the third term in Eq.~\eqref{eq:-logP_correlated_errors}, however, finding the MWMs in the $X$ and $Z$ decoding graphs separately does not necessarily produce the most likely physical error, let alone the logical error.

\subsection{Approximate the MLD with $K$ matchings}
\label{sec: Approximate the MLD with $K$ matchings}

In order to improve the approximation of the coset probability in Eq.~\eqref{eq:P_l_correlated} from a graph perspective, we first follow Sec.~\ref{sec: The MWM decoder} to set up the $X$ and $Z$ model graphs. For a given syndrome $\boldsymbol{s}\equiv[\boldsymbol{s}^X; \boldsymbol{s}^Z]$, the consistent physical errors satisfy the following conditions
\begin{align}
\label{eq:anticommutation_relation_physical_error_correlated}
    H\begin{bmatrix}
        0_N & I_N\\
        I_N & 0_N
    \end{bmatrix}
    \boldsymbol{\eta} = \boldsymbol{s},
\end{align}
where $H\in\mathbb{F}_2^{N'\times 2N}$, and $I_N$ is the $N\times N$ identity matrix. Similar to Eq.~\eqref{eq:anticommutation_relation_physical_error}, the row vectors of the check matrix $H$ in Eq.~\eqref{eq:anticommutation_relation_physical_error_correlated} correspond to the stabilizer generators. For those relations with $s_a=1$ in Eq.~\eqref{eq:anticommutation_relation_physical_error_correlated}, we highlight the corresponding vertex in either the $X$ or $Z$ model graph based on if the corresponding row of $H$ is a $Z$-type or $X$-type stabilizer. Further, for the $i$-th edge in the $Z$ and $X$ decoding graphs, we assign them the weights ${\log(\epsilon_i^I/\epsilon_i^X)}$ and ${\log(\epsilon_i^I/\epsilon_i^Z)}$ respectively (see Eq.~\eqref{eq:weight_e_k}). 

One version of the $K$-MWM decoding works by first finding $K$ MWMs of one decoding graph followed by finding the $K$ MWMs of another graph with modified weights. More concretely, let $\left\{\mathcal{M}^Z\right\}$ denote the set of $K$ MWMs for the $Z$ decoding graph. Since our objective is to minimize the quantity in Eq.~\eqref{eq:-logP_correlated_errors}, for each $\mathcal{M}^Z$, we first define a vector $\boldsymbol{\eta}^Z\in\mathbb{F}_2^N$ according to Eq.~\eqref{eq:def_physical_error_from_matching}, followed by modifing the weights of the $i$-th edge in the $X$ decoding graph as
\begin{align}
    \log\left(\frac{{\epsilon_i^I}}{\epsilon_i^X}\right) + \eta_i^Z\log\left(\frac{{\epsilon^X_i\epsilon^Z_i}}{\epsilon_i^Y{\epsilon^I_i}}\right).
\end{align}
The modified weights in the $X$ decoding graph correspond to the term in the square brackets in Eq.~\eqref{eq:-logP_correlated_errors}.
We then find $\left\{\mathcal{M}^X\right\}$, a set of $K$ MWMs, for the modified graph, with which a set of vectors $\left\{\boldsymbol{\eta}^X\right\}$ can be identified via Eq.~\eqref{eq:def_physical_error_from_matching}. In total, this process yields $K^2$ physical errors, which can then be used to approximate the logical error probability.

We note that the vector $\boldsymbol{\eta}\equiv[\boldsymbol{\eta}^X; \boldsymbol{\eta}^Z]$, which is obtained from the matchings $\mathcal{M}^{X, Z}$, is indeed a physical error consistent with the syndrome. This can be seen by following the same argument in Sec.~\ref{sec: Approximate the MLD with $K$ MWMs}, which shows that $\boldsymbol{\eta}$ satisfies the condition in Eq.~\eqref{eq:anticommutation_relation_physical_error_correlated}. Conversely, for a given $\boldsymbol{\eta}$ that is consistent with the syndrome, the above process can indeed find the matchings $\mathcal{M}^{X, Z}$, with large enough $K$, such that $P(\boldsymbol{\eta})=\exp(-w(\mathcal{M}^{Z}) -w(\mathcal{M}^{X}))$.

In this work, instead of the above approach, which finds $K^2$ matchings in total, we implement a more practically efficient approach which finds $K$ matchings for approximating the MLD \cite{latticealgorithms}. We will explain the latter approach in the context of GKP code in App.~\ref{sec: Details of approximate MLD for the surface-GKP code}.

\section{$K$-MWM decoding for GKP codes}
\label{sec: $K$-MWM decoding for GKP codes}

In this section, we describe the application of the $K$-MWM decoding to GKP codes. Throughout this paper, we focus on concatenated-GKP codes, which are constructed by concatenating $N$ one-mode GKP codes, each of which encodes a single qubit, to an $[[N,k]]$ stabilizer code. The resultant concatenated-GKP code encodes $k$ qubits in $N$ modes. 
Below we provide an overview of the $K$-MWM decoding for concatenated-GKP codes, using surface-square and surface-hexagonal GKP codes as examples. More complete descriptions will be presented in  App.~\ref{sec: Preliminary} and \ref{sec:Maximum likelihood decoder for concatenated GKP codes: A lattice perspective}.

\subsection{Brief review of GKP codes}
\label{sec: Brief review of GKP codes}
A GKP code is a stabilizer code that encodes quantum information using the quadrature operators of $N$ bosonic modes $\hat{\boldsymbol{x}} \equiv [\hat{x}_{1} , \hat{x}_{2} , \cdots , \hat{x}_{2N} ]^{T} \equiv [\hat{q}_{1} , \cdots, \hat{q}_{N} , \hat{p}_{1} , \cdots, \hat{p}_{N} ]^{T}$. These quadrature operators satisfy the following commutation relations
\begin{align}
    [\hat{x}_j, \hat{x}_k] = i\Omega_{jk},
\end{align}
where the symplectic form $\Omega$ is a $2N\times 2N$ matrix
\begin{align}
\label{eq:def_Omega_0}
    \Omega 
    =  \begin{bmatrix} 0 & 1 \\ -1 & 0\end{bmatrix} \otimes I_N =  \begin{bmatrix} 0_N & I_N \\ -I_N & 0_N\end{bmatrix}.
\end{align}
As a stabilizer code, the stabilizer group of a GKP code is isomorphic to a $2N$-dimensional lattice 
\begin{align}
    \Lambda(M) \equiv \lbrace M^T\boldsymbol{a} ~|~ \boldsymbol{a}= [a_{1},\cdots, a_{2N}]^{T} \in \mathbb{Z}^{2N} \rbrace,
\end{align}
where $M\in\mathbb{R}^{2N\times 2N}$ is the generator matrix of the lattice that satisfies the condition
\begin{align}
\label{eq:def_symplectic_gram}
    A\equiv M\Omega M^T\in\mathbb{Z}^{2N\times 2N}.
\end{align}
In other words, the symplectic Gram matrix $A$ needs to be {integer valued} such that the stabilizer group elements commute with each other. For each lattice $\Lambda$, we can define its symplectic dual lattice as
\begin{align}
\label{eq:def_lambda_perp_0}
    \Lambda^\perp \equiv  \left\{\boldsymbol{u} ~|~ \boldsymbol{u}^T\Omega\boldsymbol{v}\in\mathbb{Z}, ~ \forall \boldsymbol{v}\in\Lambda(M)\right\},
\end{align}
which consists of all vectors that have integer symplectic inner product with all vectors in $\Lambda$. As a result of Eq.~\eqref{eq:def_symplectic_gram}, $\Lambda$ is a sublattice of $\Lambda^\perp$ which implies that the logical operators of the GKP code $\boldsymbol{l}\in\mathbb{R}^{2N}$ are encoded in the quotient group $\sqrt{2\pi}(\Lambda^\perp/\Lambda)$.

In this work, we consider the Gaussian random displacement error for GKP codes where the quadrature variables are subject to independent and identically distributed (iid) additive errors 
\begin{align}
\label{eq:def_shift_errors_0}
    \hat{\boldsymbol{x}}\rightarrow\hat{\boldsymbol{x}}' = 
    \hat{\boldsymbol{x}} + \boldsymbol{\xi}.
\end{align}
Here $\boldsymbol{\xi}\equiv[\xi^{(\hat{q})}_1, \cdots , \xi^{(\hat{q})}_N, \xi^{(\hat{p})}_1, \cdots , \xi^{(\hat{p})}_N]\sim_\text{iid}\mathcal{N}(0,\sigma^2)$ are random displacements that follow the Gaussian distribution
\begin{align}
\label{eq:def_Gaussian_0}
    P_\sigma(\boldsymbol{\xi}) \equiv \frac{1}{\sqrt{2\pi\sigma^2}}\exp\left(-\frac{||\boldsymbol{\xi}||^2}{2\sigma^2}\right),
\end{align}
where $||\boldsymbol{\xi}||$ denotes the Euclidean norm of the vector. Because of its random nature, the error $\boldsymbol{\xi}$ is not known a priori, and we can only learn its effect through the syndrome defined as
\begin{align}
\label{eq:def_syndrome_0}
    \boldsymbol{s} \equiv \sqrt{2\pi} M \Omega^{-1}\boldsymbol{\xi} ~\mod~ 2\pi,
\end{align}
where the modulo operation is applied element-wise. We will derive Eq.~\eqref{eq:def_syndrome_0} in App.~\ref{sec:Maximum likelihood decoder for GKP codes}.

For the purpose of introducing MLD for GKP codes, let
\begin{align}
\label{eq:def_[xi]_0}
    [\boldsymbol{\xi}]\equiv
    \left\{\boldsymbol{\xi}-\boldsymbol{u} ~|~ \boldsymbol{u}\in \sqrt{2\pi}\Lambda\right\}
\end{align}
denotes a coset of real valued vectors that differ from $\boldsymbol{\xi}$ only by a lattice vector in $\sqrt{2\pi}\Lambda$, the scaled GKP lattice. We deliberately use the same notation as Eq.~\eqref{def:coset} to denote physical errors that are different only by a stabilizer, but it is important to note that the stabilizer group of a GKP code is infinite dimensional. The MLD for GKP codes aims to solve the same equation as Eq.~\eqref{eq:MLD} but with the coset probability replaced by
\begin{align}
\label{eq:P_l_gkp}
    P_{\boldsymbol{l}} \equiv P([\boldsymbol{\eta_s}-\boldsymbol{l}]) = \sum_{\boldsymbol{u}\in\sqrt{2\pi}\Lambda}P_\sigma(\boldsymbol{\eta_s}-\boldsymbol{l}-\boldsymbol{u}),
\end{align}
where $\boldsymbol{l}\in\sqrt{2\pi}(\Lambda^\perp/\Lambda)$ is a logical operator and 
\begin{align}
    \label{eq:def_eta_s_0}
    \boldsymbol{\eta_s}\equiv\frac{1}{\sqrt{2\pi}}\Omega M^{-1}\boldsymbol{s}
\end{align}
is a physical error that is consistent with the syndrome.

The MLD can be approximated by considering the limit where the noise variance is nearly zero.
In this limit, we only need to keep the largest contribution in the summation in Eq.~\eqref{eq:P_l_gkp}, and the MLD reduces to 
\begin{equation}
\label{eq:closest_point_problem_0}
    \begin{aligned}
    &\text{argmax}_{\boldsymbol{l}\in\sqrt{2\pi}(\Lambda^\perp/\Lambda)}\sum_{\boldsymbol{u}\in\sqrt{2\pi}\Lambda}P_\sigma(\boldsymbol{\eta_s}-\boldsymbol{l}-\boldsymbol{u}) \\
\approx&\text{argmax}_{\boldsymbol{l}\in\sqrt{2\pi}(\Lambda^\perp/\Lambda)}\text{max}_{\boldsymbol{u}\in\sqrt{2\pi}\Lambda}P_\sigma(\boldsymbol{\eta_s}-\boldsymbol{l}-\boldsymbol{u}) \\
=&\text{argmax}_{\boldsymbol{l}^\perp\in\sqrt{2\pi}\Lambda^\perp}P_\sigma(\boldsymbol{\eta_s}-\boldsymbol{l}^\perp) \\
=&\text{argmin}_{\boldsymbol{l}^\perp\in\sqrt{2\pi}\Lambda^\perp}||\boldsymbol{\eta_s}-\boldsymbol{l}^\perp||.
    \end{aligned}
\end{equation}
The problem in Eq.~\eqref{eq:closest_point_problem_0} is known as the closest point search problem in the mathematical literature \cite{agrell2002closest}, where we are asked to find a lattice point $\boldsymbol{l}^\perp$ that is closest to the given vector $\boldsymbol{\eta_s}$ compared to all other lattice points in the lattice $\sqrt{2\pi}\Lambda^\perp$.

Although the closest point problem in Eq.~\eqref{eq:closest_point_problem_0} is generally hard to solve \cite{van1981another,micciancio2001hardness}, in Ref.~\cite{lin2023closest}, we introduced several efficient {closest} point decoding techniques for structured GKP codes, particularly a matching algorithm for the surface-square GKP code (see App.~\ref{sec:Efficient closest point decoder with decoding graph} for a review), which is the starting point of our $K$-MWM decoder. 

\subsection{Surface-square GKP code: A lattice perspective of $K$-MWM decoding}
\label{sec: Surface-square GKP code: A lattice perspective of $K$-MWM decoding}

A surface-square GKP code is obtained by concatenating the surface code with $N$ one-mode square GKP codes. For this code, because the Gaussian random displacement error is a graphlike error, we can focus on one subspace, say the $N$-dimensional $\hat{p}$ subspace. In particular, it was shown in Ref.~\cite{lin2023closest} that the closest point solution to Eq.~\eqref{eq:closest_point_problem_0} corresponds to the MWM of the decoding graph where the weights of the edges are defined with respect to $\boldsymbol{\eta_s}^{(\hat{p})}$ (see Eq.~\eqref{eq:weight_e_k_gkp}).
In this work, we generalize such correspondence to that between the MWMs in the decoding graph and certain coset representatives of the lattice $\sqrt{2\pi}\Lambda^{\perp,(\hat{p})}$, providing a lattice perspective for the $K$-MWM decoding when applied to the surface-square GKP code. Below, we will omit the superscript for the $\hat{p}$ subspace for clarity.

Because the surface-square GKP code is constructed on top of the surface code, its lattice is highly structured and we can show that $2Z_{N}\subset\sqrt{2}\Lambda^\perp$ where $Z_{N}$ is the $N$-dimensional integer lattice (see Eq.~\eqref{eq:chain_subsets}). As a result, we can define the following coset of equivalent lattice points (not to confuse with the coset defined in Eq.~\eqref{eq:def_[xi]_0})
\begin{align}
\label{eq:coset_u_v1_0}
    [[\boldsymbol{\chi}]] = \left\{\boldsymbol{\chi}-\boldsymbol{v} ~|~ \boldsymbol{v}\in2Z_{N}\right\},
\end{align}
and collect all the different coset representatives into an ordered set as
\begin{align}
\label{eq:def_ordered_representatives_0}
    \mathcal{U} = \left\{\boldsymbol{\chi}_1, \boldsymbol{\chi}_2, \boldsymbol{\chi}_3, \cdots, \boldsymbol{\chi}_{|\mathcal{U}|}\right\},
\end{align}
where $\boldsymbol{\chi}_i\notin[[\boldsymbol{\chi}_j]]$ if $i\neq j$. The elements in $\mathcal{U}$ are ordered based on their distances to $\boldsymbol{\eta_s}$, i.e., $||\boldsymbol{\eta}_{\boldsymbol{s}}-\sqrt{\pi}\boldsymbol{\chi}_i|| \leq ||\boldsymbol{\eta}_{\boldsymbol{s}}-\sqrt{\pi}\boldsymbol{\chi}_j||$ if $i\leq j$. In other words, there is a one-to-one mapping between the cosets of $\sqrt{2}\Lambda^\perp$ and the elements in $\mathcal{U}$. 
For convenience, we will assume $\boldsymbol{\chi}$ {is closer} to $\boldsymbol{\eta_s}$ than any other elements in $[[\boldsymbol{\chi}]]$, hence $\boldsymbol{\chi}_1$ is essentially the closest point to $\boldsymbol{\eta_s}$. It is important to note that $\boldsymbol{\chi}_2$ needs not be the second closest point to $\boldsymbol{\eta}_s$ because a certain point in $[[\boldsymbol{\chi}_1]]$ could be closer to $\boldsymbol{\eta}_s$ than $\boldsymbol{\chi}_2$. Nevertheless, as will be shown below, it is not necessary to find all the points in a coset because the coset probability can be straightforwardly calculated once the coset representative is identified. Hence we are only interested in finding the coset representatives in $\mathcal{U}$.

To see the benefit of introducing the coset in Eq.~\eqref{eq:coset_u_v1_0}, let us partition the set $\mathcal{U}$ based on {the logical errors of its elements} 
\begin{align}
\label{eq:def_U_l}
    \mathcal{U} \equiv \bigcup_{\boldsymbol{l}\in\sqrt{2\pi}(\Lambda^\perp/\Lambda)}\mathcal{U}_{\boldsymbol{l}},
\end{align}
where $\mathcal{U}_{\boldsymbol{l}}$ contains the coset representatives that result in the logical operator $\boldsymbol{l}$. Then the MLD can be approximated as
\begin{equation}
\label{eq:mld_conc_sq_0}
    \begin{aligned}
        &\text{argmax}_{\boldsymbol{l}\in\sqrt{2\pi}(\Lambda^\perp/\Lambda)}\sum_{\boldsymbol{u}\in\sqrt{2\pi}\Lambda}P_\sigma(\boldsymbol{\eta_s}-\boldsymbol{l}-\boldsymbol{u}) \\
        \approx& \text{argmax}_{\boldsymbol{l}\in\sqrt{2\pi}(\Lambda^\perp/\Lambda)} \sum_{\boldsymbol{\chi}\in \mathcal{U}_{\boldsymbol{l}}}P([[\boldsymbol{\chi}]]),
    \end{aligned}
\end{equation}
where the coset probability $P([[\boldsymbol{\chi}]])$ is given by 
\begin{align}
    \label{eq:def_P_chi_0}
    P([[\boldsymbol{\chi}]])  
        =&\sum_{\boldsymbol{v}\in2Z_{N}}P_\sigma(\boldsymbol{\eta}_{\boldsymbol{s}}-\sqrt{\pi}(\boldsymbol{\chi}+\boldsymbol{v}))\\
        =&\frac{1}{\sqrt{2\pi\sigma^2}}\prod_{i=1}^{N}\sum_{{v}_i\in\mathbb{Z}}\exp\left\{-\frac{(\eta_{s, i}-\sqrt{\pi}({\chi}_i+2v_i))^2}{2\sigma^2}\right\}.\nonumber
\end{align}
From the definition of the coset representatives $\boldsymbol{\chi}\in\mathcal{U}$, we have $0\leq\eta_{s, i}-\sqrt{\pi}{\chi}_i\leq1$ for the summand in Eq.~\eqref{eq:def_P_chi_0}. We further note that, for the typical noise strength $\sigma=0.6$,
\begin{align*}
    \exp\left(-\frac{\pi}{2\sigma^2}(2\tilde{v})^2\right)
\end{align*}
is of the order $10^{-31}$ and $10^{-69}$ for $\tilde{v}=2$ and $3$ respectively. Hence, the coset probability $P([[\boldsymbol{\chi}]])$ can be efficiently calculated and we only need to include $N\times N_v$ terms in the summation of Eq.~\eqref{eq:def_P_chi_0} for $N_v\leq 4$.

With that, for approximating the MLD for surface-square GKP codes, we could identify the first $K$ \emph{closest coset representatives} in $\mathcal{U}$, calculating their contributions as in Eq.~\eqref{eq:def_P_chi_0}, followed by adding their contribution to the corresponding logical error probability as in Eq.~\eqref{eq:mld_conc_sq_0}. As will be proved in detail in App.~\ref{sec: $K$-MWM decoder for surface-square GKP codes}, finding $K$ closest coset representatives in $\mathcal{U}$ is equivalent to finding the $K$ MWMs in the decoding graph, which will be addressed in Sec.~\ref{sec:Finding the $K$-th matching for the decoding graph}.

\subsection{Surface-hexagonal GKP code}
The $K$-MWM decoding can be adopted to approximate the MLD for general surface-GKP code where the inner codes are non-square lattices, such as the hexagonal lattice. The corresponding code will be referred to as the surface-hexagonal GKP code. Similar to the surface-square GKP code, we can define a lattice coset structure for the surface-hexagonal GKP lattice as
\begin{align}
    \left\{\boldsymbol{\chi}-\boldsymbol{v} ~|~ \boldsymbol{v}\in2\Lambda\left(S\right)\right\}
\end{align}
where $S$ is the direct sum of $N$ identical symplectic matrices $S_\text{hex}\in\mathbb{R}^{2\times2}$ that characterizes the one-mode hexagonal GKP code (See Eq.~\eqref{eq:def_S}).
Hence approximating the MLD for surface-hexagonal GKP code reduces to finding the lattice coset representatives of the corresponding lattice. 

However, because the Gaussian random displacement error acts as a correlated error for the surface-hexagonal GKP code, unlike the surface-square GKP code, there is no efficient method to identify the closest point solution for Eq.~\eqref{eq:closest_point_problem_0}. To resolve this issue, we could adopt a similar approach as outlined in Sec.~\ref{sec: Approximate the MLD with $K$ matchings} by finding the first $K$ coset representatives in the $\hat{p}$ subspace, followed by finding the first $K$ coset representatives with modified candidate errors in the $\hat{q}$ subspace. In App.~\ref{sec: Coset structures for general surface-GKP codes}, we will explain another more efficient approach, which is adopted in the numerical {simulations} in Sec.~\ref{sec: Surface-hexagonal GKP code}. 

\subsection{Surface code subject to graphlike errors as a special case of surface-square GKP code}
\label{sec: Surface code subject to graphlike errors as a special case of surface-square GKP code}

Here we illustrate that the MLD for the surface code subject to graphlike errors can be treated as a special case of that for the surface-square GKP code. A similar relation can be drawn between the surface code subject to correlated errors and general surface-GKP code, which will be illustrated in App.~\ref{sec: Surface code subject to correlated errors as a special case of general surface-GKP code}. Such connections allow us to implement the $K$-MWM decoder for concatenated-GKP codes and use it to decode the corresponding qubit stabilizer codes \cite{latticealgorithms}.

Recall that the MLD for surface code subject to $Z$-type errors requires evaluating the following coset probability
\begin{align}
\label{eq:coset_prob_qubit}
    P([\boldsymbol{\eta}]) = \sum_{\boldsymbol{g}^Z\in\mathcal{G}^Z}\prod_{i=1}^N\left(\frac{\epsilon_i}{1-\epsilon_i}\right)^{\eta_i\ominus g^Z_i},
\end{align}
where $\boldsymbol{\eta},\boldsymbol{g}^Z\in\mathbb{F}_2^N$ denote the candidate error and $Z$-type stabilizers respectively (see Eq.~\eqref{eq:prob_physical_error} and \eqref{eq:coset_prob}). In order to decode the surface code as a surface-square GKP code, we consider correcting errors in the $\hat{p}$ subspace and  the following coset probability
\begin{align}
\label{eq:MLD_4}
\sum_{\boldsymbol{u}\in\sqrt{2\pi}\Lambda}P_\sigma(\boldsymbol{\eta_s}-\boldsymbol{u}),
\end{align}
where $\Lambda\equiv\Lambda^{(\hat{p})}$ and $\boldsymbol{\eta_s}\in\mathbb{R}^N$. As it is shown in App.~\ref{sec: Concatenated GKP code},  because $2Z_{N}\subset\sqrt{2}\Lambda$ (see Eq.~\eqref{eq:chain_subsets}), for any $\boldsymbol{u}\in\sqrt{2\pi}\Lambda$, it can be written as $\boldsymbol{u}=\sqrt{\pi}(\boldsymbol{g}^Z+\boldsymbol{v})$ where  $\boldsymbol{v}\in2Z_{N}$. As a result, the summation in Eq.~\eqref{eq:MLD_4} can be written as \cite{conrad2022gottesman} (we have omitted the irrelevant factor from the Gaussian distribution)
\begin{equation}
\label{eq:MLD_5}
\begin{aligned}
    &\sum_{\boldsymbol{g}^Z\in\mathcal{G}^Z}\sum_{\boldsymbol{v}\in2Z_N}P_\sigma(\boldsymbol{\eta}_{\boldsymbol{s}}-\boldsymbol{g}-\boldsymbol{v})\\
    =&\sum_{\boldsymbol{g}^Z\in\mathcal{G}^Z}\prod_{i=1}^{N}\sum_{{v}_i\in\mathbb{Z}}\exp\left(-\frac{(\eta_{s,i}-\sqrt{\pi}(g_i+2v_i))^2}{2\sigma^2}\right).
\end{aligned}
\end{equation}
Upon comparing Eq.~\eqref{eq:coset_prob_qubit} to Eq.~\eqref{eq:MLD_5}, we notice that they take a similar form which hints that it is possible to approximate the MLD for surface code with graphlike errors by considering the surface-square GKP code.
Indeed, suppose the displacement errors for the surface-square GKP code are discrete and defined as
\begin{align}
\label{eq:discrete_displacement_error}
    \boldsymbol{\eta}_{\boldsymbol{s}}=\sqrt{\pi}\boldsymbol{\eta}'
\end{align}
for certain $\boldsymbol{\eta}'\in\mathbb{F}_2^N$, then by setting $v_i=0$ and removing the summation in Eq.~\eqref{eq:MLD_5}, the coset probability for the surface-square GKP code reduces to
\begin{equation}
\label{eq:MLD_5_v2}
\begin{aligned}
    \sum_{\boldsymbol{g}^Z\in\mathcal{G}^Z}\prod_{i=1}^{N}\exp\left(-\frac{\pi(\eta'_{i}\ominus g^Z_i)}{2\sigma^2}\right).
\end{aligned}
\end{equation}
Here we have used the fact that $(\eta'_{i}- g^Z_i)^2 = (\eta'_{i}\ominus g^Z_i)$ because $\eta'_{i}, g_i^Z\in\mathbb{F}_2$. 
Comparing Eq.~\eqref{eq:MLD_5_v2} to Eq.~\eqref{eq:coset_prob_qubit} , it suggests that if the noise variance of the surface-square GKP code satisfies
\begin{align}
\label{eq:mapping_sigma_epsilon}
    \exp\left(-\frac{\pi}{2\sigma^2}\right) = \frac{\epsilon}{1-\epsilon},
\end{align}
then the GKP code would behave identically to its underlying qubit code, provided the former is subject to discrete displacement errors defined in Eq.~\eqref{eq:discrete_displacement_error}. We have used such mapping for simulating surface code in Sec.~\ref{sec: Qubit surface code with Z noise}.  

{We emphasize that the mappings presented in this section and App.~\ref{sec: Surface code subject to correlated errors as a special case of general surface-GKP code} should be regarded as a useful tool for decoding concatenated GKP codes and their corresponding qubit stabilizer codes within a unified framework. However, it does not imply that fidelities can be directly translated between the two families of codes, since their underlying error models are qualitatively different.
}

\section{Finding $K$ minimum weight  matchings for a decoding graph}
\label{sec:Finding the $K$-th matching for the decoding graph}

In this section, we demonstrate an efficient algorithm to find $K$ MWMs for a given decoding graph $G = (\mathcal{V}, \mathcal{E}, \mathcal{W}, \mathcal{V}_h)$. Specifically, we are interested in finding distinct matchings $\mathcal{M}_1, \mathcal{M}_2, .., \mathcal{M}_K$ such that 
\begin{align}
\label{eq:sequence_M}
    w(\mathcal{M}_1) \leq w(\mathcal{M}_2) \leq \cdots \leq w(\mathcal{M}_K) \leq w(\mathcal{M}) ,
\end{align}
for $\forall \mathcal{M}\neq \mathcal{M}_1, \cdots, \mathcal{M}_K$. The weight of a matching is defined in Eq.~\eqref{eq:weight_highlighted_matching_0}. Throughout this section, we will use $\mathcal{M}_i(G)$ (or simply $\mathcal{M}_i$ if there is no confusion from the context) to denote the $i$-th matching for the graph $G$. 
We emphasize that $\mathcal{M}_i$ and $\mathcal{M}_j$ are two distinct matchings if $i\neq j$, despite {the fact that} they may have the same weight. 
By the $k$-th MWM, we mean specifically a matching $\mathcal{M}_k$ that has weight larger than or equal to $(k-1)$ MWMs. 
Hence if $\mathcal{M}_i$ has the same weight as $\mathcal{M}_j$, we can refer to the first as the $j$-th matching and the latter as the $i$-th matching.
In other words, if two matchings in the sequence in Eq.~\eqref{eq:sequence_M} have the same weight, which also means the matchings between them have the same weight as them, then we can swap the two matchings, or any other matchings between them.
We also assume that we have access to an algorithm which outputs the MWM $\mathcal{M}_1(G)$ for the graph $G$ \cite{higgott2023sparse, wu2023fusion}.

We will first present how to find the second MWM for a decoding graph in Sec.~\ref{sec:Finding the second matching for the syndrome graph}, and then generalize it to $K$ MWMs in Sec.~\ref{sec:Finding K matchings for the decoding graph}. The correctness of the algorithm for finding the second MWM is presented in Sec.~\ref{sec: Proof for finding the second matching of a decoding graph}.

\subsection{Finding the second MWM}
\label{sec:Finding the second matching for the syndrome graph}

\begin{figure}
\centering
\includegraphics[width=\linewidth]{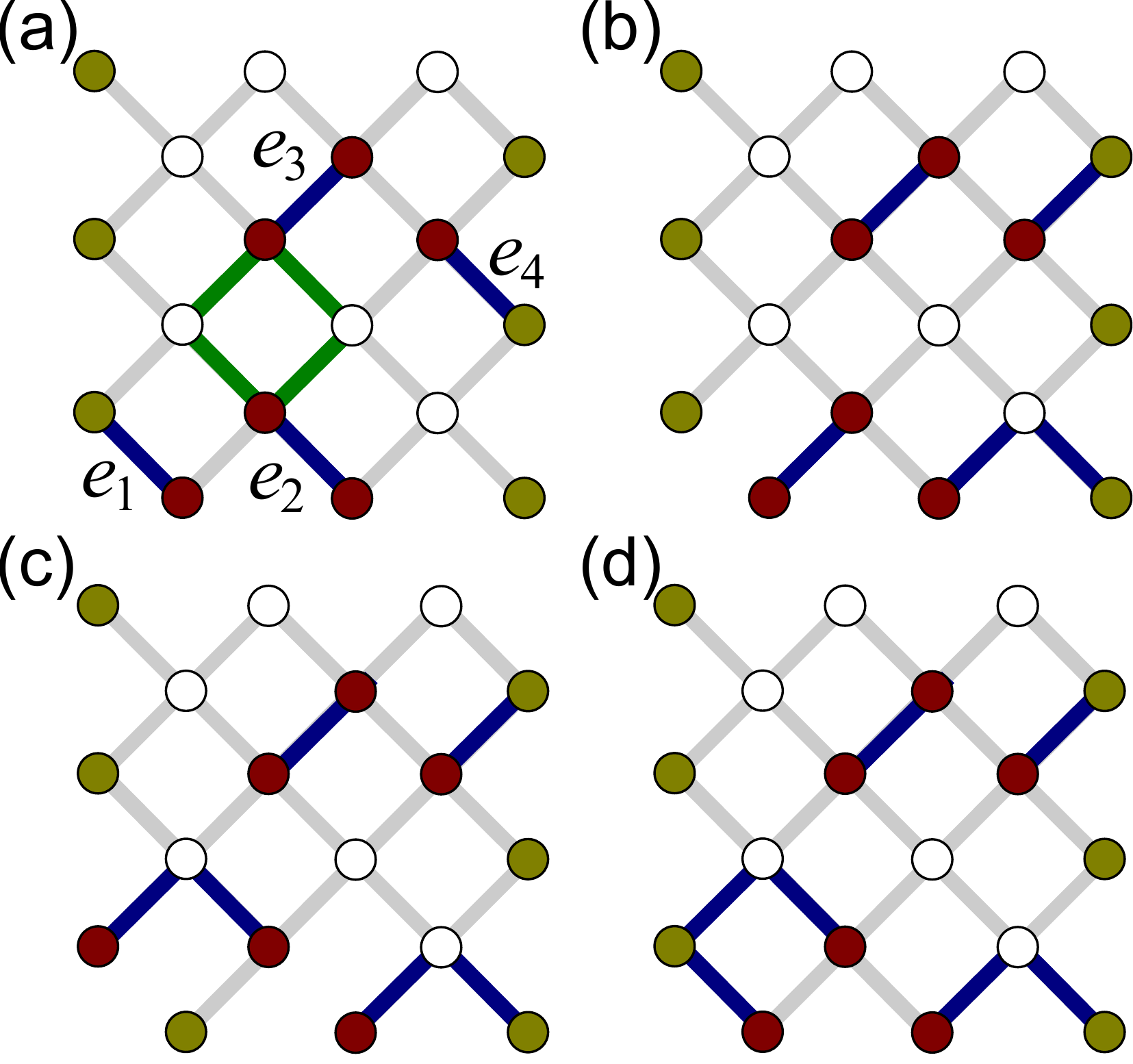}
 \caption{Illustrations of reduced decoding graphs, and their matchings.
 (a) Reproduction of the decoding graph $G$ in Fig.~\ref{fig:MWM}(d) with an ordering of the edges in $\mathcal{M}_1(G)$ included. 
 The set of green edges represents a cycle $\mathcal{C}$, 
 and $\mathcal{C}\cup \mathcal{M}_1(G)$ is one candidate for $\mathcal{M}_2(G)$.
 (b) An illustration of the reduced decoding graph $G^{(1)}$ defined in Eq.~\eqref{eq:def_G_j_1}-\eqref{eq:def_G_j_2}, obtained by removing $e_1$ from $G$ while retaining the same set of highlighted vertices.
 The set of blue edges represents $\mathcal{M}_1(G^{(1)})$, the MWM of the reduced graph $G^{(1)}$, which is another candidate for $\mathcal{M}_2(G)$.
 (c) An illustration of the reduced decoding graph $G^{(2)}$, which is obtained by removing $e_1$ and $e_2$ from $G$ and flipping the states of vertices in $e_1$.
 The set of blue edges represents $\mathcal{M}_1(G^{(2)})$, the MWM of the reduced graph $G^{(2)}$.
 (d) Another candidate for $\mathcal{M}_2(G)$ is obtained by taking the union of $e_1$ and $\mathcal{M}_1(G^{(2)})$, which matches the highlighted vertices in $G$ pairwise.
}
 \label{fig:reduced_graphs}
\end{figure}

We start by finding the second MWM for the decoding graph $G$, and the method outlined here will be used repeatedly to find subsequent MWMs.
Since $\mathcal{M}_1(G)$ is a subset of $\mathcal{E}$, we may write it as an ordered set of edges 
\begin{align}
\label{eq:def_M1}
    \mathcal{M}_1 = \left\{e_1, e_2, \cdots, e_{|\mathcal{M}_1|}\right\}.
\end{align}
{The ordering in Eq.~\eqref{eq:def_M1} is not essential to our discussion and we treat a matching as an ordered set merely to facilitate proper indexing of its elements.}
To determine $\mathcal{M}_2(G)$, we will need to examine a set of candidate matchings. Let $\mathcal{X}$ denotes such a set with $|\mathcal{M}_1|+1$ candidates.
For finding the $j$-th candidate, where $1\leq j\leq |\mathcal{M}_1|+1$, we construct the following reduced decoding graph
\begin{align}
\label{eq:def_G_j_1}
    G^{(j)} = (\mathcal{V}, \mathcal{E}^{(j)}, \mathcal{W}^{(j)}, \mathcal{V}_h^{(j)}),
\end{align}
which has the same set of vertices as $G$, but
\begin{align}
\label{eq:def_G_j_2}
    \mathcal{E}^{(j)} &=\begin{cases}
        \mathcal{E}\setminus\left\{e_1, \cdots, e_j\right\} & j \leq |\mathcal{M}_1|\\
        \mathcal{E}\setminus\mathcal{M}_1 & j = |\mathcal{M}_1|+1
    \end{cases},\nonumber\\
    \mathcal{W}^{(j)} &= \left\{w(e) ~|~ e\in \mathcal{E}^{(j)} \right\},\\
    \mathcal{V}_h^{(j)} &= \mathcal{V}_h  \oplus (\oplus_{l=1}^{j-1}e_l).\nonumber
\end{align}
In other words, $G^{(j)}$ is obtained by removing the first $j$ edges in $\mathcal{M}_1$, keeping the weights of the remaining edges unchanged, followed by modifying the highlighted vertices using the first $j-1$ edges. 
We note that both $G^{(|\mathcal{M}_1|)}$ and $G^{(|\mathcal{M}_1|+1)}$ have all the edges from $\mathcal{M}_1$ removed, but $\mathcal{V}_h^{(|\mathcal{M}_1|)}=e_{|\mathcal{M}_1|}$ whereas $\mathcal{V}_h^{(|\mathcal{M}_1|+1)}=\emptyset$.
We also note that $\mathcal{V}_h^{(1)}=\mathcal{V}_h$.
In Fig.~\ref{fig:reduced_graphs}(b-c), we illustrate the first two reduced graphs for the decoding graph in Fig.~\ref{fig:MWM}(d), which is reproduced in Fig.~\ref{fig:reduced_graphs}(a) with the ordering of the matching included. 
With the $j$-th reduced graph set up, 
the $j$-th candidate for $\mathcal{M}_2(G)$ is defined as
\begin{align}
\label{eq:M_1_G_j}
    \mathcal{M}_2^{(j)} = \mathcal{M}_1(G^{(j)})\cup (\cup_{l=1}^{j-1}e_l),
\end{align}
which is the union of the MWM of $G^{(j)}$ with the first $j-1$ edges in $\mathcal{M}_1(G)$.
We note that for the very last candidate with $j=|\mathcal{M}_1|+1$, the corresponding reduced graph has no highlighted vertex, and its matchings satisfy the condition
\begin{align}
\label{eq:def_cycle}
    \oplus_{e\in\mathcal{M}}e = \emptyset,
\end{align}
for $\mathcal{M}\equiv\mathcal{M}_1(G^{(|\mathcal{M}_1|+1)})$.
A matching satisfying Eq.~\eqref{eq:def_cycle} will be called a \emph{cycle} of the graph.
We have made an important choice of convention in  Eq.~\eqref{eq:M_1_G_j} that \emph{the MWM of a decoding graph with no highlighted vertex corresponds to its minimum weight cycle (MWC)}. 
Another choice would be to treat the MWM of the decoding graph with no highlighted vertex as an empty set, and the second MWM as the MWC, but it turns out that the convention chosen in Eq.~\eqref{eq:M_1_G_j} is more convenient for subsequent discussions.
The first two candidates for $\mathcal{M}_2(G)$ are illustrated in Fig.~\ref{fig:reduced_graphs}(b) and (d) respectively, and we show an example of a cycle in Fig.~\ref{fig:reduced_graphs}(a).

We claim that the second MWM $\mathcal{M}_2(G)$ has to be in the following set of candidates
\begin{align}
\label{eq:M2_candidates_main}
    \mathcal{X} = \left\{\mathcal{M}_2^{(1)}, \cdots, \mathcal{M}_2^{(|\mathcal{M}_1|)}, \mathcal{M}_2^{(|\mathcal{M}_1|+1)}\right\}.
\end{align}
This is one of the main results of the paper, and we will provide the proof in Sec.~\ref{sec: Proof for finding the second matching of a decoding graph}. Here we will use a simple case with $\mathcal{M}_1 = \left\{e_1\right\}$ to illustrate its validity. The essential idea of finding  $\mathcal{M}_2(G)$ is that it either shares or does not share the edge $e_1$ with $\mathcal{M}_1(G)$, and we will {discuss} each case separately below.

If $e_1\notin\mathcal{M}_2(G)$, we have $\mathcal{M}_2(G)=\mathcal{M}_1(G^{(1)})$.
To see that, we note that any matching of $G^{(1)}$, including $\mathcal{M}_1(G^{(1)})$, is a matching of $G$, because the two graphs share the same set of highlighted vertices by definition. 
This leads to $w(\mathcal{M}_1(G^{(1)}))\geq w(\mathcal{M}_1(G))$ by the fact that $\mathcal{M}_1(G)$ is the MWM of $G$.
Conversely, $\mathcal{M}_2(G)$ is also a matching of $G^{(1)}$ because, by our assumption of $\mathcal{M}_2(G)$, the former matches all the highlighted vertices of $G^{(1)}$ without using the edge $e_1$. Combining these two facts, we have
\begin{align*}
    w(\mathcal{M}_2(G))\geq w(\mathcal{M}_1(G^{(1)}))\geq w(\mathcal{M}_1(G)).
\end{align*}
Because $e_1\notin\mathcal{M}_1(G^{(1)})$, it is a distinct matching of $G$ compared to $\mathcal{M}_1(G)$, which implies that $\mathcal{M}_2(G)$ has to be an identical matching as  $\mathcal{M}_1(G^{(1)})$, otherwise it will contradict the fact that $\mathcal{M}_2(G)$ is the second MWM of $G$. 
Hence we have $\mathcal{M}_2(G)=\mathcal{M}_1(G^{(1)})$, the first element in $\mathcal{X}$ (see Eq.~\eqref{eq:M_1_G_j} and \eqref{eq:M2_candidates_main}). We note that if $\mathcal{M}_1(G^{(1)})$ is not unique, or multiple matchings of $G^{(1)}$ have the same weight as $\mathcal{M}_1(G^{(1)})$, we can simply select one of them as $\mathcal{M}_2(G)$.

If $e_1\in\mathcal{M}_2(G)$, then it implies $\mathcal{M}_1(G)\subset\mathcal{M}_2(G)$ {because} $\mathcal{M}_1 = \left\{e_1\right\}$. Let 
\begin{align}
    \mathcal{C} \equiv  \mathcal{M}_2(G)\setminus\mathcal{M}_1(G)
\end{align}
be the set of edges that are in $\mathcal{M}_2(G)$ but not in $\mathcal{M}_1(G)$.
Since both $\mathcal{M}_1(G)$ and $\mathcal{M}_2(G)$ are matchings for the decoding graph, by definition (see Eq.~\eqref{eq:def_matching}) we have
\begin{align}
     \oplus_{e\in \mathcal{C}}e 
     = \oplus_{e\in \mathcal{M}_2 \setminus \mathcal{M}_1}e 
     = (\oplus_{e\in \mathcal{M}_2}e)\oplus(\oplus_{e\in \mathcal{M}_1}e)
     = \emptyset,
\end{align}
which suggests that $\mathcal{C}$ is a cycle in $G$. Since $\mathcal{M}_2(G)=\mathcal{C}\cup\mathcal{M}_1(G)$, and $e_1\notin\mathcal{C}$, in order to minimize $w(\mathcal{M}_2(G))$, we arrive that $\mathcal{C}$ is the MWC of the reduced graph $G^{(2)}$, i.e., $\mathcal{C}=\mathcal{M}_1(G^{(2)})$.
In Appendix \ref{sec: Minimum weight cycle of a graph}, we illustrate an algorithm to find the MWC of a graph, which can be thought of as a subroutine for the algorithm MWM($G$) when the decoding graph has no highlighted vertex.

Upon combining the results for these two cases, indeed, we see that $\mathcal{M}_2(G)$ has to be in the set $\mathcal{X} = \left\{\mathcal{M}_2^{(1)}, \mathcal{M}_2^{(2)}\right\}$, as defined in Eq.~\eqref{eq:M2_candidates_main}, for the simple case $\mathcal{M}_1 = \left\{e_1\right\}$.

Before proceeding, we remark that because a cycle consists of at least three edges, we have $|\mathcal{M}_1(G)|\geq1$ for arbitrary decoding graphs, regardless of whether $\mathcal{V}_h=\emptyset$. Further, our constructions of $G^{(j)}$ and $\mathcal{M}_2^{(j)}$ in Eq.~\eqref{eq:def_G_j_1}-\eqref{eq:M_1_G_j} remain valid even if $G$ has no highlighted vertex. The only difference is that if $\mathcal{V}_h=\emptyset$, then $\mathcal{V}_h^{1}=\mathcal{V}_h^{(|\mathcal{M}_1(G)|+1)}=\emptyset$, whereas if $\mathcal{V}_h\neq\emptyset$, then we have only $\mathcal{V}_h^{(|\mathcal{M}_1(G)|+1)}=\emptyset$. 

\subsection{Generalization to $K$ MWMs}
\label{sec:Finding K matchings for the decoding graph}

\begin{figure}[!ht]
\centering
\includegraphics[width=\linewidth]{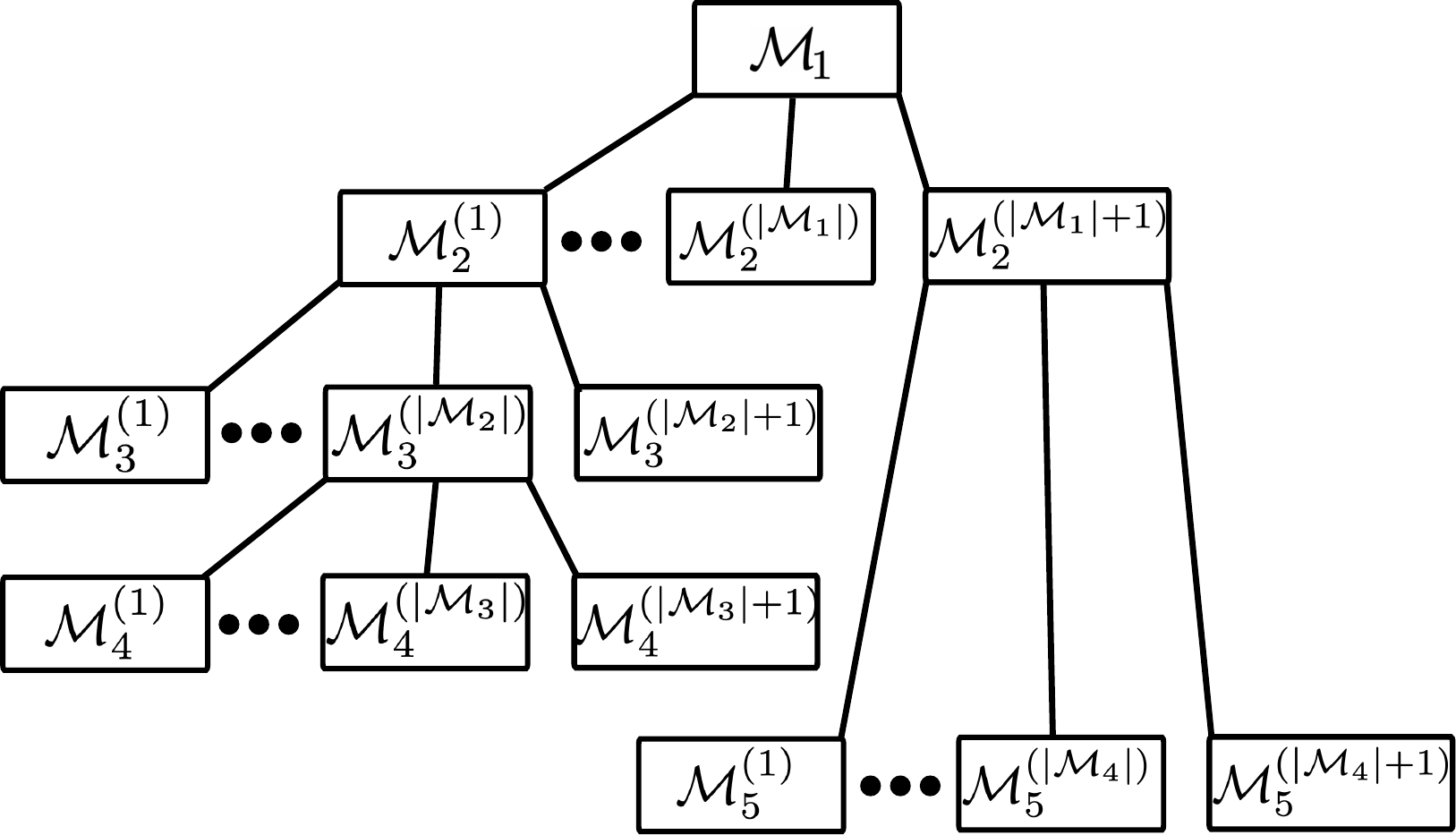}
 \caption{An example decoding tree for a decoding graph $G$. 
 The root of the decoding tree is $\mathcal{M}_1(G)$, the MWM of the graph, and all of its {descendants} are subsequent  matchings of the graph.
 As defined in Eq.~\eqref{eq:general_M}, each node has three pieces of information: the corresponding matching $\mathcal{M}$, the corresponding reduced graph $G'$ and the edges $\mathcal{E}''$ needed for matching completion.
 {To identify the children of a node, which are defined in Eq.~\eqref{eq:def_children_M}, we follow the protocol in Sec.~\ref{sec:Finding the second matching for the syndrome graph} to find the candidates of $\mathcal{M}_2(G')$ followed by using $\mathcal{E}''$ for matching completion. The number of children of a node is given by the number of candidates of $\mathcal{M}_2(G')$.
 Each level of the tree consists only of the child nodes from a parent node and the $k$-th level of the tree has at most $|\mathcal{M}_{k-1}(G)|+1$ nodes.
 }
 The leaf nodes up to the $k$-th level (nodes with no {descendants} if we cut the tree between the $k$-th and $(k+1)$-th levels) form the set of all the candidates for $\mathcal{M}_k(G)$. 
}
 \label{fig:decoding_tree}
\end{figure}

The method of finding the second MWM can be adopted to find all subsequent MWMs for the decoding graph $G$. 
Interestingly, finding subsequent matchings for a decoding graph $G$ can be visualized using a tree structure, as shown in Fig.~\ref{fig:decoding_tree}, which will be referred to as the \emph{decoding tree}. 
The root of the decoding tree is $\mathcal{M}_1(G)$ of the  graph, and its children are the candidates for $\mathcal{M}_2(G)$ defined in Eq.~\eqref{eq:M2_candidates_main}. 
The decoding tree will have $K$ levels after the first $K$ MWMs have been found. 
In view of Eq.~\eqref{eq:M_1_G_j}, each node of the tree corresponds to a matching of the form
\begin{align}
\label{eq:general_M}
   \mathcal{M} =  \mathcal{M}_1(G')\cup\mathcal{E}'',
\end{align}
where $\mathcal{M}_1(G')$ is the MWM for a certain reduced graph 
\begin{align}
\label{eq:def_G'_decoding_tree}
    G'=(\mathcal{V}, \mathcal{E}', \mathcal{W}', \mathcal{V}'_h),    
\end{align}
and $\mathcal{E}''$ is a subset of edges with $\mathcal{E}'\cap\mathcal{E}''=\emptyset$.
Because {$\mathcal{M}_1(G')$} is a matching of $G'$, instead of the original $G$, we take the union of $\mathcal{M}_1(G')$ and $\mathcal{E}''$ to match the highlighted vertices in $G$ pairwise.  
In other words, we have
\begin{align}
    \mathcal{V}_h'\oplus(\oplus_{e\in\mathcal{E}''}e) = \mathcal{V}_h.
\end{align}
We will refer to the process of adding edges to another matching as \emph{matching completion}. As an example, we remark that $\mathcal{M}_1(G)$ is a special case of Eq.~\eqref{eq:general_M} with $(G', \mathcal{E}'')=(G, \emptyset)$. Similarly, Eq.~\eqref{eq:M_1_G_j} is also a special case of Eq.~\eqref{eq:general_M} because we need to add a set of edges to a matching of a reduced decoding graph to obtain a valid matching for the original graph.

For a node in the decoding tree as defined in Eq.~\eqref{eq:general_M}, its children are identified by finding the second MWM of $G'$. Let
\begin{align}
    \mathcal{M}_1(G') = \left\{e_1, e_2, \cdots, e_{|\mathcal{M}_1(G')|}\right\},
\end{align}
we define the $j$-th child of the node as
\begin{align}
\label{eq:def_children_M}
    \mathcal{M}^{(j)} = \left(\mathcal{M}_1(G'^{(j)})\cup (\cup_{l=1}^{j-1}e_l)\right)\cup\mathcal{E}'',
\end{align}
where $1\leq j\leq |\mathcal{M}_1(G')|+1$ and $G'^{(j)}$ is the $j$-th reduced graph of $G'$ as defined in Eq.~\eqref{eq:def_G_j_2}.
We note that the term in the parentheses of $\mathcal{M}^{(j)}$ is nothing but the $j$-th candidate for $\mathcal{M}_2(G')$, per Eq.~\eqref{eq:M_1_G_j}.
In other words, the children of a node $\mathcal{M}$, as defined in Eq.~\eqref{eq:general_M}, can be obtained by finding the candidates for $\mathcal{M}_2(G')$ followed by taking the union with $\mathcal{E}''$ individually.
$\mathcal{M}^{(j)}$ is indeed a node in the decoding tree because it takes the same matching completion form as the definition in Eq.~\eqref{eq:general_M}.
{From the above discussion, we see that the nodes in the decoding tree (except its root) always have weights larger than or equal to their parents, a desired property that enables searching the $K$ MWMs efficiently.}

To proceed, let $\mathcal{X}_K=\left\{\mathcal{M}_1(G), \cdots, \mathcal{M}_K(G)\right\}$ be the set of $K$ MWMs of $G$ and $\mathcal{X}$ be the set of explored matchings for identifying $\mathcal{X}_K$. The set $\mathcal{X}$ contains all the children of $\mathcal{M}_k\in\mathcal{X}_K$, for $1\leq k\leq K-1$, that are not in $\mathcal{X}_K$. 
Taking $K=2$ as an example, we have $\mathcal{X}_K=\left\{\mathcal{M}_1, \mathcal{M}_2\right\}$ and $\mathcal{X}=\left\{\mathcal{M}_2^{(1)}, \cdots, \mathcal{M}_2^{(|\mathcal{M}_1|)}, \mathcal{M}_2^{(|\mathcal{M}_1|+1)}\right\}\setminus\left\{\mathcal{M}_2\right\}$, per Eq.~\eqref{eq:M2_candidates_main}, which contains all the children of $\mathcal{M}_1$ that are not in $\mathcal{X}_K$.
In order to find $\mathcal{M}_{K+1}(G)$, we add the children of $\mathcal{M}_{K}(G)$, denoted as $\mathcal{M}_{K+1}^{(j)}(G)$ into $\mathcal{X}$, if they are not already in $\mathcal{X}$ or $\mathcal{X}_K$. 
Hence $\mathcal{X}$ is updated as 
\begin{align}
\label{eq:def_X'}
    \mathcal{X}\rightarrow \mathcal{X}' \equiv \mathcal{X} \cup\left\{\mathcal{M}_{K+1}^{(j)}\right\}\setminus\mathcal{X}_K.
\end{align}
{We note that it is possible that some children of $\mathcal{M}_K$ have appeared in the decoding tree, and it is important to exclude them when expanding the tree, as done in Eq.~\eqref{eq:def_X'}.}
We claim that $\mathcal{M}_{K+1}(G)$ must be in the set $\mathcal{X}'$, which is the main result of this section. 

The proof of the claim is provided below. Given the definition of a node in Eq.~\eqref{eq:general_M}, $\mathcal{M}_{K+1}(G)$ must take the following form 
\begin{align}
    \mathcal{M}_{K+1}(G) =  \mathcal{M}_1(G')\cup\mathcal{E}''
\end{align}
for a certain $G'$ as defined in Eq.~\eqref{eq:def_G'_decoding_tree}. We are interested in the case where $G'\neq G$ and hence a non-empty $\mathcal{E}''$ is needed for matching {completion}. Suppose $\mathcal{E}''$ contains at least one edge, say $e'_1$.
Let us now construct a graph $\tilde{G}=(\mathcal{V}, \tilde{\mathcal{E}}, \tilde{\mathcal{W}}, \tilde{\mathcal{V}}_h)$ with
\begin{align}
    \tilde{\mathcal{E}} &=\left\{e'_1\right\}\cup\mathcal{E}',\nonumber\\
    \tilde{\mathcal{W}} &= \left\{w(e) ~|~ e\in \tilde{\mathcal{E}} \right\},\\
    \tilde{\mathcal{V}}_h &= \mathcal{V}'_h,\nonumber
\end{align}
and consider the following matching
\begin{align}
    \mathcal{M}_K^*\equiv\mathcal{M}_1(\tilde{G})\cup\mathcal{E}''.
\end{align}
We notice that the first reduced graph $\tilde{G}^{(1)}$, as defined in Eq.~\eqref{eq:def_G_j_2} with respect to $\tilde{G}$, is precisely $G'$. Using Eq.~\eqref{eq:def_children_M}, we find that the first child of $\mathcal{M}_K^*$ is precisely $\mathcal{M}_{K+1}(G)$. Hence it must be the case that $w(\mathcal{M}_K^*)\leq w(\mathcal{M}_{K+1}(G))$ and $\mathcal{M}_K^*$  must be in the set of the first $K$ MWMs, i.e., $\mathcal{M}_K^*\in\mathcal{X}_K$. Because $\mathcal{M}_{K+1}(G)$ is a child of an element in $\mathcal{X}_K$, by the construction of $\mathcal{X}_K$ and $\mathcal{X}'$, $\mathcal{M}_{K+1}(G)$ has to be in either of these two sets. 
Because $\mathcal{X}_K$ only has the first $K$ MWMs, i.e., $\mathcal{M}_{K+1}(G)\notin\mathcal{X}_K$, we have $\mathcal{M}_{K+1}(G)\in\mathcal{X}'$, as claimed.

\begin{algorithm}
\SetAlgoLined
\LinesNumbered
\caption{MWMs($G, K$)}
\label{alg: MWMs_v1}
{\bf Input: } The decoding graph $G$ and $K\in\mathbb{N}$; \\ 
{\bf Output: } $K$ MWMs $\mathcal{X}_K \equiv\left\{\mathcal{M}_1, \cdots, \mathcal{M}_K\right\}$; \\ 
$\mathcal{M}_1 \leftarrow $ \text{MWM}($G$) \\
$\mathcal{X}_{K} \leftarrow \left\{\mathcal{M}_1\right\}$ // The set of found MWMs \\ 
$\mathcal{X} \leftarrow \left\{(\mathcal{M}_1, G, \emptyset)\right\}$ // The set of explored candidates\\ 
\For{$2\leq k \leq K$}{
 $(\mathcal{M}_{k-1}, G', \mathcal{E}'') \leftarrow \text{Pop $\mathcal{M}_{k-1}$ from } \mathcal{X}$ \\
$\mathcal{M}_1(G') \leftarrow \mathcal{M}_{k-1}\setminus\mathcal{E}''$ // Defined in Eq.~\eqref{eq:general_M}\\
$\mathcal{V}', \mathcal{E}', \mathcal{W}', \mathcal{V}'_h\leftarrow G'$ \\
$\left\{e_1, \cdots, e_{|\mathcal{M}_1(G')|}\right\}\leftarrow \mathcal{M}_1(G')$ \\
\For{$1\leq j \leq |\mathcal{M}_1(G')|+1$}{
    \eIf{$1\leq j\leq|\mathcal{M}_1(G')|$}
    {
        $\mathcal{E}'^{(j)}\leftarrow \mathcal{E}'\setminus\left\{e_1, \cdots, e_j\right\}$ \\
    }
    {
        $\mathcal{E}'^{(j)}\leftarrow \mathcal{E}'\setminus\mathcal{M}_1(G')$ \\
    }
    $\mathcal{W}'^{(j)}\leftarrow \left\{w(e) ~|~ e\in \mathcal{E}'^{(j)} \right\}$ \\
    $\mathcal{V}_h'^{(j)}\leftarrow \mathcal{V}'_h  \oplus (\oplus_{l=1}^{j-1}e_l)$ \\
    $G'^{(j)} \leftarrow (\mathcal{V}', \mathcal{E}'^{(j)}, \mathcal{W}'^{(j)}, \mathcal{V}_h'^{(j)})$ \\
    $\mathcal{E}''^{(j)}\leftarrow\mathcal{E}''\cup (\cup_{l=1}^{j-1}e_l)$ // Edges for completion\\
    $ \mathcal{M}^{(j)}\leftarrow\text{MWM}(G'^{(j)})\cup \mathcal{E}''^{(j)}$ // Defined in Eq.~\eqref{eq:def_children_M}\\
    \If{$\mathcal{M}^{(j)}$ \text{exists and} $\mathcal{M}^{(j)}\notin\mathcal{X}_{K}\cup\mathcal{X}$}
    {
        $\text{Put } (\mathcal{M}^{(j)}, G'^{(j)}, \mathcal{E}''^{(j)}) \text{ into } \mathcal{X}$
    }
 }
 $\mathcal{M}_{k} \leftarrow \text{The element in } \mathcal{X} \text{ with minimum weight}$ \\
 $\text{Put } \mathcal{M}_{k} \text{ into } \mathcal{X}_{K}$
 }
\end{algorithm}

We have described a procedure to grow a decoding tree, starting from the MWM of a decoding graph as its root, for the purpose of finding $K$ MWMs {that satisfy Eq.~\eqref{eq:sequence_M} (equivalently Eq.~\eqref{eq:sequence_M_0}). 
The intuition behind the procedure is that the $(K+1)$-th MWM has to be either a child of the $K$-th MWM or one of the MWMs already explored during the search process of the first $K$ MWMs.
By construction, the nodes in the tree are distinct from each other and they always have weights larger than or equal to those of their parent nodes. Due to these observations, the $K$ MWMs of interest are guaranteed to be in the decoding tree, which can be identified efficiently.
}

The above discussion is summarized in Algorithm \ref{alg: MWMs_v1}, where we present MWMs($G, K$) for finding the first $K$ MWMs of a given decoding graph $G$. In the algorithm, the explored candidates are saved in a priority queue $\mathcal{X}$ and each element in $\mathcal{X}$ consists of three pieces of information, namely the candidate matching $\mathcal{M}$, its reduced graph $G'$ and the edges $\mathcal{E}''$ needed for matching completion as defined in Eq.~\eqref{eq:general_M}. 
In terms of the decoding tree shown in Fig.~\ref{fig:decoding_tree}, each node corresponds to $(\mathcal{M}, G', \mathcal{E}'')$.
We note that the MWM $\mathcal{M}_1(G)$ of a given decoding graph can be found using either the algorithms in \cite{higgott2023sparse, wu2023fusion}, if the decoding graph has a non-empty set of highlighted vertices, or MWC($G$) in Appendix \ref{sec: Minimum weight cycle of a graph}, if the graph has no highlighted vertex.
In both cases, finding $\mathcal{M}_1(G)$ has time complexity that is bounded by polynomials in the number of edges and vertices, hence the time complexity of finding the first $K$ MWMs is given by
\begin{align}
\label{eq:complexity_mwms}
    O(K\times \text{poly}(|\mathcal{V}|, |\mathcal{E}|)).
\end{align}
{As an example, in In App.~\ref{sec: An example of decoding with $K$ MWMs} we apply Algorithm \ref{alg: MWMs_v1} to a six-qubit code, verifying that it correctly finds all MWMs of the decoding graph.}

\subsection{Proof of the algorithm for finding the second MWM}
\label{sec: Proof for finding the second matching of a decoding graph}

Here we prove the main result of this paper, which is the algorithm in Sec.~\ref{sec:Finding the second matching for the syndrome graph}.
For simplicity, let us refer to the algorithm as SecondMWM($G, \mathcal{M}_1(G)$) which finds the second MWM for a given $G$ and its first MWM. We will prove the algorithm via induction by showing that it is valid regardless of the size of $\mathcal{M}_1$.

Recall that we have chosen the convention that if $\mathcal{V}_h=\emptyset$ for $G$, then the MWM of the graph is its MWC. Hence we have  $|\mathcal{M}_1(G)|\geq1$ regardless of whether there is highlighted vertex in the decoding graph. 
Since we have shown the validity of SecondMWM($G, \mathcal{M}_1$) for the case $|\mathcal{M}_1(G)|=1$ in Sec.~\ref{sec:Finding the second matching for the syndrome graph},  let us assume SecondMWM($G$, $\mathcal{M}_1$) is valid for $|\mathcal{M}_1(G)|= m\geq1$. We aim to show that the algorithm is also valid for $|\mathcal{M}_1(G)|=m+1$. In this case, $\mathcal{M}_1(G)$ is non-empty and it has at least one edge $e_1$. 
Similar to the argument in Sec.~\ref{sec:Finding the second matching for the syndrome graph}, the essential idea of finding $\mathcal{M}_2(G)$ is that it either shares or does not share the edge $e_1$ with $\mathcal{M}_1(G)$. 
For the case where $e_1\notin\mathcal{M}_2(G)$, the argument is exactly the same as that presented in Sec.~\ref{sec:Finding the second matching for the syndrome graph}, and it leads to $\mathcal{M}_2(G)=\mathcal{M}_1(G^{(1)})$ which is the first element in $\mathcal{X}$, as defined in Eq.~\eqref{eq:M2_candidates_main}. 

Suppose $e_1\in\mathcal{M}_2(G)$, let us first define an auxiliary graph $G'=(\mathcal{V}, \mathcal{E}', \mathcal{W}', \mathcal{V}'_h)$ with
\begin{align}
\label{eq:def_G'}
    \mathcal{E}' &= \mathcal{E}\setminus\left\{e_1\right\}, \nonumber \\
    \mathcal{W}' &= \left\{w(e) ~|~ e\in \mathcal{E}' \right\},\\
    \mathcal{V}_h’ &= \mathcal{V}_h  \oplus e_1 .\nonumber
\end{align}
We emphasize that $G'$ is not the same as the reduced graph $G^{(1)}$ or $G^{(2)}$ defined in Eq.~\eqref{eq:def_G_j_1}-\eqref{eq:def_G_j_2}. We assert that  
\begin{align}
    \mathcal{M}_1(G) &=\left\{e_1\right\} \cup \mathcal{M}_1(G') \label{eq:m1},\\
    \mathcal{M}_2(G) &=\left\{e_1\right\} \cup \mathcal{M}_2(G') \label{eq:m2}.
\end{align}
We will prove these assertions below, and for now let us understand their implications to the case where $e_1\in\mathcal{M}_2(G)$. 
From Eq.~\eqref{eq:m1} and Eq.~\eqref{eq:def_M1}, we have that 
\begin{align}
\label{eq:M_1_G'}
    \mathcal{M}_1(G')=\left\{e_2, \cdots, e_{|\mathcal{M}_1|}\right\}  \equiv\left\{e'_1, \cdots, e'_{|\mathcal{M}_1(G')|}\right\}
\end{align}
where $|\mathcal{M}_1(G')|=m$. In Eq.~\eqref{eq:M_1_G'}, we have relabelled the edges in $\mathcal{M}_1(G')$ which will be used below.
With the assumption that SecondMWM($G$, $\mathcal{M}_1(G)$) is valid for graphs with $|\mathcal{M}_1(G)|=m$, we can apply the algorithm to the graph $G'$, and SecondMWM($G'$, $\mathcal{M}_1(G')$) produces a set of candidates for $\mathcal{M}_2(G')$ defined as (see Eq.~\eqref{eq:M_1_G_j})
\begin{equation}
\label{eq:M2_candidates_2_v2}
{\mathcal{M}'_2}^{(j)} = \mathcal{M}_1(G'^{(j)})\cup (\cup_{l=1}^{j-1}e'_l), ~\text{for}~ 1\leq j\leq m+1.
\end{equation}
Here $G'^{(j)} = (\mathcal{V}, {\mathcal{E}'}^{(j)}, {\mathcal{W}'}^{(j)}, {\mathcal{V}'}_h^{(j)})$ is a reduced decoding graphs defined with respect to $G'$. Using Eq.~\eqref{eq:def_G_j_2}-\eqref{eq:def_cycle} and Eq.~\eqref{eq:def_G'}-\eqref{eq:m1}, we have 
\begin{align}
\label{eq:def_G_j_2_app_v2}
    \mathcal{E}'^{(j)} &=\begin{cases}
        \mathcal{E}'\setminus\left\{e'_1, \cdots, e'_j\right\} & j \leq m\\
        \mathcal{E}'\setminus\mathcal{M}_1(G') & j = m+1
    \end{cases}\nonumber\\
    &=\begin{cases}
        \mathcal{E}'\setminus\left\{e_2, \cdots, e_{j+1}\right\} & j \leq m\\
        \mathcal{E}'\setminus\mathcal{M}_1(G') & j = m+1
    \end{cases}\nonumber\\
    &=\begin{cases}
        \mathcal{E}\setminus\left\{e_1, \cdots, e_{j+1}\right\} & j \leq m\\
        \mathcal{E}\setminus\mathcal{M}_1(G) & j = m+1
    \end{cases}\nonumber\\
    &=\mathcal{E}^{(j+1)},\nonumber\\
    {\mathcal{W}'}^{(j)} &= \left\{w(e) ~|~ e\in {\mathcal{E}'}^{(j)} \right\}= \left\{w(e) ~|~ e\in {\mathcal{E}}^{(j+1)} \right\},\nonumber\\
    {\mathcal{V}'_h}^{(j)} &= \mathcal{V}'_h  \oplus e'_1 \oplus e'_2 \oplus \cdots \oplus e'_{j-1}\nonumber\\
    &= \mathcal{V}'_h  \oplus e_2 \oplus e_3 \oplus \cdots \oplus e_{j}\nonumber\\
    &= \mathcal{V}_h  \oplus e_1 \oplus e_2 \oplus \cdots \oplus e_{j}\nonumber\\
    &={\mathcal{V}_h}^{(j+1)},\nonumber
\end{align}
which implies that $G'^{(j)}$ is the same graph as $G^{(j+1)}$ defined in Eq.~\eqref{eq:def_G_j_2} with $1\leq j\leq m+1=|\mathcal{M}_1(G)|$. 
As a result, we have, from Eq.~\eqref{eq:M2_candidates_2_v2}, that 
\begin{equation}
    \begin{aligned}
        {\mathcal{M}'_2}^{(j)} &= \mathcal{M}_1(G^{(j+1)})\cup (\cup_{l=2}^{j}e_l)\\
        &={\mathcal{M}_2}^{(j+1)}\setminus\left\{e_1\right\},
    \end{aligned}
\end{equation}
where ${\mathcal{M}_2}^{(j+1)}$ is defined in Eq.~\eqref{eq:M_1_G_j}.
In other words, if $e_1\in\mathcal{M}_2(G)$, the candidates for the latter are the second to the $(|\mathcal{M}_1(G)|+1)$-th candidates in the set $\mathcal{X}$ defined in Eq.~\eqref{eq:def_G_j_2}-\eqref{eq:M_1_G_j}, which can be further found using SecondMWM($G'$, $\mathcal{M}_1(G')$), followed by taking the union with $\left\{e_1\right\}$ individually, per Eq.~\eqref{eq:m2}.

Combining the two cases discussed above, we have proved the following: If SecondMWM($G$, $\mathcal{M}_1(G)$) can find $\mathcal{M}_2(G)$ in the case of $|\mathcal{M}_1|=m\geq1$, then it can also find $\mathcal{M}_2(G)$ in the case of $|\mathcal{M}_1|=m+1$. Since the case with $|\mathcal{M}_1|=1$ is proved in Sec.~\ref{sec:Finding the second matching for the syndrome graph}, using induction, we have that the algorithm is valid for all cases of $|\mathcal{M}_1|$.

To complete the proof above, we now prove the assertions Eq.~\eqref{eq:m1}-\eqref{eq:m2}.
Let 
\begin{align}
    \mathcal{M}'_i &\equiv \mathcal{M}_i(G) \setminus \left\{e_1\right\},\label{eq:M'_1}\\
    \mathcal{M}''_i &\equiv \mathcal{M}_i(G') \cup \left\{e_1\right\},\label{eq:M''_1}
\end{align}
where $i=1,2$.
We note that $\mathcal{M}'_i$ are matchings for the graph $G'$ because
\begin{equation}
\label{eq:M_1'_is_a_matching_G'}
\begin{aligned}
    \oplus_{e\in\mathcal{M}'_i}e \stackrel{\text{Eq.~\eqref{eq:M'_1}}}{=} \oplus_{e\in\mathcal{M}_i(G) \setminus \left\{e_1\right\}}e 
    = \mathcal{V}_h \oplus e_1 
    \stackrel{\text{Eq.~\eqref{eq:def_G'}}}{=} \mathcal{V}_h',
\end{aligned}    
\end{equation}
which implies, by the definition of $\mathcal{M}_1(G')$, that
\begin{align}
\label{eq:M_1' > M_1G'}
    w(\mathcal{M}_i') &\geq w(\mathcal{M}_1(G')), \quad i=1,2.
\end{align}
Upon setting $i=1$ in Eq.~\eqref{eq:M'_1}-\eqref{eq:M_1' > M_1G'}, we have
\begin{align}
\label{eq:M_1_M_1''}
    w(\mathcal{M}_1(G)) 
    &\stackrel{\text{Eq.~\eqref{eq:M'_1}}}{=} w(\mathcal{M}_1') + w(e_1) \nonumber\\
    &\stackrel{\text{Eq.~\eqref{eq:M_1' > M_1G'}}}{\geq} w(\mathcal{M}_1(G')) + w(e_1) \\
    &\stackrel{\text{Eq.~\eqref{eq:M''_1}}}{=} w(\mathcal{M}_1'').\nonumber
\end{align}
We further notice that $\mathcal{M}_i''$ are also matchings of $G$ because
\begin{align}
\label{eq:M_1''_is_a_matching_G}
    \oplus_{e\in\mathcal{M}''_i}e \stackrel{\text{Eq.~\eqref{eq:M''_1}}}{=} \oplus_{e\in\mathcal{M}_i(G') \cup \left\{e_1\right\}}e = \mathcal{V}_h’ \oplus e_1 \stackrel{\text{Eq.~\eqref{eq:def_G'}}}{=} \mathcal{V}_h,
\end{align}
which implies, by the definition of $\mathcal{M}_1(G)$, that
\begin{align}
\label{eq:M_1' > M_1G}
    w(\mathcal{M}_i'') &\geq w(\mathcal{M}_1(G)), \quad i=1,2.
\end{align}
Upon combining Eq.~\eqref{eq:M_1' > M_1G} with \eqref{eq:M_1_M_1''}, we have $w(\mathcal{M}_1'') = w(\mathcal{M}_1(G))$ and hence we can identify $\mathcal{M}_1''$ as $\mathcal{M}_1(G)$, and setting $i=1$ in Eq.~\eqref{eq:M''_1} reproduces precisely Eq.~\eqref{eq:m1}. We note that the MWM of a graph needs not be unique, and if there are multiple (first) MWMs for $G$, Eq.~\eqref{eq:m1} is one of them.

To prove Eq.~\eqref{eq:m2}, we notice that because $\mathcal{M}_1(G')$ and $\mathcal{M}_2(G')$ are two different matchings for $G'$, upon comparing Eq.~\eqref{eq:m1} to \eqref{eq:M''_1} ($i=2$), we have that $\mathcal{M}_1(G)$ and $\mathcal{M}_2''$ are two different matchings for $G$. 
Hence if $w(\mathcal{M}_2(G)) > w(\mathcal{M}_2'')$, together with Eq.~\eqref{eq:M_1' > M_1G} for $i=2$, it would imply $\mathcal{M}_2(G)$ has weight larger than two distinct matchings in $G$, which contradicts the fact that $\mathcal{M}_2(G)$ is the second MWM of $G$. Hence we arrive at $w(\mathcal{M}_2(G)) \leq w(\mathcal{M}_2'')$. 
But if $w(\mathcal{M}_2(G)) < w(\mathcal{M}_2'')$, it implies
\begin{align}
\label{eq:M_2G > M_2G'+e1}
    w(\mathcal{M}_2(G)) < w(\mathcal{M}_2'') \stackrel{\text{Eq.~\eqref{eq:M''_1}}}{=} w(\mathcal{M}_2(G')) + w(e_1),
\end{align}
or
\begin{equation} 
\begin{aligned}
\label{eq:m2_m1g'}
    w(\mathcal{M}_2(G')) &\stackrel{\text{Eq.~\eqref{eq:M_2G > M_2G'+e1}}}{>} w(\mathcal{M}_2(G)\setminus\left\{e_1\right\}) \\
    &\stackrel{\text{Eq.~\eqref{eq:M'_1}}}{=} w(\mathcal{M}_2') \\
    &\stackrel{\text{Eq.~\eqref{eq:M_1' > M_1G'}}}{\geq} w(\mathcal{M}_1(G')) .
\end{aligned}
\end{equation}
Again, because $\mathcal{M}_1(G)$ and $\mathcal{M}_2(G)$ are two different matchings for $G$, upon comparing Eq.~\eqref{eq:m1} to \eqref{eq:M'_1} ($i=2$), we have that $\mathcal{M}_1(G')$ and $\mathcal{M}_2'$ are two different matchings for $G'$. Hence Eq.~\eqref{eq:m2_m1g'} contradicts the fact that $\mathcal{M}_2(G')$ is the second MWM of $G'$, which leads to 
\begin{align}
    w(\mathcal{M}_2(G)) = w(\mathcal{M}_2'') \stackrel{\text{Eq.~\eqref{eq:M''_1}}}{=} w(\mathcal{M}_2(G') \cup \left\{e_1\right\}).
\end{align}
This implies $\mathcal{M}_2(G)\setminus\left\{e_1\right\}$ can be treated as the second MWM of $G'$, as claimed in Eq.~\eqref{eq:m2}.

\section{Numerical results}
\label{sec:Numerical results}

In this section, we show the numerical results of applying the $K$-MWM decoder to the surface-square GKP code, the surface-hexagonal GKP code, and the qubit stabilizer code subject to the independent graphlike errors.

\subsection{Surface-square GKP code}
\label{sec: numerical result for Surface-square GKP code}

We first consider the surface-square GKP code subject to independent and identically distributed additive errors as defined in Eq.~\eqref{eq:def_shift_errors_0}. Because the $\hat{q}$ and $\hat{p}$ subspaces of the code are decoupled from each other, we only need to apply the $K$-MWM decoder to one of the subspaces, and the fidelity reported below is the square of the fidelity from the subspace considered.
The surface-square GKP code we consider has a square topology (which is not related to the shape of the inner GKP code) as shown in Fig.~\ref{fig:MWM}(a), and a surface-square code of distance $d$ has $d^2$ GKP qubits. For such a family of code, it turns out that we can perform MLD exactly and efficiently. This was first demonstrated by Bravyi, Suchara and Vargo (BSV) in Ref.~\cite{bravyi2014efficient} for the ``unrotated'' qubit surface code.  
In Ref.~\cite{lin2024exploring}, we show that it is possible to adopt the exact MLD to the (rotated) surface code, and generalize it to the case where each qubit is a square GKP qubit, which allows us to study the quantum capacity of the Gaussian random displacement channel.

\begin{widetext}

\begin{figure}
\centering
\includegraphics[width=\linewidth]{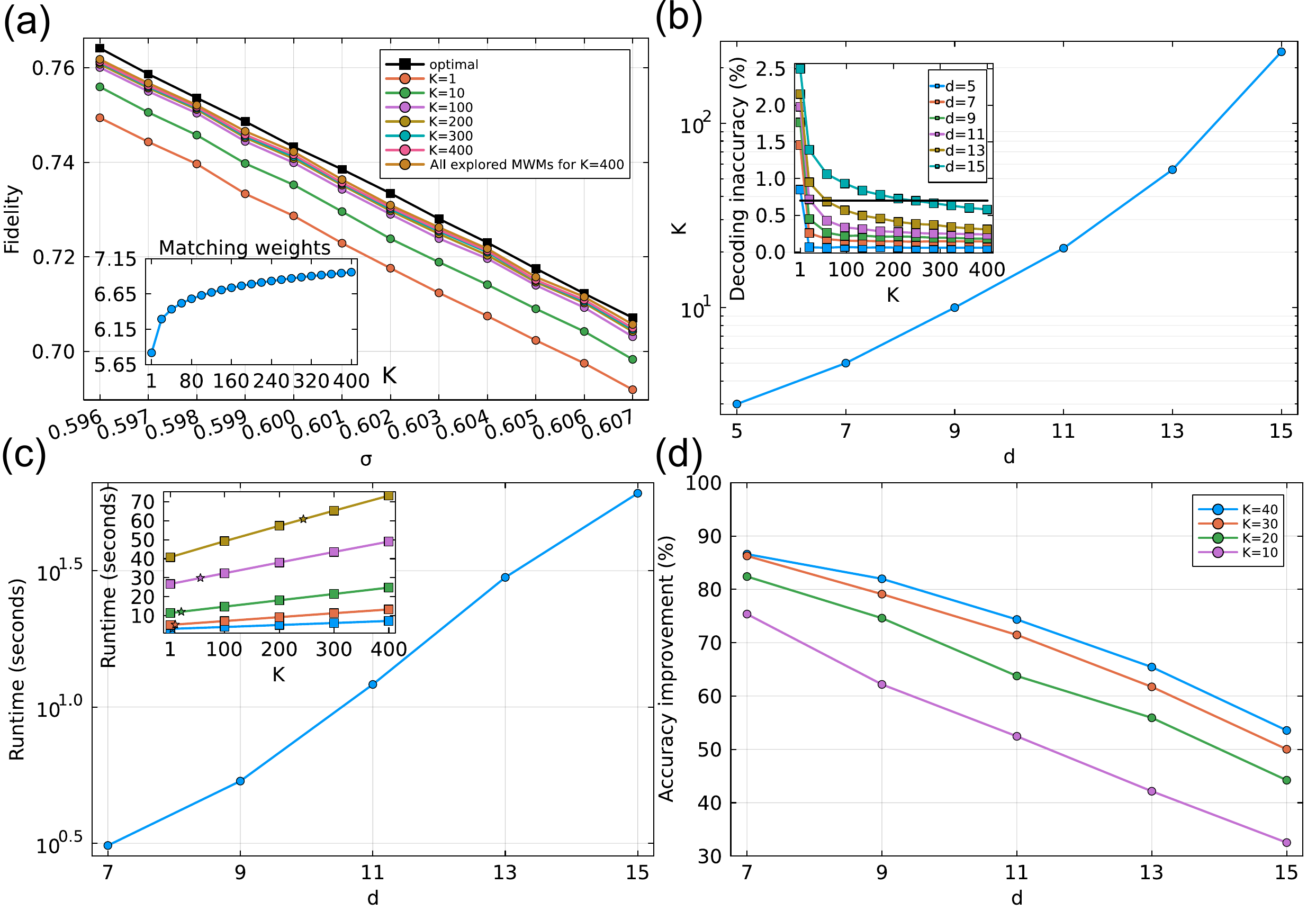}
 \caption{Results of $K$-MWM decoder for surface-square GKP code.
 (a) The fidelity for $d=13$ as a function of noise variance $\sigma$. The MLD is approximated with $K=400$ MWMs. The fidelity achieved for various values of $K$, ranging from $K=1$ to $K=400$ are shown for $\sigma=0.596, 0.597, \cdots, 0.607$. Each data point is obtained from $10^6$ Monte-Carlo samples. When finding the first $400$ MWMs, we explored much more candidate MWMs. The fidelity for including all explored MWMs is also shown, which is slightly higher than that without the additional MWMs. The black squares are the optimal fidelity achieved via the BSV decoder {with $10^7$ samples}.
 The inset shows the weights of the first 400 MWMs for $d=13$, averaged over 480 Monte-Carlo samples. The weight differences between the matchings become smaller as we increase $K$, which is consistent with the observation from the main plot that the fidelity improvement also becomes smaller as $K$ increases.
 (b) The minimum value of $K$ needed to achieve decoding inaccuracy $0.7\%$ as a function of distance {at $\sigma=0.607$ (in logarithmic scale), which increases slightly faster than exponentially with the distance}, agreeing with the fact that exact MLD is in general a NP-hard problem. The inset shows the decoding inaccuracy for various values of $K$ for $d=15$ with the horizontal solid line indicates $0.7\%$ inaccuracy.
 (c) The runtime needed to achieve decoding inaccuracy $0.7\%$ as a function of distance {(in logarithmic scale)}, which also scales exponentially with $d$ consistent with that found in (b). The inset shows that the runtime scales linearly as a function of the number of MWMs for a given distance, as expected. The stars label the values of $K$ plotted in (b).
 (d). The accuracy improvement as a function of distance for various values of $K$.
}
 \label{fig:surf_sq}
\end{figure}

\end{widetext}

The possibility of efficiently performing MLD also enables us to benchmark the efficacy of our $K$-MWM decoder.
In Fig.~\ref{fig:surf_sq}(a), we show the fidelity for $d=13$ with noise variance $\sigma$ {ranging} from $0.596$ to $0.607$ where $10^6$ Monte-Carlo samples are taken for each value of $\sigma$. The black squares indicate the fidelity achieved by the exact MLD, and the circles indicate the fidelity achieved by the $K$-MWM decoder with various values of $K$.
As expected, the fidelity achieved by the $K$-MWM decoder gradually approaches that of the exact MLD when we increase $K$, the number of MWMs used to approxiate the MLD. 
As explained in Sec.~\ref{sec:Finding the $K$-th matching for the decoding graph}, during the process of searching for the MWMs, we have explored more candidate MWMs. 
One obvious and cheap way to improve the fidelity of the $K$-MWM decoder is to include all the MWMs explored. This is also shown in Fig.~\ref{fig:surf_sq}(a), where a slight improvement for the fidelity is indeed observed.
In the inset, we show the weights of the matchings as a function of $K$, averaged over $480$ Monte Carlo samples.
We observe that the weight difference between consecutive matchings is getting smaller as $K$ increases.
Similarly, we notice that the improvement of the fidelity is getting smaller as $K$ is increased to larger values. For example, the fidelity improvement from $K=1$ to $K=10$ is much larger than that between $K=300$ and $400$. 
This suggests that we have to include much more matchings to fill the gap between the fidelity achieved by $K$-MWM decoder and the optimal fidelity, as seen in Fig.~\ref{fig:surf_sq}(a).

To quantify the efficacy of the $K$-MWM decoder compared to the exact MLD, we introduce \emph{decoding inaccuracy},
\begin{align}
    |f_\text{optimal} - f_\text{MWM}^{(K)}|/ f_\text{optimal},
\end{align}
where $f_\text{optimal}$ and $f_\text{MWM}^{(K)}$ are the fidelity of the exact MLD and the $K$-MWM decoder respectively. 
This metric indicates how far away is the approximate MLD to the exact MLD when $K$ MWMs are used.
For example, when we use $K=400$ MWMs for the surface-square GKP code with $d=15$, the decoding inaccuracy is around 0.7\% for the range of noise variance considered. 
In Fig.~\ref{fig:surf_sq}(b), we show the minimum number of MWMs needed in order to achieve decoding inaccuracy 0.7\% at $\sigma=0.607$, for various sizes of surface-square GKP codes. It turns out that for $d\leq 9$, we only need less than $15$ MWMs to approximate the MLD with high accuracy, but the number of MWMs needed increases quickly for larger sizes of the code.
In the inset, we show the decoding inaccuracy as a function of $K$ for different sizes of the code at $\sigma=0.607$, and the horizontal solid line indicates the decoding inaccuracy 0.7\%.
For smaller values of noise variance, we expected that less number of MWMs would be needed to achieve the same decoding inaccuracy because errors are less likely to occur for smaller $\sigma$. However, for any finite $\sigma$, the number of MWMs needed to achieve certain decoding inaccuracy would still grow exponentially with respect to the size of the codes, as shown in Fig.~\ref{fig:surf_sq}(b).

The fact that the number of MWMs required for certain decoding inaccuracy grows exponentially with $d$ suggests that the runtime to achieve the same decoding inaccuracy also scales exponentially, as shown in Fig.~\ref{fig:surf_sq}(c).
This is because the runtime for finding $K$ MWMs scales linearly with $K$ as shown in the inset of Fig.~\ref{fig:surf_sq}(c). We stress that this does not mean it would take twice the runtime if we were to double the size of the MWMs searched. The runtime shown in the inset include not only the runtime for searching MWMs, but also that for other parts of the algorithm, such as setting up the model graph. In fact, as evident from the inset, for $d=15$, the runtime with $K=400$ is less than twice of that with $K=1$.

One should not be surprised that it takes exponential runtime to approximate the MLD with certain accuracy because MLD itself is NP hard in general. 
Although the exact MLD can be implemented for the surface-square GKP code considered, this is generally impossible for surface-square GKP code defined on other graphs.
To shed light on the efficacy of the $K$-MWM decoder, we introduce another metric, \emph{accuracy improvement}, 
\begin{align}
    |f_\text{MWM}^{(K)} - f_\text{MWM}^{(1)}|/ |f_\text{optimal} - f_\text{MWM}^{(1)}|,
\end{align}
which quantifies the fidelity improvement achieved with $K$ MWMs, compared to that with just one MWM, or the conventional MWM decoder. 
In Fig.~\ref{fig:surf_sq}(d), we plot the accuracy improvement for various size of surface-square GKP code at $\sigma=0.607$. 
From our discussions of Fig.~\ref{fig:surf_sq}(b), it is not surprised that with $K=40$, the accuracy improvements are as large as 80\% for $d\leq 9$. However, it is remarkable that for $d=15$, the largest code we explore, the accuracy improvements is larger than 30\% with just $10$ MWMs. 
In particular, for $d=15$, the accuracy improvement with $K=40$ is nearly 60\% whereas the increase of runtime is relatively insignificant, as evident from Fig.~\ref{fig:surf_sq}(c).
This shows that even for surface-square GKP codes of moderate or large sizes, $K$-MWM decoder can still effectively improve the decoding accuracy with just a handful of MWMs.

\begin{figure}[!ht]
\centering
\includegraphics[width=\linewidth]{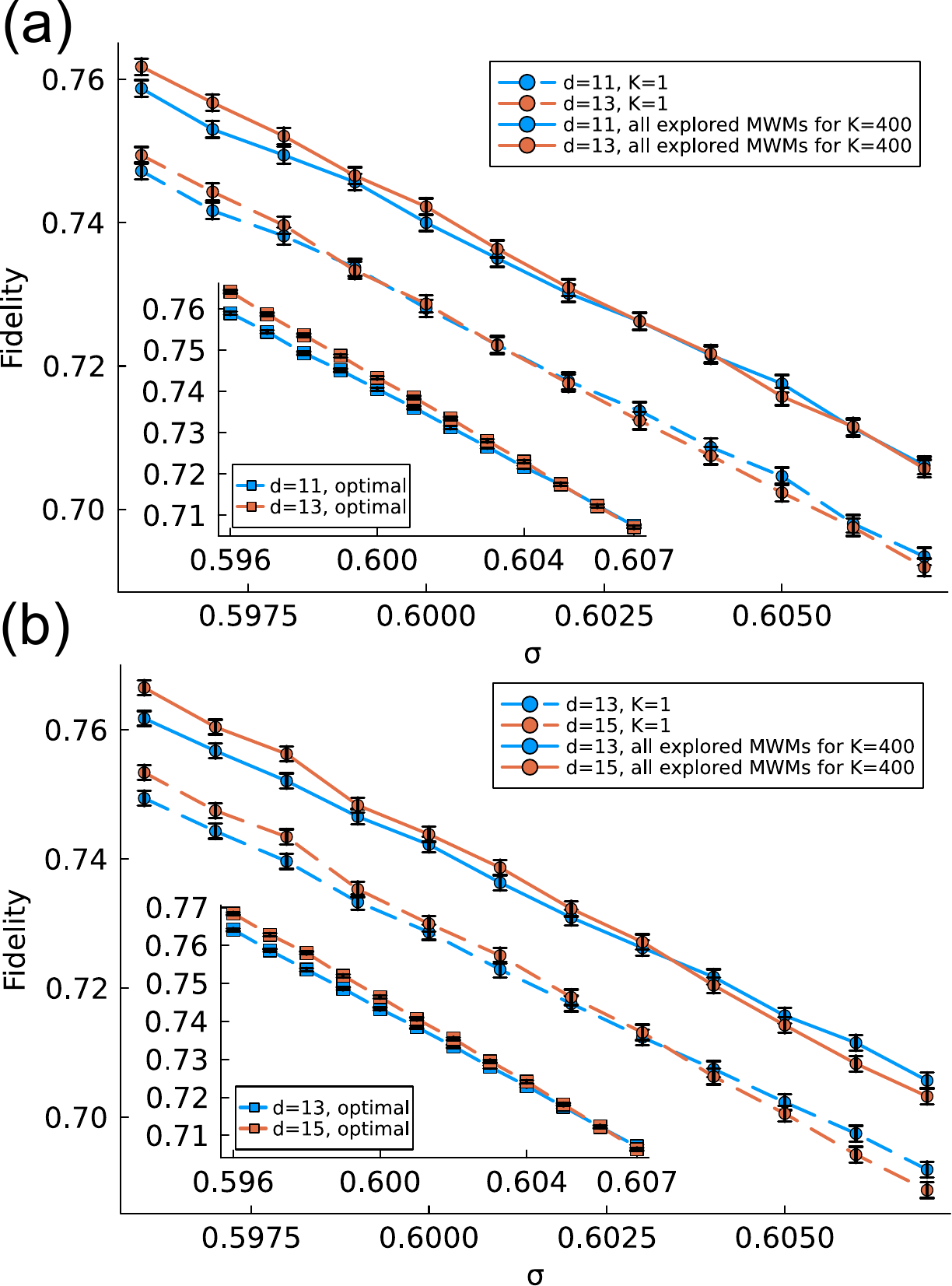}
 \caption{
 Threshold behavior of the $K$-MWM decoder. (a) The fidelity for $d=11$ and $d=13$ as a function of noise variance and the number of MWMs used. Error bars represent two standard errors, approximately $0.001$ with $10^6$ samples. The inset shows results obtained with the optimal BSV decoder, where the error bars are approximately $0.0003$ based on $10^7$ samples. (b) Logical fidelities for $d = 13$ and $15$ under the $K$-MWM and BSV decoders.}
 \label{fig:threshold_surf_sq}
\end{figure}

Let us comment on the threshold behavior observed with the $K$-MWM decoder. To estimate the threshold for a given $K$, we identify the crossing point $\sigma^*(K, d)$ at which the logical fidelity for distance $d+2$ exceeds that for distance $d$ when decoding with $K$ MWMs. The threshold is then defined as  $\sigma^*(K) \equiv \lim_{d \rightarrow \infty} \sigma^*(K, d)$. We expect that $\sigma^*(K=1) \approx 0.602$ and $\lim_{K \rightarrow \infty} \sigma^*(K) \approx 0.6065$, corresponding respectively to the thresholds of the standard MWM decoder \cite{lin2023closest} and the optimal decoder \cite{lin2024exploring}.
Figure~\ref{fig:threshold_surf_sq}(a) presents the logical fidelity as a function of $\sigma$ for code distances $d=11$ and $13$, with error bars representing two standard errors, approximately $0.001$. For $K=1$, the crossing point lies between $0.600$ and $0.602$, while for $K=400$, using all explored MWMs, the crossing shifts to between $0.603$ and $0.605$. For comparison, the crossing obtained from the optimal BSV decoder, shown in the inset of Fig.~\ref{fig:threshold_surf_sq}(a), lies between $0.605$ and $0.606$, with smaller error bars of about $0.0003$ due to the use of $10^7$ Monte Carlo samples.
In Fig.~\ref{fig:threshold_surf_sq}(b), we also show the crossings for $d=13$ and $15$, where the threshold appears to remain between $0.602$ and $0.603$ despite increasing the number of MWMs. From these results, we are currently unable to reliably extract $\sigma^*(K, d)$ or verify our expectation that the threshold improves with increasing $K$.
A key challenge is that the difference between the thresholds of the conventional MWM and optimal decoders occurs at the third decimal place, requiring significantly more Monte Carlo samples to resolve. Indeed, Refs.\cite{lin2023closest, lin2024exploring} determine thresholds using $10^7$ samples, which is an order of magnitude more than that used for the $K$-MWM decoder. While increasing the sample size by a factor of $10$ is feasible with our current implementation, it remains computationally expensive. As discussed in Sec.\ref{sec:Discussion}, we propose a possible speedup strategy for future work, and thus leave a more precise investigation of the $K$-MWM decoder threshold to future studies.

\subsection{Surface code subject to independent graphlike errors}
\label{sec: Qubit surface code with Z noise}

Next, we demonstrate the efficacy the $K$-MWM decoder for the qubit surface code subject to independent $Z$ and $X$-type errors. 
As explained in Sec.~\ref{sec: Surface code subject to graphlike errors as a special case of surface-square GKP code}, we consider the code as a special case of the surface-square GKP code with noise variance given by Eq.~\eqref{eq:mapping_sigma_epsilon} and discrete displacement errors sampled from {Bernoulli} distribution with error rate $\epsilon$. 
Here we consider surface code up to distance $d=15$ with $\epsilon$ {ranging} from $0.105$ to $0.115$.

\begin{figure}[!ht]
\centering
\includegraphics[width=\linewidth]{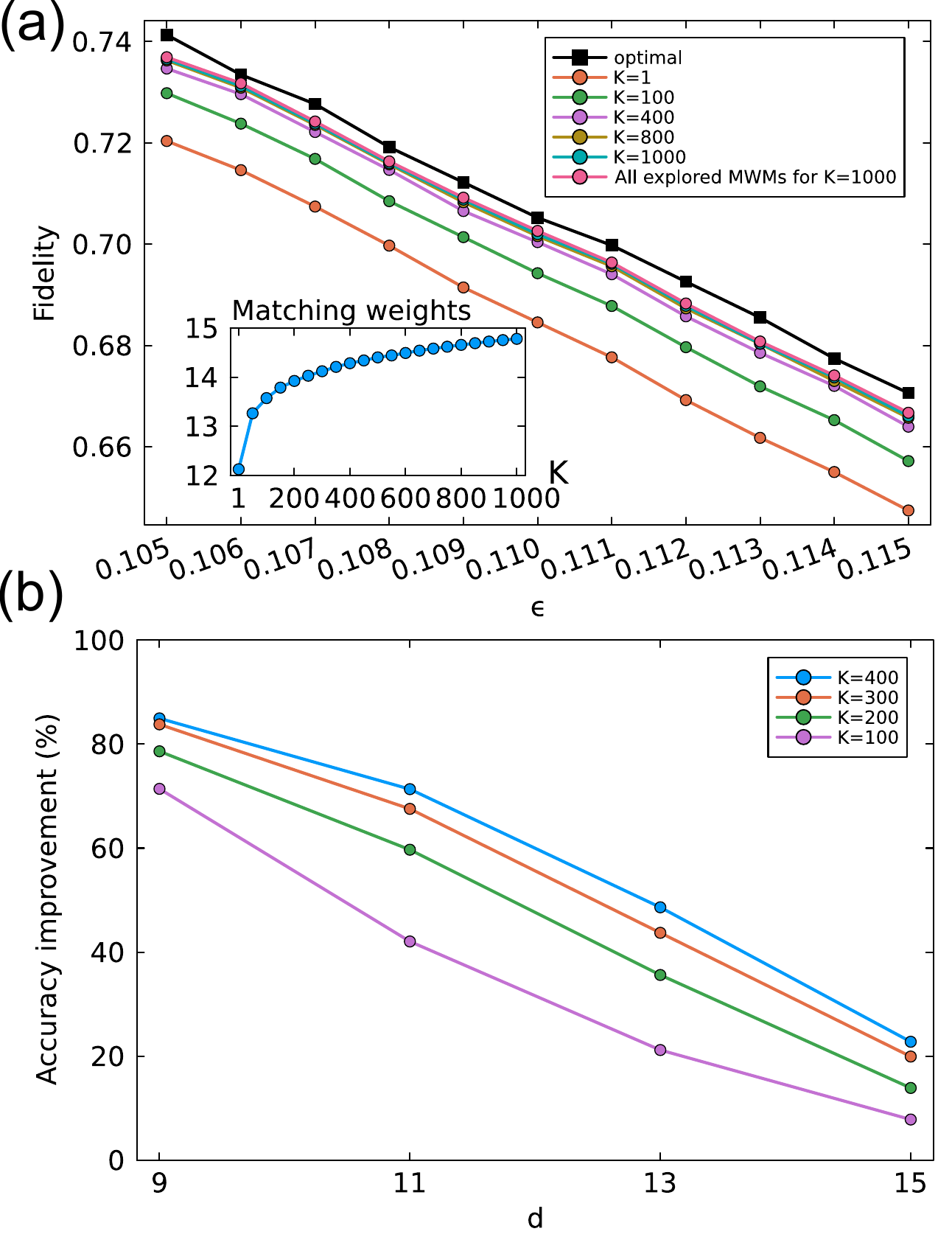}
 \caption{Results of $K$-MWM decoder for the qubit surface code subject to independent $X$ and $Z$-type errors. 
 (a) The fidelity for $d=11$ as a function of error rate $\epsilon$ with $K=1000$ MWMs. The black squares are the optimal fidelity achieved via the BSV decoder.
 The inset shows the weights of the MWMs as a function of $K$ for $\epsilon=0.114$.
 (b). The accuracy improvement as a function of distance for various values of $K$. 
}
 \label{fig:qubit-surf}
\end{figure}

In Fig.~\ref{fig:qubit-surf}(a), we benchmark the fidelity obtained by the $K$-MWM decoder against that of the exact MLD, where we use up to $K=1000$ MWMs for $d=11$. We observe that the fidelity achieved by the $K$-MWM decoder indeed steadily approaches that of the exact MLD as $K$ increases.
Similar to that observed in Fig.~\ref{fig:surf_sq}(a), the improvement of fidelity is getting smaller as $K$ increases, consistent with the fact that the difference of the matching weights are getting smaller as shown in the inset.
In Fig.~\ref{fig:qubit-surf}(b), we plot the accuracy improvement as a function of distance for various values of $K$. It is found that for $d=15$, the largest surface code considered, the accuracy improvement is approximately 22\% with $K=400$.

\subsection{Surface-hexagonal GKP code}
\label{sec: Surface-hexagonal GKP code}

\begin{figure}[!ht]
\centering
\includegraphics[width=\linewidth]{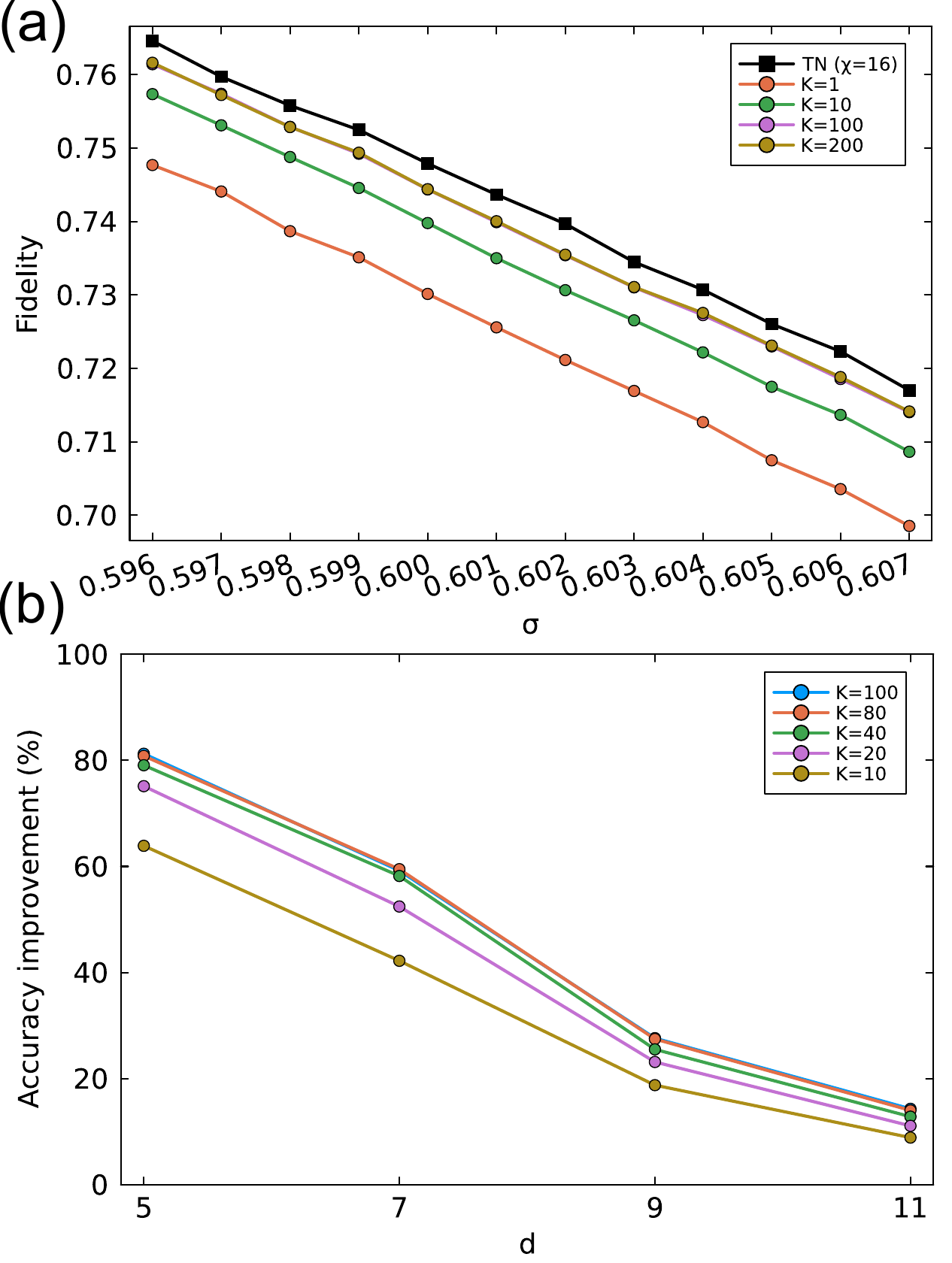}
 \caption{Results of $K$-MWM decoder for surface-hexagonal GKP code. (a) The fidelity for $d=7$ as a function of error rate $\sigma$ with $K=200$ MWMs. The black squares are fidelity achieved via a tensor-network decoder with bond dimenstion $\chi=16$.
 (b). The accuracy improvement as a function of distance for various values of $K$.
}
 \label{fig:surf-hex}
\end{figure}

The third example we considered is the surface-hexagonal GKP code subject to the same independent and identically distributed additive errors as defined in Eq.~\eqref{eq:def_shift_errors_0}. In contrast to the surface-square GKP code, the BSV decoder cannot be applied to the surface-hexagonal GKP code because its  $\hat{q}$ and $\hat{p}$ subspaces are coupled. To benchmark the performance of our $K$-MWM decoder, we use a variant of the tensor-network decoder adapted to the GKP codes \cite{bravyi2014efficient, chubb2021general, SweepContractor_v0_1_7}.

In Fig.~\ref{fig:surf-hex}(a), we compare the fidelity obtained by the $K$-MWM decoder and the tensor-network decoder with bond dimension $\chi=16$, where we use up to $K=200$ MWMs for $d=7$. We observe that the fidelity achieved by the $K$-MWM decoder approaches to that of the tensor-network decoder as $K$ increases.
We note that the improvement of the fidelity from $K=100$ to $K=200$ is relatively small, which is consistent with the similar finding in the other two examples.
The accuracy improvement of the $K$-MWM decoder is shown in Fig.~\ref{fig:surf-hex}(b) as a function of distance for various values of $K$. It is found that for $d=11$, the largest surface code considered, the accuracy improvement is approximately 15\% with $K=100$. 

We note that this example with surface-hexagonal code shows that the $K$-MWM decoder can effectively approximate MLD for the case where the $X$ and $Z$ errors are correlated. 
Indeed, as explained in App.~\ref{sec: Surface code subject to correlated errors as a special case of general surface-GKP code}, the qubit surface code subject to the depolarizing noise can be viewed as an example of the surface-hexagonal code (see Eq.~\eqref{eq:para_depolarizing_noise}). 
Hence, given the results obtained for surface-hexagonal code, we expect that the $K$-MWM decoder can also be applied to approximate the MLD, {and leave a comparative performance analysis between the $K$-MWM decoder and other decoders \cite{james2012high,bombin2012strong,ohzeki2012error} for future work.}

\section{Discussion}
\label{sec:Discussion}

In this work, we show that MLD for a wide {family} of QEC codes can be approximated by considering multiple MWMs in their decoding graphs, which can be found systematically via the introduced $K$-MWM decoder. The decoder is characterized by a single parameter, the number of MWMs used to approximate the MLD, which enables a trade-off between efficiency and accuracy.
We first apply the decoder to surface-square GKP code, demonstrating that the fidelity obtained by the $K$-MWM decoder steadily approaches the optimal fidelity as $K$ is increased. 
{Although} it requires exponentially many MWMs to reach the decoding accuracy of the MLD, because MLD is a NP-hard problem, we numerically show that a handful of MWMs would be enough for a significant fidelity improvement, for distances up to $d=15$. 
This highlights one appealing advantage of our decoder, which allows one to spend a little more compute resource to gain a much better performance, a feature that is absent in most decoding strategies such as the traditional MWM decoder.
It is well known that the MWM can be found efficiently for the surface-square GKP code. To show that the $K$-MWM decoder can still be applied in cases where the MWM cannot be found efficiently, we further consider the surface-hexagonal GKP code. 
Our numerical result suggests that the fidelity of the $K$-MWM decoder steadily approaches that obtained via the tensor-network decoder, although the convergence rate is slower than that for the surface-square GKP code. This is expected because the decoding hypergraph of the former is twice {as large as} that of the latter. It is also important to note that the $K$-MWM decoder needs not find the true MWM of the decoding hypergraph, {yet its efficacy is evident from the numerical simulations.} 
Although we have described the $K$-MWM decoder in the context of GKP codes, we prove that a qubit stabilizer code can be decoded by considering the corresponding concatenated-GKP code subject to discrete displacement errors. In particular, we show that the noise variance and the shape of the inner GKP code of the concatenated-GKP code can be determined from the error model of the qubit stabilizer code. Using such a mapping, the $K$-MWM decoder is shown to reproduce the optimal fidelity of the qubit surface code subject to the pure $Z$-error. The decoder can also be applied to qubit surface codes subject to other noise model. For example, a qubit surface code subject to depolarizing noise can be decoded by approximating the MLD for the surface-hexagonal GKP code.

\subsection{Comparison with prior works}

The idea of modifying a decoding graph and aggregating the MWMs of the resulting graphs to enhance decoding accuracy has been previously investigated in the literature \cite{delfosse2014adecoding, criger2018multi, yuan2022modified, iolius2023performance, higgott2023improved, delfosse2023splitting, shutty2024efficient, jones2024improved,hammar2022error}. 
Let us compare to some prior works and highlight their connections with the current work.

Recently, several studies \cite{shutty2024efficient, jones2024improved} have explored improving decoding fidelity by including the contributions of an ensemble of matchings. These approaches
generate candidate matchings through 
%
%
randomly perturbing the edge weights in a hypergraph and applying a standard matching decoder to find the approximate MWM of the hypergraph. The process is repeated multiple times and the resulting matchings are combined by likelihood summation, voting, or synthetic reconstruction that incorporates the locality of the hypergraph. 
While these heuristic methods were shown to approach MLD performance for the surface code with circuit-level noise, they are inherently stochastic, and their efficiency depends on the quality and diversity of the generated ensemble. 
In contrast, the $K$-MWM decoder is based on a theoretically grounded enumeration of the matchings, which yields a deterministic improvement toward MLD as $K$ increases. 
Nevertheless, the $K$-MWM decoder and ensembling based methods share two notable similarities, namely that they are both well suited to parallel implementation and the efficiency and the accuracy is controlled by a single parameter ($K$ and ensemble size), allowing a smooth trade-off between runtime and performance.
In this regard, both $K$-MWM decoder and ensembling based methods are similar to the tensor-network decoder whose performance is also controlled by a single parameter, the bond dimension.
Given these distinctions and similarities, hybrid strategies are a natural possibility. For example, one can apply $K$-MWM method to the candidates generated by heuristic perturbations, or use belief-based reweighting method \cite{higgott2023improved} before deterministic enumeration. 
We leave it as an exciting research direction to combine the strengths of deterministic enumeration, probabilistic ensembling, and tailored weighting.

In \cite{hammar2022error}, the authors adopt a different approach to approximate the coset probability by counting the number and effective weight
of the most likely physical errors in each equivalence.
As the authors already noted, this approximation relies on certain assumptions which could lead to degraded performance at larger code distances. 
Moreover, it is unclear whether the approximation remains effective in cases where the edge weights in a graph are non-uniform, such as in GKP codes.
Conversely, our $K$-MWM decoder is inherently ineffective for graphs with uniform edge weights, since it would have to enumerate multiple matchings of identical weight.
Given that efficient algorithms such as the Fisher–Kasteleyn–Temperley (FKT) algorithm exist for counting perfect matchings in planar graphs, an interesting research direction would be to develop an FKT-based variant to accelerate our $K$-MWM decoder in the uniform-weight setting.

\subsection{Outlook}
{Beyond the directions already discussed in the previous section, our work opens several additional interesting avenues for future research.} We first note that it is possible to improve significantly the efficiency of our $K$-MWM decoder. Because the decoder approximates MLD by finding $K$ MWMs from a decoding graph, its efficiency is determined by how fast we can find subsequent MWMs in the decoding graph. 
Specifically, to find the second MWM from the decoding graph $G$, we define a series of reduced graphs $G^{(j)}$ (see Eq.~\eqref{eq:def_G_j_1}), followed by using an existing algorithm \cite{higgott2023sparse} to find the MWM of each reduced graph from scratch.
However, this needs not be the most efficient approach, because from Eq.~\eqref{eq:def_G_j_2}, we see that $G^{(j)}$ can be obtained from $G^{(j-1)}$ (assuming the original graph $G\equiv G^{(0)}$) by removing a single edge and modifying the corresponding highlighted vertices.
Since the two graphs are very similar, we expect that if the MWM of $G^{(j-1)}$ has been found, there should be a more efficient method to find the MWM for $G^{(j)}$, compared to searching it from scratch.
Similar idea has been adopted in the Chegireddy-Hamacher's paper \cite{chegireddy1987algorithms}, where two approaches are described to find the first $K$ MWPMs of a complete weighted graph.
The first approach, which motivates our $K$-MWM decoder, also defines a series of reduced graphs first, followed by using the blossom algorithm \cite{edmonds1965paths} to find the MWPM of the graphs from scratch. 
The time complexity of this algorithm is estimated to be $O(Kn^4)$ for a graph with $n$ nodes. 
Remarkably, Chegireddy and Hamacher presented a second algorithm, taking advantage of the intuition described above, with time complexity reduced to $O(Kn^3)$, which is one order of magnitude better than the first algorithm. 
Unfortunately, we cannot directly apply Chegireddy-Hamacher's algorithm, because converting a problem of finding MWMs on a decoding graph to finding MWPMs on a complete graph typically requires the very expensive Dijkstra's shortest path algorithm, as noted in Ref.~\cite{higgott2023sparse, wu2023fusion}, or even Yen's $K$ shortest path algorithm \cite{yen1971finding}.
Nevertheless, the result from Chegireddy and Hamacher is encouraging and it suggests that it is possible to reduce the runtime of finding each subsequent MWM of the decoding graph.
Such reduction of runtime would be substantial for determining the threshold of QEC code or decoding circuit level noise, where it typically requires much more MWMs to reach certain desired decoding accuracy.
We will leave this interesting topic as a future research direction.

Another interesting research direction is to generalize the $K$-MWM decoder to other families of QEC codes and error models. As demonstrated in \cite{delfosse2014decoding, kubica2023efficient}, a color code can be decoded by independently decoding two copies of the surface code; it is therefore of interest to investigate the performance of the $K$-MWM decoder in this setting. Furthermore, recent work has introduced two minimum-weight decoding algorithms for quantum low-density parity-check (qLDPC) codes \cite{beni2025tesseract, ott2025decision}, motivating the question of whether the $K$-MWM method can be adapted to this class of codes. In addition, although efficient MLD algorithms are known for the surface code with purely erasure errors \cite{delfosse2020linear, stace2009thresholds}, it is worthwhile to explore their extension to channels exhibiting both erasure and Pauli errors \cite{sean2010fault}, and to determine whether the $K$-MWM decoder can improve upon the performance of the union–find decoder \cite{delfosse2021almost} in such scenarios.

In previous sections, we have described the $K$-MWM decoder as an interpolation between the conventional MWM decoder and MLD. Interestingly, a closely related interpolation can also be made between quantum error correction and \emph{detection}. For example, in \cite{smith2024mitigating}, the authors introduce an exclusive decoder that decides whether to apply a correction or to abort based on a comparison of coset probabilities for a given syndrome.
The algorithm was shown to reduce the resources needed for magic state distillation. 
Since the $K$-MWM decoder produces a ranked list of candidate matchings and associated weights, we expect it can be combined with the exclusive decoding paradigm to enhance confidence testing in error detection, a promising direction for future work.

Lastly, we comment on the exact BSV decoder used to benchmark our $K$-MWM decoder for the surface-square GKP code, and qubit surface code subject to the pure $Z$-error. 
Despite being optimal, the exact BSV decoder can only be applied to surface code defined on a square lattice. On the other hand, $K$-MWM decoder can be applied to surface code defined on other lattices, such as irregular graphs \cite{chubb2021general}, or even circuit level noise where no exact MLD is known for even the pure $Z$-noise.
It would be interesting to compare the performance of the $K$-MWM decoder (after investigating the possible improvement of efficiency mentioned above) to other decoders based on 2D and 3D tensor networks in such applications.

\section{Acknowledgements}

I would like to thank Kyungjoo Noh for his discussion and input during the early phase of this project. 
I would like to thank Yue Wu for his detailed feedback and point out a mistake in the early version of this manuscript.
The anonymous referees have provided multiple valuable feedbacks and ideas which are greatly appreciated.
I would like to also thank the discussion with April Zeng.
I would like to thank Péter Kómár and Eric Kessler for their supports of the project. It is a pleasure to thank Connor Stewart and the Day 1 Science Mentorship (D1SM) program at Amazon where these projects were initiated. I would like to acknowledge the AWS EC2 resources which were used for part of the simulations performed in this work.

\appendix

\section{Review of GKP code}
\label{sec: Preliminary}

In this section, we expand the discussion in Sec.~\ref{sec: Brief review of GKP codes} and provide more details on the key concepts for GKP codes that will be used for formulating the $K$-MWM decoder from a lattice perspective in App.~\ref{sec:Maximum likelihood decoder for concatenated GKP codes: A lattice perspective}. For more detailed introduction of GKP codes, particularly from the lattice perspective, we refer readers to some recent works \cite{lin2023closest, conrad2022gottesman, royer2022encoding} and the original work \cite{gottesman2001encoding}. 

\subsection{Multimode GKP code}

We start by reviewing the stabilizer group structure of a GKP code.
For an $N$-mode GKP code, the $2N$ independent stabilizer generators are the displacement operators in the $2N$-dimensional phase space given by
\begin{align}
\label{eq:def_S_j}
    \hat{{S}}_{j} \equiv \hat{D}(\sqrt{2\pi}\boldsymbol{v}_{j}) \equiv \exp[ -i \sqrt{2\pi}\boldsymbol{v}_j^{T} \Omega \hat{\boldsymbol{x}} ], 
\end{align}
where $j\in\left\{1,\cdots,2N\right\}$, and $\boldsymbol{v}_j$ is the translation vector for the $j$-th generator. 
The full stabilizer group is given by
\begin{align}
    \mathcal{S} = \lbrace  \hat{{S}}_{1}^{a_1}\cdots \hat{{S}}_{2N}^{a_{2N}} ~|~ \boldsymbol{a} = [a_{1},\cdots, a_{2N}]^{T} \in \mathbb{Z}^{2N} \rbrace,
\end{align}
and since the stabilizer generators commute with each other, a generic stabilizer group element reads
\begin{align}
\label{eq:stabilizer_as_lattice_point}
    \hat{S} &= \hat{D}(\sqrt{2\pi}M^T\boldsymbol{a}) = \exp[ i  \sqrt{2\pi} ( \boldsymbol{a}^{T} M )  \Omega^{-1} \hat{\boldsymbol{x}}  ], 
\end{align}
where $M$ is a $2N\times2N$ matrix with the $j$-th row corresponding to $\boldsymbol{v}_j^T$. Eq.~\eqref{eq:stabilizer_as_lattice_point} suggests that the stabilizer group of the GKP code is isomorphic to a lattice with the generator matrix $M$
\begin{align}
    \Lambda(M) \equiv \lbrace M^T\boldsymbol{a} ~|~ \boldsymbol{a}= [a_{1},\cdots, a_{2N}]^{T} \in \mathbb{Z}^{2N} \rbrace,
\end{align}
where the stabilizer group element $\hat{S}$ is mapped to the lattice point $\sqrt{2\pi}M^T\boldsymbol{a}$. 
Since the stabilizers form an Abelian group, the symplectic Gram matrix
\begin{align}
\label{eq:def_A}
    A\equiv M\Omega M^T
\end{align}
is required to have only integer entries. Lattices with this property are called symplectic integral lattices \cite{conway2013sphere}.

For a given lattice $\Lambda(M)$, we can define its symplectic dual lattice as
\begin{align}
\label{eq:def_lambda_perp}
    \Lambda^\perp \equiv \Lambda(M^{\perp}) = \left\{\boldsymbol{u} ~|~ \boldsymbol{u}^T\Omega\boldsymbol{v}\in\mathbb{Z}, ~ \forall \boldsymbol{v}\in\Lambda(M)\right\},
\end{align}
where the $2N\times2N$ generator matrix is denoted as $M^\perp$.
By definition, $\Lambda^\perp$ consists of all vectors that have integer symplectic inner product with all vectors in $\Lambda(M)$.
One possible representation for the generator matrix of $\Lambda(M^{\perp})$ reads
\begin{align}
\label{eq:def_M_perp}
    M^{\perp} = \Omega A^{-1}M = -\Omega (M^T)^{-1}\Omega,
\end{align}
because $\boldsymbol{b}^TM^{\perp}\Omega M^T\boldsymbol{a}=\boldsymbol{b}^T\Omega\boldsymbol{a}$ is an integer for all $\boldsymbol{a}, \boldsymbol{b}\in\mathbb{Z}^{2N}$.

In the following, we will be interested in scaled lattice{s}, which {are} obtained by multiplying the generator matrix by a constant $c$. The scaled lattice will be denoted as $c\Lambda\equiv \Lambda(cM)$, and its symplectic dual lattice is given by $\Lambda(c^{-1}M^\perp)=c^{-1}\Lambda^\perp$.

\subsection{Concatenated GKP code}
\label{sec: Concatenated GKP code}

In this work, we mainly focus on the concatenated GKP codes, which are constructed by concatenating $N$ one-mode GKP codes, each of which encodes a single qubit, to an $[[N, k]]$ qubit stabilizer code. The resultant {concatenated GKP} code encodes $k$ qubits in $N$ modes. 
We note that each stabilizer generator for the $[[N, k]]$ qubit code corresponds to a binary vector with $2N$ components, where the first and second $N$ entries represent the presence of $X$ and $Z$ operators respectively. 
For concatenated-square GKP codes, where the inner GKP codes are identical one-mode square GKP codes, their generator matrices can be written as \cite{lin2023closest}
\begin{align}
\label{eq:M_conc}
&M_\text{conc-sq} = \nonumber\\ 
&\frac{1}{\sqrt{2}}
\begin{array}{@{} c @{}}
    \left [
      \begin{array}{ *{6}{c} }
        {\hspace{0.23cm}I} \hspace{0.cm}& {\hspace{0.5cm}A_1} \hspace{0.3cm}& {\hspace{0.23cm}A_2} \hspace{0.3cm}& {\hspace{0.13cm}B} \hspace{0.1cm}& {\hspace{0.53cm}0} \hspace{0.4cm}& {\hspace{0.2cm}C}\\
        {\hspace{0.23cm}0} \hspace{0.cm}& {\hspace{0.5cm}0} \hspace{0.3cm}& {\hspace{0.23cm}0} \hspace{0.3cm}& {\hspace{0.13cm}D} \hspace{0.1cm}& {\hspace{0.53cm}I} \hspace{0.4cm}& {\hspace{0.2cm}E}\\
        {\hspace{0.23cm}0} \hspace{0.cm}& {\hspace{0.5cm}0} \hspace{0.3cm}& {\hspace{0.23cm}2I} \hspace{0.3cm}& {\hspace{0.13cm}0} \hspace{0.1cm}& {\hspace{0.53cm}0} \hspace{0.4cm}& {\hspace{0.2cm}0}\\
        {\hspace{0.23cm}0} \hspace{0.cm}& {\hspace{0.5cm}0} \hspace{0.3cm}& {\hspace{0.23cm}0} \hspace{0.3cm}& {\hspace{0.13cm}2I} \hspace{0.1cm}& {\hspace{0.53cm}0} \hspace{0.4cm}& {\hspace{0.2cm}0}\\
        {\hspace{0.23cm}0} \hspace{0.cm}& {\hspace{0.5cm}2I} \hspace{0.3cm}& {\hspace{0.23cm}0} \hspace{0.3cm}& {\hspace{0.13cm}0} \hspace{0.1cm}& {\hspace{0.53cm}0} \hspace{0.4cm}& {\hspace{0.2cm}0}\\
        \undermat{r}{\hspace{0.23cm}0} \hspace{0.cm}& \undermat{N-k-r}{\hspace{0.5cm}0} \hspace{0.3cm}& \undermat{k}{\hspace{0.23cm}0} \hspace{0.3cm}& \undermat{r}{\hspace{0.13cm}0} \hspace{0.1cm}& \undermat{N-k-r}{\hspace{0.53cm}0} \hspace{0.4cm}& \undermat{k}{\hspace{0.2cm}2I}
      \end{array}
    \right ]
    \begin{array}{@{} r @{}}
      \}r\hspace{1.36cm}\\
      \}N-k-r \\
      \}k\hspace{1.36cm}\\
      \}r\hspace{1.36cm} \\
      \}N-k-r\\
      \}k\hspace{1.36cm}
    \end{array}    
    \vspace{0.5cm}
    \mathstrut
  \end{array},
\end{align}
where $I$ denotes identity matrix and  $A_1, A_2, B, C, D$ and $E$ are binary matrices with entries being zero or one. The first $N-k$ rows of $\sqrt{2}M_\text{conc-sq}$ represent the standard form of the $[[N, k]]$ qubit stabilizer code \cite{nielsen2002quantum}. One could confirm that the symplectic Gram matrix $M_\text{conc-sq}\Omega M_\text{conc-sq}^T$ is indeed integer valued with $\text{det}(M_\text{conc-sq})=2^k$, hence $\Lambda(M_\text{conc-sq})$ is indeed a GKP code that encodes $k$ qubits.

We note that $\sqrt{2}M_\text{conc-sq}$ is an {integer valued} matrix and hence $\sqrt{2}\Lambda(M_\text{conc-sq})$ is a sublattice of the $2N$-dimensional integer lattice, denoted as $Z_{2N}\equiv\Lambda(I_{2N})$. 
Further, the scaled integer lattice $2Z_{2N}$ is a sublattice of $\sqrt{2}\Lambda(M_\text{conc-sq})$. This can be seen by noticing that the last $N+k$ rows of $\sqrt{2}M_\text{conc-sq}$ form a subset of the basis vectors of $2Z_{2N}$, and the remaining basis vectors can be obtained by multiplying the first $N-k$ rows by 2 followed by subtracting certain rows in the last $N+k$ rows to eliminate the block matrices.
Because $\Lambda(M_\text{conc-sq})$ is a sublattice of its symplectic dual lattice, we have
\begin{align}
\label{eq:chain_subsets}
    2Z_{2N} \subset \sqrt{2}\Lambda(M_\text{conc-sq}) \subset \sqrt{2}\Lambda(M^\perp_\text{conc-sq})  \subset Z_{2N}.
\end{align}
To see the last part of Eq.~\eqref{eq:chain_subsets}, we note that with Eq.~\eqref{eq:M_conc}, the generator matrix $M_\text{conc-sq}$ can be rewritten as
\begin{align}
    M_\text{conc-sq} = \frac{1}{\sqrt{2}}
    \begin{bmatrix}
        I & 0 & A_2 & B & A_1 & C\\
        0 & I & 0 & D & 0 & E\\
        0 & 0 & 2I & 0 & 0 & 0\\
        0 & 0 & 0 & 2I & 0 & 0\\
        0 & 0 & 0 & 0 & 2I & 0\\
        0 & 0 & 0 & 0 & 0 & 2I
    \end{bmatrix}T,
\end{align}
where $T$ is a permutation matrix that swaps the first $N-k-r$ rows with the second $N-k-r$ rows in Eq.~\eqref{eq:M_conc}. As one can check, we have
\begin{align*}
    M_\text{conc-sq}^{-1} &= \frac{1}{\sqrt{2}}T^{-1}\begin{bmatrix}
        2I & 0 & -A_2 & -B & -A_1 & -C\\
        0 & 2I & 0 & -D & 0 & -E\\
        0 & 0 & I & 0 & 0 & 0\\
        0 & 0 & 0 & I & 0 & 0\\
        0 & 0 & 0 & 0 & I & 0\\
        0 & 0 & 0 & 0 & 0 & I
    \end{bmatrix}\\
    &=\frac{1}{\sqrt{2}}\begin{bmatrix}
        2I & 0 & -A_2 & -B & -A_1 & -C\\
        0 & 0 & 0 & 0 & I & 0\\
        0 & 0 & I & 0 & 0 & 0\\
        0 & 0 & 0 & I & 0 & 0\\
        0 & 2I & 0 & -D & 0 & -E\\
        0 & 0 & 0 & 0 & 0 & I
    \end{bmatrix}.
\end{align*}
Since $\sqrt{2}M_\text{conc-sq}^{-1}$ is integer valued, upon plugging $M_\text{conc-sq}^{-1}$ into Eq.~\eqref{eq:def_M_perp}, we conclude that $\sqrt{2}M_\text{conc-sq}^\perp$ is also {integer valued} and hence $\sqrt{2}\Lambda(M_\text{conc-sq}^\perp)$ is a sublattice of $Z_{2N}$. 
The relations in Eq.~\eqref{eq:chain_subsets} will be critical for developing the MLD from a lattice perspective, which will be explained in App.~\ref{sec:Maximum likelihood decoder for concatenated GKP codes: A lattice perspective}. 

For more general concatenated GKP codes, their generator matrices can be written as
\begin{align}
\label{eq:M_general}
    M = M_\text{conc-sq}S^T \equiv M_\text{conc-sq}\begin{bmatrix}
        \Gamma_1 & \Gamma_2 \\ \Gamma_3  & \Gamma_4 
    \end{bmatrix}^T,
\end{align}
where $S$ is a symplectic matrix satisfying 
\begin{align}
\label{eq:symplectic_condition}
    S\Omega S^T=\Omega.
\end{align}
$\Gamma_{1,2,3,4}$ are $N\times N$ diagonal matrices that satisfy the condition $\Gamma_1\Gamma_4-\Gamma_2\Gamma_3=I_N$.
For example, if the inner GKP qubits are identical hexagonal GKP code, we can choose a basis such that
\begin{align}
\label{eq:basis_hex}
    \Gamma_1 = \frac{\sqrt{2}}{3^{\frac{1}{4}}}I_N, \quad \Gamma_2=-\frac{1}{2}\Gamma_1, \quad \Gamma_3 = 0_N, \quad \Gamma_4 = \Gamma_1^{-1}.
\end{align}

\subsection{Maximum likelihood decoder for GKP codes}
\label{sec:Maximum likelihood decoder for GKP codes}

In this section, we follow \cite{conrad2022gottesman} and review MLD, the optimal strategy to correct the shift error for the GKP codes.
We start by introducing some notations. Let (see Eq.~\eqref{eq:def_[xi]_0})
\begin{align}
\label{eq:def_[xi]}
    [\boldsymbol{\xi}]\equiv
    \left\{\boldsymbol{\xi}-\boldsymbol{u} ~|~ \boldsymbol{u}\in \sqrt{2\pi}\Lambda\right\}
\end{align}
denotes a coset of real valued vectors that differ from $\boldsymbol{\xi}$ only by a lattice vector in $\sqrt{2\pi}\Lambda$, the scaled GKP lattice. 
Because of its random nature, the actual shift error $\boldsymbol{\xi}$ is not known a priori, and we can only learn its effect through the stabilizer measurements. Recall from Eq.~\eqref{eq:def_S_j} that the commuting GKP stabilizers are given by 
$\hat{S}_{j} = \exp[ i\sqrt{2\pi} \boldsymbol{v}_{j}^{T} \Omega^{-1}\hat{\boldsymbol{x}} ]$, %
where $\boldsymbol{v}_{j}^{T}$ is the $j$-th row of the generator matrix $M$, measuring the stabilizers is equivalent to measuring their exponents simultaneously.
Let $s_j$ denote {the} error syndrome from the homodyne measurements of $\hat{S}_j$, then it differs from $ \sqrt{2\pi} \boldsymbol{v}_{j}^{T} \Omega^{-1}{\boldsymbol{\xi}}$ by an integer multiple of $2\pi$, i.e.,
\begin{align}
\label{eq:def_syndrome}
    \boldsymbol{s} \equiv \sqrt{2\pi} M \Omega^{-1}\boldsymbol{\xi} ~\mod~ 2\pi,
\end{align}
where the modulo operation is applied element-wise. Given a syndrome, we can guess a physical error that is consistent with the syndrome as   
\begin{align}
    \label{eq:def_eta_s}
    \boldsymbol{\eta_s}\equiv\frac{1}{\sqrt{2\pi}}\Omega M^{-1}\boldsymbol{s}.
\end{align}

We note that errors in the same coset $[\boldsymbol{\eta_s}]$ give the same syndrome because
\begin{align}
\label{eq:def_maximal_set_0}
    \sqrt{2\pi}M\Omega^{-1}\boldsymbol{u}\equiv 0 ~\mod~ 2\pi,
\end{align}
for all $\boldsymbol{u}\in\sqrt{2\pi}\Lambda$. However, $[\boldsymbol{\eta_s}]$ is not the maximal set of errors that give the same syndrome because Eq.~\eqref{eq:def_maximal_set_0} holds also for $\boldsymbol{u}\in\sqrt{2\pi}\Lambda^\perp$.
Hence, we have that the maximal set of consistent errors is given by 
\begin{align}
\label{eq:def_maximal_set}
    \left\{\boldsymbol{\eta_s}-\boldsymbol{v} ~|~ \boldsymbol{v}\in \sqrt{2\pi}\Lambda^\perp\right\}.
\end{align}
Further, if the GKP code encodes $k$ logical qubits, there are $2^{k+1}$ disjoint cosets in $\Lambda^\perp$, each of which is labeled by a logical operator
\begin{align}
\label{eq:def_local_op}
    \boldsymbol{l}\in\sqrt{2\pi}\Lambda^\perp/\sqrt{2\pi}\Lambda\equiv\sqrt{2\pi}(\Lambda^\perp/\Lambda),
\end{align}
where there are $2^{k+1}$ elements in the quotiont group for the $k$ encoded qubits.
Hence the the maximal set in Eq.~\eqref{eq:def_maximal_set} can also be divided into $2^{k+1}$ cosets $[\boldsymbol{\eta_s}-\boldsymbol{l}_i]$. 
Although the actual $\boldsymbol{\xi}$ is unknown, it has to be within the maximal set in Eq.~\eqref{eq:def_maximal_set} and hence one of the cosets $[\boldsymbol{\eta_s}-\boldsymbol{l}_i]$. 
The MLD {strategy} is to find the most probable coset, conditioned on the given syndrome $\boldsymbol{s}$, that could contain $\boldsymbol{\xi}$. 
In other words, MLD aims to solve the following problem
\begin{align}
\label{eq:MLD_1}  
    \text{argmax}_{\boldsymbol{l}\in\sqrt{2\pi}(\Lambda^\perp/\Lambda)}P([\boldsymbol{\eta_s}-\boldsymbol{l}]|\boldsymbol{s}),
\end{align}
where $P([\boldsymbol{\eta_s}]|\boldsymbol{s})$ denotes {the} probability of $[\boldsymbol{\eta_s}]$ conditioned on the given syndrome $\boldsymbol{s}$. 
With Bayes' theorem, which states $P(A|B)P(B) = P(B|A) P(A)$ for two events A and B with probabilities $P(A)$ and $P(B)$ respectively, we have
\begin{equation}
\begin{aligned}
\label{eq:MLD_2}
&\text{argmax}_{\boldsymbol{l}\in\sqrt{2\pi}(\Lambda^\perp/\Lambda)}P([\boldsymbol{\eta_s}-\boldsymbol{l}]|\boldsymbol{s}) \\
=& \text{argmax}_{\boldsymbol{l}\in\sqrt{2\pi}(\Lambda^\perp/\Lambda)}\frac{P([\boldsymbol{\eta_s}]-\boldsymbol{l})}{P(\boldsymbol{s})}  \\
=& \text{argmax}_{\boldsymbol{l}\in\sqrt{2\pi}(\Lambda^\perp/\Lambda)}\sum_{\boldsymbol{u}\in\sqrt{2\pi}\Lambda}P_\sigma(\boldsymbol{\eta_s}-\boldsymbol{l}-\boldsymbol{u}) .
\end{aligned}
\end{equation}
Here we have used the fact that $P(\boldsymbol{s}|[\boldsymbol{\eta_s}-\boldsymbol{l}])=1$. We have also omitted the factor $P(\boldsymbol{s})$, which is the same for all the logical operators, and hence irrelevant for the problem.
After the optimal solution $\boldsymbol{l}_*$ is found, we apply the corresponding  displacement operator $\hat{D}(-\boldsymbol{\eta}^*)\equiv\hat{D}(-\boldsymbol{\eta}_s+\boldsymbol{l}_*)$ onto the {corrupted} GKP code.
The net result of the error $\boldsymbol{\xi}$ and the attempted error correction is the displacement operator
\begin{align}
    \hat{D}(-\boldsymbol{\eta}_s+\boldsymbol{l}_*)\hat{D}(\boldsymbol{\xi}) \propto \hat{D}(-\boldsymbol{\eta}_s+\boldsymbol{l}_*+\boldsymbol{\xi}),
\end{align}
up to an irrelevant global phase. Because the coset $[\boldsymbol{\eta_s}-\boldsymbol{l}_*]$ is the most probable coset that contains $\boldsymbol{\xi}$, the probability of $-\boldsymbol{\eta}_s+\boldsymbol{l}_*+\boldsymbol{\xi}$ being a stabilizer has been maximized. 
Hence MLD is the optimal strategy to minimize the chance of getting a logical error after the attempted error correction.

\subsection{Efficient closest point decoder with decoding graph}
\label{sec:Efficient closest point decoder with decoding graph}

As discussed in Sec.~\ref{sec: Brief review of GKP codes}, in the limit of $\sigma\rightarrow0$, the MLD in Eq.~\eqref{eq:MLD_2} can be approximated by solving the following closest point problem
\begin{align}
    \label{eq:closest_point_problem}
    &\text{argmax}_{\boldsymbol{l}\in\sqrt{2\pi}(\Lambda^\perp/\Lambda)}\sum_{\boldsymbol{u}\in\sqrt{2\pi}\Lambda}P_\sigma(\boldsymbol{\eta_s}-\boldsymbol{l}-\boldsymbol{u}) \nonumber\\
    \approx&\text{argmin}_{\boldsymbol{l}^\perp\in\sqrt{2\pi}\Lambda^\perp}||\boldsymbol{\eta_s}-\boldsymbol{l}^\perp||.
\end{align}
Here, we will review the efficient closest point decoder implemented in Ref.~\cite{lin2023closest}, using the surface-square GKP code as an example. As an illustration, we can view each black dot in Fig.~\ref{fig:MWM}(a) as a one-mode square GKP code that encode a single qubit. The first step of the closest point decoder is to construct the model graph following the procedure described in Sec.~\ref{sec: The MWM decoder}.

For a given syndrome $\boldsymbol{s}$, a {decoding graph} \cite{wu2023fusion} is constructed, as shown in Fig.~\ref{fig:MWM}(c),
by highlighting those vertices with nontrivial stabilizer measurement outcomes.
To define the decoding graph mathematically, we consider an equivalent problem to the closest point problem in Eq.~\eqref{eq:closest_point_problem}, namely
\begin{align}
\label{eq: closest_point_problem_3}
\boldsymbol{\chi}_1\equiv\text{argmin}_{\boldsymbol{l}\in\sqrt{2}\Lambda^\perp}||\boldsymbol{\eta}'_{\boldsymbol{s}}-\boldsymbol{l}^\perp||,
\end{align}
where the scaled candidate error is defined as
\begin{align}
\label{eq:scaled_candidate_error}
    \boldsymbol{\eta}'_{\boldsymbol{s}} = \frac{\boldsymbol{\eta_s}}{\sqrt{\pi}}.
\end{align}
The solution to Eq.~\eqref{eq:closest_point_problem} is simply $\sqrt{\pi}\boldsymbol{\chi}_1$. From Eq.~\eqref{eq:chain_subsets}, because $2Z_{2N}\subset\sqrt{2}\Lambda^\perp\subset Z_{2N}$, $\boldsymbol{\chi}_1$ has to be an {integer valued} vector and satisfies
\begin{align}
\label{eq:u_1j}
    {\chi}_{1, j} = f_1(\eta_{s,j}') ~\text{or}~ f_2(\eta_{s,j}'),
\end{align}
where $\chi_{1, j}$ and $\eta_{s,j}'$ are the $j$-th component of $\boldsymbol{\chi}_1$ and $\boldsymbol{\eta}'_{\boldsymbol{s}}$ respectively. 
Here $f_1(x)$ and $f_2(x)$ are the first and second closest integers for $x\in\mathbb{R}$.
In other words, the closest point to $\boldsymbol{\eta}'_{\boldsymbol{s}}$ can be found by rounding each component to either its closest or second closest integer.
This, however, does not suggest that we have to consider $2^{2N}$ integer-value vectors because some of them are not even in the lattice $\sqrt{2}\Lambda^\perp$. In particular, by the definition of symplectic dual lattice (see Eq.~\eqref{eq:def_lambda_perp}), an integer valued vector $\boldsymbol{v}\in\sqrt{2}\Lambda^\perp$ if and only if \cite{lin2023closest}
\begin{align}
\label{eq:conditions_for_in_Lambda_perp}
    \text{mod}(\boldsymbol{g}_l^T\Omega\boldsymbol{v}, 2)=0, \quad 1\leq l\leq (N-k),
\end{align}
where $\left\{\boldsymbol{g}_l\right\}$ are the first $(N-k)$ rows in $\sqrt{2}M_\text{conc-sq}$ defined in Eq.~\eqref{eq:M_conc}. These binary vectors $\left\{\boldsymbol{g}_l\right\}$ correspond to the stabilizers of the concatenated GKP code. 

To define the decoding graph, we consider the following {integer valued} vector
\begin{align}
\label{eq:ansatz}
    \boldsymbol{\chi}'\equiv [f_1(\eta_{s,{1}}'), \cdots, f_1(\eta_{s,{2N}}')]^T.
\end{align}
It is clear that $\boldsymbol{\chi}'$ is the closest {integer valued} vector for $\boldsymbol{\eta_s}$, but it needs not be a solution of Eq.~\eqref{eq: closest_point_problem_3} because $\boldsymbol{\chi}'$ needs not satisfy the conditions in Eq.~\eqref{eq:conditions_for_in_Lambda_perp}. Despite that, $\boldsymbol{\chi}'$ serves as an ansatz for solving Eq.~\eqref{eq: closest_point_problem_3} with a decoding graph.
Specifically, after the model graph is set up, its $l$-th vertex will be highlighted if $\text{mod}(\boldsymbol{g}_l^T\Omega\boldsymbol{\chi}', 2)=1$.
The virtual vertex will be highlighted if and only {an odd number of white vertices are highlighted}, such that there are always even number of highlighted vertices in the decoding graph.
One of the key differences between the qubit surface code and the surface-square GKP code is that we will assign non-uniform weights to the edges, {that depends} on the candidate error $\boldsymbol{\eta}'_{\boldsymbol{s}}$.
In particular, for the edge $e_j$ that corresponds to the $j$-th GKP qubit,  we assign the weight \cite{lin2023closest}
\begin{align}
\label{eq:weight_e_k_gkp}
    w(e_j) \equiv (f_2(\eta'_{s, j}) - \eta'_{s, j})^2 - (f_1(\eta'_{s, j}) - \eta'_{s, j})^2.
\end{align}
In other words, the weight of the $j$-th edge given in Eq.~\eqref{eq:weight_e_k_gkp} is \emph{the distance squared penalty of rounding $\eta'_{s, j}$ in the wrong way}.

After the decoding graph $G = (\mathcal{V}, \mathcal{E}, \mathcal{W}, \mathcal{V}_h)$ is set up, the matchings of the graph are defined to be a subset of edges that satisfy the condition (see Eq.~\eqref{eq:def_matching})
\begin{align}
\label{eq:def_matching_2}
    \mathcal{V}_h = \oplus_{e\in\mathcal{M}}e.
\end{align}
For each matching $\mathcal{M}$, we can associate to it a $2N$-component vector $\boldsymbol{\chi}_\mathcal{M}$ where its $j$-th component is defined as
\begin{align}
\label{eq:components_u_M}
    \chi_{\mathcal{M}, j} = \begin{cases}
f_1(\eta_{s, j}') & \text{if } e_j\notin \mathcal{M} \\
f_2(\eta_{s, j}') & \text{if } e_j\in \mathcal{M}
\end{cases}.
\end{align}
By the definition of a matching, we have $\text{mod}(\boldsymbol{g}_l^T\Omega\boldsymbol{\chi}_\mathcal{M}, 2)=0$ for all $1\leq l\leq (N-k)$, and hence $\boldsymbol{\chi}_{\mathcal{M}}\in\sqrt{2}\Lambda^\perp$.
Upon comparing to Eq.~\eqref{eq:ansatz}, the distance between $\boldsymbol{\chi}_{\mathcal{M}}$ and $\boldsymbol{\eta}'_{\boldsymbol{s}}$ can be written as
\begin{align}
\label{eq:weight_closest_point}
    ||\boldsymbol{\chi}_{\mathcal{M}} - \boldsymbol{\eta}'_{\boldsymbol{s}}||^2 = ||\boldsymbol{\chi}' - \boldsymbol{\eta}'_{\boldsymbol{s}}||^2 + w(\mathcal{M}),
\end{align}
where the weight of the matching is defined in Eq.~\eqref{eq:weight_highlighted_matching_0}. Because $\boldsymbol{\chi}'$ is the closest possible integer valued vector to $\boldsymbol{\eta}'_{\boldsymbol{s}}$, {although} it needs not be in $\sqrt{2}\Lambda^\perp$, Eq.~\eqref{eq:weight_closest_point} suggests that the closest point in $\sqrt{2}\Lambda^\perp$ can be found by rounding certain components of $\boldsymbol{\chi}'$ in the wrong way, and the increased distance squared is given by the weight of the MWM.
The correspondence between the closest point of the surface-square GKP code and {the} MWM is one of the major findings in Ref.~\cite{lin2023closest}.
In Sec.~\ref{sec: $K$-MWM decoder for surface-square GKP codes}, {we will generalize this correspondence to subsequent MWMs and closest points.}.

\section{A lattice perspective for $K$-MWM decoding}
\label{sec:Maximum likelihood decoder for concatenated GKP codes: A lattice perspective}

In this section, we introduce the $K$-MWM decoder for several families of GKP codes.
We start by showing that the MLD for general GKP codes can be approximated by finding the first $K$ closest points in the corresponding symplectic dual lattice. This is a generalization of the closest point problem in Eq.~\eqref{eq:closest_point_problem}, hence it is not expected to be an efficient method for general GKP codes.
Nevertheless, as shown in App.~\ref{sec:Efficient closest point decoder with decoding graph}, the closest point for certain families of GKP codes, such as the surface-square GKP code, can indeed be found efficiently, which is the MWM of the decoding graph.
Motivated by {this}, we show that the MLD for surface-square GKP code can be approximated by finding $K$ MWMs from the decoding graph.
Because the symplectic dual lattice of the surface-square GKP code (or generally concatenated-square GKP code) admits a coset structure, we show that each MWM in the decoding graph corresponds to a coset representative, and the corresponding coset probability to the logical errors can be easily calculated once the coset representative is identified.
{This} coset structure allows us to approximate MLD without the need to find each individual closest point, and we generalize this approach to broader classes of surface-GKP codes.

\subsection{MLD approximated by $K$ closest points in the symplectic dual lattice}
\label{sec:MLD approximated by K closest points in the symplectic dual lattice}

Recall from App.~\ref{sec:Maximum likelihood decoder for GKP codes} that, for a given GKP lattice $\Lambda$, MLD aims to find a logical operator $\boldsymbol{l}\in\sqrt{2\pi}(\Lambda^\perp/\Lambda)$ that maximizes the summation 
\begin{align}
\label{eq:MLD_3}
\sum_{\boldsymbol{u}\in\sqrt{2\pi}\Lambda}P_\sigma(\boldsymbol{\eta_s}-\boldsymbol{l}-\boldsymbol{u}).
\end{align}
The closest point decoder finds an approximate solution by finding the largest term in the summation. In particular, the closest point decoder searches {for} an $\boldsymbol{l}^\perp\in\sqrt{2\pi}\Lambda^\perp$ that has the closest distance to the vector $\boldsymbol{\eta_s}$.
As noted in Sec.~\ref{sec: Brief review of GKP codes}, the closest point decoder is only a good approximation {to the } MLD in the limit of $\sigma\rightarrow0$.

For finite $\sigma$, it is possible to improve the approximation by finding subsequent closest points in the lattice $\sqrt{2\pi}\Lambda^\perp$. Let $\Xi=\left\{\boldsymbol{l}^\perp_i\in\sqrt{2\pi}\Lambda^\perp, 1\leq i\leq K\right\}$ be the set of first $K$ closest lattice points in $\sqrt{2\pi}\Lambda^\perp$ such that
\begin{align*}
    ||\boldsymbol{\eta_s}-\boldsymbol{l}^\perp_1||\leq ||\boldsymbol{\eta_s}-\boldsymbol{l}^\perp_2||\leq \cdots ||\boldsymbol{\eta_s}-\boldsymbol{l}^\perp_K||\leq ||\boldsymbol{\eta_s}-\boldsymbol{l}^\perp||,
\end{align*}
for all $\boldsymbol{l}^\perp\in\sqrt{2\pi}\Lambda^\perp$ {that are} not in $\Xi$. 
We can partition the set $\Xi$ as
\begin{align}
\label{eq:partition_of_Xi}
    \Xi \equiv \bigcup_{\boldsymbol{l}\in\sqrt{2\pi}(\Lambda^\perp/\Lambda)}\Xi_{\boldsymbol{l}},
\end{align}
where $\Xi_{\boldsymbol{l}}$ contains lattice points $\boldsymbol{l}^\perp\in\sqrt{2\pi}\Lambda^\perp$ that result in the logical operator $\boldsymbol{l}\in\sqrt{2\pi}(\Lambda^\perp/\Lambda)$.
Then the MLD in Eq.~\eqref{eq:MLD_2} can be approximated as
\begin{equation}
    \begin{aligned}
        &\text{argmax}_{\boldsymbol{l}\in\sqrt{2\pi}(\Lambda^\perp/\Lambda)}\sum_{\boldsymbol{u}\in\sqrt{2\pi}\Lambda}P_\sigma(\boldsymbol{\eta_s}-\boldsymbol{l}-\boldsymbol{u}) \\
        \approx& \text{argmax}_{\boldsymbol{l}\in\sqrt{2\pi}(\Lambda^\perp/\Lambda)} \sum_{\boldsymbol{l}^\perp\in \Xi_{\boldsymbol{l}}}P_\sigma(\boldsymbol{\eta_s}-\boldsymbol{l}^\perp) .
    \end{aligned}
\end{equation}
The accuracy of the approximation can be improved further by increasing the size of $\Xi$, and the {computational} complexity increases linearly with the size of $\Xi$. 

Hence, in order to approximate the MLD, we solve the following {optimization} problem
\begin{align}
\label{eq:approximated_MLD}
    \Xi = \text{argmin}^{(K)}_{\boldsymbol{l}^\perp\in\sqrt{2\pi}\Lambda^\perp}||\boldsymbol{\eta_s}-\boldsymbol{l}^\perp||,
\end{align}
where the superscript $(K)$ indicates that we need to find the first $K$ solutions to the optimization problem. When $K=1$, Eq.~\eqref{eq:approximated_MLD} simply reduces to the closest point problem presented in Eq.~\eqref{eq:closest_point_problem_0}.

\subsection{Approximate MLD for concatenated-square GKP codes}
\label{sec: approximate MLD for concatenated-square GKP codes}

As noted in Sec.~\ref{sec: Brief review of GKP codes}, finding the closest point of a generic lattice is an NP-hard problem, let alone finding its {first} $K$ closest points. 
On the other hand, in App.~\ref{sec:Efficient closest point decoder with decoding graph} we demonstrated efficient algorithms to find the closest point for structured lattices, particularly for those similar to the surface-square GKP code. 
Hence, we anticipate that it is possible to efficiently find the $K$ closest points, or approximate the MLD for certain concatenated-square GKP codes.

This is indeed the case for surface-square GKP code as we have demonstrated in Sec.~\ref{sec: Surface-square GKP code: A lattice perspective of $K$-MWM decoding}. More generally, we can infer from Eq.~\eqref{eq:chain_subsets} that for an arbitrary concatenated-square GKP lattice $\Lambda$, its scaled dual lattice $\sqrt{2}\Lambda^\perp$ contains $2Z_{2N}$ as a sublattice. Hence a coset structure similar to the surface-square GKP code can be defined for arbitrary concatenated-square GKP codes. With that, the MLD for a concatenated-square GKP code reduces to find{ing} the first $K$ coset representatives in the ordered set $\mathcal{U}$, which is defined in Eq.~\eqref{eq:def_ordered_representatives_0}, followed by approximating the logical error probability with Eq.~\eqref{eq:def_U_l}-\eqref{eq:def_P_chi_0}.

\subsection{$K$-MWM decoder for surface-square GKP codes}
\label{sec: $K$-MWM decoder for surface-square GKP codes}

Here we show that, for concatenated-square GKP codes that are similar to the surface-square GKP code, the coset representatives are one-to-one corresponding to the MWMs of the decoding graph.

Recall from Sec.~\ref{sec:Efficient closest point decoder with decoding graph}, a decoding graph $G = (\mathcal{V}, \mathcal{E}, \mathcal{W}, \mathcal{V}_h)$ can be defined for the surface-square GKP code, and its MWM corresponds to the closest point $\boldsymbol{\chi}_1$. 
Here we generalize this correspondence. Let 
\begin{align}
    \mathcal{U}_\mathcal{M}\equiv \left\{\mathcal{M}_1(G), \mathcal{M}_2(G), \cdots, \mathcal{M}_{|\mathcal{U}_\mathcal{M}|}(G)\right\}
\end{align}
denote the set of all matchings of the decoding graph $G$, which are ordered by the weights of the matching. 
We claim that there is a one-to-one mapping between $\mathcal{U}_\mathcal{M}$ and the ordered set $\mathcal{U}$ of coset representatives defined in Eq.~\eqref{eq:def_ordered_representatives_0}.
To see that, we first note that for an arbitrary matching $\mathcal{M}(G)\in\mathcal{U}_\mathcal{M}$, an {integer valued} vector $\boldsymbol{\chi}_\mathcal{M}$ can be defined based on Eq.~\eqref{eq:components_u_M}. The vector $\boldsymbol{\chi}_\mathcal{M}$ has to satisfy the conditions in Eq.~\eqref{eq:conditions_for_in_Lambda_perp}, otherwise there will be vertices left un-matched in the decoding graph. Hence $\boldsymbol{\chi}_{\mathcal{M}}\in\sqrt{2}\Lambda^\perp$ and it must belong to certain coset $[[\boldsymbol{\chi}]]$. 
Because the components of $\boldsymbol{\chi}_{\mathcal{M}}$ are either the first or the second closest points of the corresponding components of $\boldsymbol{\eta}'_{\boldsymbol{s}}$, we have that $\boldsymbol{\chi}_{\mathcal{M}}$ must be the closest point in $[[\boldsymbol{\chi}]]$, namely $\boldsymbol{\chi}$ by our convention. Hence for an arbitrary $\mathcal{M}(G)\in\mathcal{U}_\mathcal{M}$, there is a corresponding $\boldsymbol{\chi}_{\mathcal{M}}\in\mathcal{U}$, and distinct $\mathcal{M}(G)$ are mapped to distinct $\boldsymbol{\chi}_{\mathcal{M}}$ per Eq.~\eqref{eq:components_u_M}. 
Similarly, for a given $\boldsymbol{\chi}\in\mathcal{U}$, since the $j$-th component $\chi_j$ is either $f_1(\eta_{s, j}')$ or $f_2(\eta_{s, j}')$, we can define the corresponding matching $\mathcal{M}\in\mathcal{U}_\mathcal{M}$ using Eq.~\eqref{eq:components_u_M}. The matching has to satisfy the condition in Eq.~\eqref{eq:def_matching_2}, otherwise it implies that $\boldsymbol{\chi}$ will violate one of the constraints in Eq.~\eqref{eq:conditions_for_in_Lambda_perp}, rendering the vector $\boldsymbol{\chi}$ not in the lattice $\sqrt{2}\Lambda^\perp$.

This one-to-one correspondence is the core of the $K$-MWM decoder for the surface-square GKP code. It implies that finding the first $K$ closest coset representatives, for the purpose of approximating MLD, is equivalent to finding the first $K$ MWMs in the decoding graph, which is addressed in Sec.~\ref{sec:Finding the $K$-th matching for the decoding graph}.

\subsection{Coset structures for general surface-GKP codes}
\label{sec: Coset structures for general surface-GKP codes}
As shown in Sec.~\ref{sec: Surface-square GKP code: A lattice perspective of $K$-MWM decoding}, the $K$-MWM decoder depends on the coset structure of the surface-square GKP code, which can be generalized to more general surface-GKP code. Following Eq.~\eqref{eq:M_general}, we can write the generator matrix for the surface-GKP code as
\begin{align}
\label{eq:generator_surf}
    M_\text{surf} = M_\text{surf-sq}S^T,
\end{align}
where $M_\text{surf-sq}$ is the generator matrix for the surface-square GKP code.
Since the surface-GKP code is not a concatenated-square GKP code, the coset structure defined in Sec.~\ref{sec: approximate MLD for concatenated-square GKP codes} cannot be applied in this case. Nevertheless, the surface-GKP code admits another coset structure because its scaled symplectic dual lattice $\sqrt{2}\Lambda^\perp_\text{surf}\equiv\sqrt{2}\Lambda(M^\perp_\text{surf})$ contains a sublattice $2\Lambda(S^T)$ with generator matrix $2S^T$.

To see that, consider an arbitrary lattice point $\boldsymbol{u}\in 2\Lambda(S^T)$ given by
\begin{align}
    \boldsymbol{u} = 2S\boldsymbol{a} = S(2\boldsymbol{a}).
\end{align}
Here $\boldsymbol{a}$ is an {integer valued} vector, hence $2\boldsymbol{a}\in2Z_{2N}\subset\sqrt{2}\Lambda(M_\text{surf-sq})$ because the surface-square GKP lattice contains $2Z_{2N}$ as a sublattice per Eq.~\eqref{eq:chain_subsets}. This implies that $2\boldsymbol{a}=\sqrt{2}(M_\text{surf-sq})^T\boldsymbol{b}$ for some $\boldsymbol{b}\in\mathbb{Z}^{2N}$ such that
\begin{align*}
    \boldsymbol{u} = 
    \sqrt{2}S(M_\text{surf-sq})^T\boldsymbol{b} =
    \sqrt{2}M_\text{surf}^T\boldsymbol{b} \in \sqrt{2}\Lambda_\text{surf}\subset\sqrt{2}\Lambda_\text{surf}^\perp,
\end{align*}
where we used the fact that a GKP lattice is a sublattice of its symplectic dual lattice. Hence, for a lattice point $\boldsymbol{\chi}\in\sqrt{2}\Lambda_\text{surf}^\perp$, we can define a set of equivalent lattice points as
\begin{align}
    \left\{\boldsymbol{\chi}-\boldsymbol{v} ~|~ \boldsymbol{v}\in2\Lambda(S^T)\right\},
\end{align}
which is {analogous} to Eq.~\eqref{eq:coset_u_v1_0}, the coset for concatenated-square GKP codes. Similar to the arguments in Sec.~\ref{sec: approximate MLD for concatenated-square GKP codes}, the coset structure allows us to approximate the MLD for the surface-GKP code via finding its coset representatives followed by calculating the coset probabilities, which will be illustrated below. 

\subsection{$K$-MWM decoder for the surface-rectangular GKP code}
\label{sec: approximate MLD for the surface-rectangular GKP code}

To prepare the discussion of $K$-MWM decoding for general surface-GKP code in App.~\ref{sec: Details of approximate MLD for the surface-GKP code}, we present the $K$-MWM decoder for the surface-rectangular GKP code where, instead of square GKP code, the inner codes are rectangular GKP codes. The aspect ratios of each mode could be different, and following Eq.~\eqref{eq:M_general}, the generator matrix for the surface-rectangular GKP code can be written as 
\begin{align}
\label{eq:generator_surf_rec}
    M_\text{surf-rec} = \begin{bmatrix}
        M^{(\hat{q})}_\text{surf-sq} & \\
        & M^{(\hat{p})}_\text{surf-sq} \\
    \end{bmatrix}
    \begin{bmatrix}
        \Gamma & \\  & \Gamma^{-1}
    \end{bmatrix},
\end{align}
where $M_\text{surf-sq}$, the generator matrix for the surface-square GKP code, is a direct sum of that for the $\hat{q}$ and $\hat{p}$ subspaces respectively. 
Here $\Gamma\equiv\text{diag}(\gamma_1,\cdots,\gamma_N)$ is a diagonal matrix, where $\gamma_i>0$ is the square root of the aspect ratio for the $i$-th rectangular GKP code. It is clear that Eq.~\eqref{eq:generator_surf_rec} is a special case of $M_\text{surf}$ defined in Eq.~\eqref{eq:generator_surf}, and hence the coset structure described in Sec.~\ref{sec: Coset structures for general surface-GKP codes} can be applied to the surface-rectangular GKP code.

To find the coset representatives in the scaled symplectic lattice for the surface-rectangular GKP code, we consider
\begin{equation}
    \begin{aligned}
        M_\text{surf-rec}^\perp &= -\Omega(M_\text{surf-rec}^T)^{-1}\Omega\\
        &=-\Omega\begin{bmatrix}
        M^{(\hat{q}),T}_\text{surf-sq} & \\
        & M^{(\hat{p}),T}_\text{surf-sq} \\
    \end{bmatrix}^{-1}\begin{bmatrix}
        \Gamma^{-1} & \\  & \Gamma
    \end{bmatrix}\Omega\\
    &=\begin{bmatrix}
        {M^{(\hat{p}),T}_\text{surf-sq}}^{-1}\Gamma & \\
         & {M^{(\hat{q}),T}_\text{surf-sq}}^{-1}\Gamma^{-1}
    \end{bmatrix}.
    \end{aligned}
\end{equation}
Because the $\hat{q}$ and $\hat{p}$ subspaces are decoupled from each other, we can decode each subspace separately followed by determining the most likely logical error. In particular, for the case of surface-square GKP code, where $\Gamma=I_N$, the logical error in the $\hat{q}$ subspace is found by finding the coset representatives in the lattice generated by ${M^{(\hat{p}),T}_\text{surf-sq}}^{-1}$ that are closest to the scaled candidate error ${\boldsymbol{\eta}'_{\boldsymbol{s}}}^{(\hat{q})}$. For that, we have shown an algorithm in Sec.~\ref{sec: approximate MLD for concatenated-square GKP codes}-\ref{sec: $K$-MWM decoder for surface-square GKP codes}, which finds the most likely logical error via finding $K$ MWMs in the decoding graph that corresponds to ${M^{(\hat{p}),T}_\text{surf-sq}}^{-1}$.
For the surface-rectangular GKP code, where $\Gamma\neq I_N$, we redefine the candidate error in the $\hat{q}$ subspace as
\begin{align}
    {\boldsymbol{\eta}''_{\boldsymbol{s}}}^{(\hat{q})}\equiv \Gamma^{-1} {\boldsymbol{\eta}'_{\boldsymbol{s}}}^{(\hat{q})},
\end{align}
where each component of the candidate error is scaled by the diagonal matrix $\Gamma^{-1}$. With that, the decoding graph for the surface-rectangular GKP code can be defined, which has the same connectivity as that for the surface-square code, but with weights defined as
\begin{equation}
\label{eq:weight_e_k_2}
    \begin{aligned}
        w(e_j) &\equiv \left[(\gamma_jf_2(\eta_{s, j}''^{(\hat{q})}) - \eta_{s, j}'^{(\hat{q})})^2 - (\gamma_jf_1(\eta_{s, j}''^{(\hat{q})}) - \eta_{s, j}'^{(\hat{q})})^2\right]\\
        &=\gamma_j^2\left[(f_2(\eta_{s, j}''^{(\hat{q})}) - \eta_{s, j}''^{(\hat{q})})^2 - (f_1(\eta_{s, j}''^{(\hat{q})}) - \eta_{s, j}''^{(\hat{q})})^2\right].
    \end{aligned}
\end{equation}
The weights in Eq.~\eqref{eq:weight_e_k_2} are analogous to those in Eq.~\eqref{eq:weight_e_k_gkp} defined for the surface-square GKP code. 
It is important to note that 
\begin{align}
    \gamma_if_1(\eta_{s, i}''^{(\hat{q})}) = \gamma_if_1\left(\frac{\eta_{s, i}'^{(\hat{q})}}{\gamma_i}\right) \neq f_1(\gamma_i\eta_{s, i}''^{(\hat{q})})=f_1(\eta_{s, i}'^{(\hat{q})}),
\end{align}
and one simple example is $2f_1(0.8/2) = 0\neq 1 = f_1(0.8)$.
 
We can proceed to find the $K$ MWMs for the decoding graph, following the algorithm in Sec.~\ref{sec:Finding the $K$-th matching for the decoding graph}.
For each matching $\mathcal{M}$, we can associate to it an $N$-component vector $\boldsymbol{\chi}_\mathcal{M}^{(\hat{q})}$ where its $j$-th component is defined as
\begin{align}
\label{eq:components_u_M_surf_rec}
    \chi_{\mathcal{M}, j}^{(\hat{q})} = \begin{cases}
\gamma_jf_1(\eta_{s, j}''^{(\hat{q})}) & \text{if } e_j\notin \mathcal{M} \\
\gamma_jf_2(\eta_{s, j}''^{(\hat{q})}) & \text{if } e_j\in \mathcal{M}
\end{cases}.
\end{align}
It is important to note that $\Gamma^{-1}\boldsymbol{\chi}_\mathcal{M}^{(\hat{q})}$ is essentially the same as the vector defined in Eq.~\eqref{eq:components_u_M}, which is in the $\hat{q}$-subspace of the surface-square GKP code.
In other words, $\boldsymbol{\chi}_\mathcal{M}^{(\hat{q})}$ is a vector in the $\hat{q}$-subspace of the surface-rectangular GKP code
\begin{align}
\label{eq:coset_rep_surf_rec}
    \boldsymbol{\chi}_\mathcal{M}^{(\hat{q})} = \left({M^{(\hat{p}),T}_\text{surf-sq}}^{-1}\Gamma\right)^T\boldsymbol{a} = \left(M_\text{surf-rec}^{\perp, (\hat{q})}\right)^T\boldsymbol{a}, 
\end{align}
for some integer valued vector $\boldsymbol{a}$. 
With that, the distance between $\boldsymbol{\chi}_\mathcal{M}^{(\hat{q})}$ and ${\boldsymbol{\eta}'_{\boldsymbol{s}}}^{(\hat{q})}$ is given by
\begin{align}
\label{eq:eq:coset_rep_surf_rec_dist}
    ||\boldsymbol{\chi}_\mathcal{M}^{(\hat{q})} - {\boldsymbol{\eta}'_{\boldsymbol{s}}}^{(\hat{q})}||^2 = ||\boldsymbol{\chi}'^{(\hat{q})} - {\boldsymbol{\eta}'_{\boldsymbol{s}}}^{(\hat{q})}||^2 + w(\mathcal{M}),
\end{align}
where, analogous to Eq.~\eqref{eq:ansatz},
\begin{align}
    \boldsymbol{\chi}'^{(\hat{q})} = [\gamma_1f_1(\eta_{s, 1}''^{(\hat{q})}), \cdots, \gamma_Nf_1(\eta_{s, N}''^{(\hat{q})})]^T.
\end{align}

With Eq.~\eqref{eq:components_u_M_surf_rec} and Eq.~\eqref{eq:coset_rep_surf_rec}-\eqref{eq:eq:coset_rep_surf_rec_dist}, following the argument in Eq.~\eqref{sec: $K$-MWM decoder for surface-square GKP codes}, we see that the coset representatives are one-to-one corresponding to the MWMs in the decoding graph, for the $\hat{q}$-suspace, and can be ordered by the weights of the MWMs. In order to find the most probable logical error, we need to calculate the coset probability for the coset representative $\boldsymbol{\chi}^{(\hat{q})}$, which is given by
\begin{equation}
\label{eq:def_P_chi_surf_rec}
    \begin{aligned}
        &P([[\boldsymbol{\chi}^{(\hat{q})}]])
        =\sum_{\boldsymbol{v}\in2Z_{N}}P_\sigma(\sqrt{\pi}(\boldsymbol{\eta}_{\boldsymbol{s}}'^{(\hat{q})}-\boldsymbol{\chi}^{(\hat{q})}-\Gamma\boldsymbol{v}))\\
        =&\frac{1}{\sqrt{2\pi\sigma^2}}\sum_{\boldsymbol{v}\in2Z_{N}}\prod_{i=1}^{N}\exp\left\{-\frac{\pi(\eta_{s, i}'^{(\hat{q})}-{\chi}_i^{(\hat{q})}-\gamma_iv_i)^2}{2\sigma^2}\right\}\\
        =&\frac{1}{\sqrt{2\pi\sigma^2}}\prod_{i=1}^{N}\sum_{{v}_i\in\mathbb{Z}}\exp\left\{-\frac{\pi(\eta_{s, i}'^{(\hat{q})}-{\chi}_i^{(\hat{q})}-2\gamma_iv_i)^2}{2\sigma^2}\right\},
    \end{aligned}
\end{equation}
which generalizes that for the surface-square GKP code, as defined in Eq.~\eqref{eq:def_P_chi_0}. 
The coset representatives in the $\hat{p}$-subspace can be found similarly, and their coset probability is similar to Eq.~\eqref{eq:def_P_chi_surf_rec}, except $\gamma_i$ is replaced by $\gamma_i^{-1}$.

\subsection{$K$-MWM decoder for general surface-GKP codes}
\label{sec: Details of approximate MLD for the surface-GKP code}

Here, we describe the $K$-MWM decoder for general surface-GKP codes. Recall that the generator matrix for the surface-GKP code reads
\begin{align}
\label{eq:generator_surf_app}
    M_\text{surf} = M_\text{surf-sq}S^T = \begin{bmatrix}
        M^{(\hat{q})}_\text{surf-sq} & \\
        & M^{(\hat{p})}_\text{surf-sq} \\
    \end{bmatrix}
    \begin{bmatrix}
        \Gamma_1 & \Gamma_2 \\ \Gamma_3 & \Gamma_4
    \end{bmatrix}^T.
\end{align}
As we have shown in Sec.~\ref{sec: Coset structures for general surface-GKP codes}, the scaled symplectic dual lattice of the surface-GKP code, $\sqrt{2}\Lambda^\perp_\text{surf}$, contains a sublattice $2\Lambda(S^T)$, which allows us to approximate the MLD for the surface-GKP code via finding its coset representatives followed by calculating the coset probabilities.

The key step of the $K$-MWM decoder is to identify the coset representatives for the symplectic dual lattice $\sqrt{2}\Lambda_\text{surf}^\perp$. We write its generator matrix as
\begin{equation}
\label{eq:M_surf_perp_general_surf_gkp_old}
    \begin{aligned}
        M_\text{surf}^\perp &= -\Omega(M_\text{surf}^T)^{-1}\Omega\\
        &=-\Omega\begin{bmatrix}
        M^{(\hat{q}),T}_\text{surf-sq} & \\
        & M^{(\hat{p}),T}_\text{surf-sq} \\
    \end{bmatrix}^{-1}\begin{bmatrix}
        \Gamma_4 & -\Gamma_2 \\ -\Gamma_3 & \Gamma_1
    \end{bmatrix}\Omega\\
    &=\begin{bmatrix}
        M^{(\hat{p}),T}_\text{surf-sq} & \\
        & M^{(\hat{q}),T}_\text{surf-sq} \\
    \end{bmatrix}^{-1}S,
    \end{aligned}
\end{equation}
where we have used the fact that $\Gamma_1\Gamma_4-\Gamma_2\Gamma_3=I_N$ (see Eq.~\eqref{eq:symplectic_condition}).
For the convience below, let us notice from Eq.~\eqref{eq:generator_surf_app} that $S$ can be written as a block diagonal matrix
\begin{align}
\label{eq:def_S}
    S \equiv \oplus_{i=1}^NS_i \equiv \oplus_{i=1}^N\begin{bmatrix}
        \gamma_{1,i} & \gamma_{2,i}\\
        \gamma_{3,i} & \gamma_{4,i}
    \end{bmatrix},
\end{align}
where we used the fact that $\Gamma_i=\text{diag}(\gamma_{1,1},\cdots,\gamma_{1,N})$ is a diagonal matrix, similarly for $\Gamma_{2,3,4}$.
For each qubit, we can always find an orthogonal matrix $O_i$ to rotate the basis such that
\begin{align}
    \tilde{S}_i \equiv S_i O_i \equiv \begin{bmatrix}
        \tilde{\gamma}_{1,i} & \tilde{\gamma}_{2,i}\\
        0 & \tilde{\gamma}_{4,i}
    \end{bmatrix}.
\end{align}
The full basis rotation for the surface-GKP code is the direct sum 
\begin{align}
    O\equiv\oplus_{i=1}^NO_i,
\end{align}
which can be applied to the symplectic dual lattice $\sqrt{2}\Lambda_\text{surf}^\perp$. The generator matrix in the rotated basis reads
\begin{equation}
\label{eq:M_surf_perp_general_surf_gkp}
    \begin{aligned}
        \tilde{M}_\text{surf}^\perp &= {M}_\text{surf}^\perp O
        &=\begin{bmatrix}
        M^{(\hat{p}),T}_\text{surf-sq} & \\
        & M^{(\hat{q}),T}_\text{surf-sq} \\
    \end{bmatrix}^{-1}\begin{bmatrix}
        \tilde{\Gamma}_1 & \tilde{\Gamma}_2 \\ 0_N & \tilde{\Gamma}_4
    \end{bmatrix}^T
    \end{aligned},
\end{equation}
where $\tilde{\Gamma}_1=\text{diag}(\tilde{\gamma}_{1,1},\cdots,\tilde{\gamma}_{1,N})$, similarly for $\tilde{\Gamma}_{2,4}$. 
Essentially, the basis rotation transforms the symplectic matrix to a upper triangular form with $\tilde{\Gamma}_{3}=0_N$.
In this basis, a lattice point $\boldsymbol{l}^\perp\in\sqrt{2}\Lambda_\text{surf}^\perp$ can be written as
\begin{equation*}
    \begin{aligned}
    \boldsymbol{l}^\perp &= \sqrt{2}(\tilde{M}_\text{surf}^\perp)^T\boldsymbol{a} = \begin{bmatrix}
        \tilde{M}_{qq} & \tilde{M}_{qp}\\
        0_N & \tilde{M}_{pp}
    \end{bmatrix}
    \begin{bmatrix}
        \boldsymbol{a}^{(\hat{q})}\\
        \boldsymbol{a}^{(\hat{p})}
    \end{bmatrix},
    \end{aligned}
\end{equation*}
where the block matrices can be determined from Eq.~\eqref{eq:M_surf_perp_general_surf_gkp} as 
\begin{equation}
    \begin{aligned}
        \tilde{M}_{qq} & = \sqrt{2}\tilde{\Gamma}_1 M^{(\hat{p}),-1}_\text{surf-sq}\\
        \tilde{M}_{qp} & = \sqrt{2}\tilde{\Gamma}_2 M^{(\hat{q}),-1}_\text{surf-sq}\\
        \tilde{M}_{pp} & = \sqrt{2}\tilde{\Gamma}_4 M^{(\hat{q}),-1}_\text{surf-sq}
    \end{aligned}.
\end{equation}
Here $\boldsymbol{a}$ is an integer valued vector, $\boldsymbol{a}^{(\hat{q})}$ and $\boldsymbol{a}^{(\hat{p})}$ indicate its supports in the $\hat{q}$ and $\hat{p}$ subspaces respectively. 
To identify the coset representatives, let us consider the distance between $\boldsymbol{l}^\perp$ and the rotated candidate error
\begin{align}
    \tilde{\boldsymbol{\eta}}_{\boldsymbol{s}}'\equiv O^T\boldsymbol{\eta}_{\boldsymbol{s}}',
\end{align}
which reads
\begin{equation}
\label{eq:dist_eta_s'}
\begin{aligned}
    &||\tilde{\boldsymbol{\eta}}_{\boldsymbol{s}}' -\boldsymbol{l}^\perp ||^2\\
    =&||\tilde{\boldsymbol{\eta}}_{\boldsymbol{s}}'^{(\hat{q})} -(\tilde{M}_{qq}\boldsymbol{a}^{(\hat{q})}+\tilde{M}_{qp}\boldsymbol{a}^{(\hat{p})}) ||^2  + ||\tilde{\boldsymbol{\eta}}_{\boldsymbol{s}}'^{(\hat{p})} - \tilde{M}_{pp}\boldsymbol{a}^{(\hat{p})}||^2.
\end{aligned}
\end{equation}
We note that the norm of a vector is invariant under basis rotation.
For surface-square GKP code, or more generally surface-rectangular GKP code, because $\tilde{M}_{qp}=0_N$, their coset representatives can be found independently for the two subspaces. As shown in Appendix \ref{sec: approximate MLD for the surface-rectangular GKP code}, if $\tilde{M}_{qp}=0_N$, we can first identify the coset representative in the $\hat{p}$ subspace by minimizing the second term in Eq.~\eqref{eq:dist_eta_s'}, followed by repeating the same process for the first term for finding the coset representative in the $\hat{q}$ subspace. The procedures for the two subspaces are exactly identical which use the algorithm presented in Sec.~\ref{sec: approximate MLD for concatenated-square GKP codes}-\ref{sec: $K$-MWM decoder for surface-square GKP codes}.

This is, however, not the case for general surface-GKP codes, and we have to consider the interplay between the coset representatives in the two subspaces. 
In particular, because $\tilde{M}_{qp}\neq 0_N$, the coset representative found in the $\hat{p}$-subspace, corresponding to certain $\boldsymbol{a}^{(\hat{p})}$ in the second term of Eq.~\eqref{eq:dist_eta_s'}, would manifest in the $\hat{q}$ subspace, as apparent from Eq.~\eqref{eq:dist_eta_s'}. 

To resolve the above issue, we could first identify a set of $K$ coset representatives in the $\hat{p}$ subspace that are closest to $\tilde{\boldsymbol{\eta}}_{\boldsymbol{s}}'^{(\hat{p})}$. These coset representatives can be associated to a set of vectors $\boldsymbol{a}^{(p), i}$ for $1\leq i\leq K$. 
Per Eq.~\eqref{eq:dist_eta_s'}, for each $\boldsymbol{a}^{(p), i}$, we define a modified candidate error
\begin{align}
\label{eq:modified_candidate_error_q}
    \tilde{\boldsymbol{\eta}}_{\boldsymbol{s}}''^{(\hat{q}), i} \equiv \tilde{\boldsymbol{\eta}}_{\boldsymbol{s}}'^{(\hat{q})} - \tilde{M}_{qp}\boldsymbol{a}^{(\hat{p}), i},
\end{align}
followed by using the algorithm in Sec.~\ref{sec: $K$-MWM decoder for surface-square GKP codes} to find the $K$ coset representatives in the $\hat{q}$ subspace.
We note that the weights in the decoding graph need to be redefined as described in Appendix \ref{sec: approximate MLD for the surface-rectangular GKP code} (see Eq.~\eqref{eq:weight_e_k_2}). Hence different $\boldsymbol{a}^{(p), i}$ correspond to decoding graphs with the same set of edges, vertices and highlighted vertices but different weights. 
The above process in total yields $K^2$ coset representatives, after which the inverse rotation $O$ will be applied to rotate them back to the original basis.

For each coset representative $\boldsymbol{\chi}$, the corresponding coset probability is given by 
\begin{align}
    P([[\boldsymbol{\chi}]]) \equiv \sum_{\boldsymbol{u}\in2\Lambda(S^T)}P_\sigma(\sqrt{\pi}(\boldsymbol{\eta}'_{\boldsymbol{s}} - \boldsymbol{\chi} - \boldsymbol{u})).
\end{align}
Because $\Lambda(S^T)$ is a direct sum of $N$ sublattices, we can rewrite the coset probability as
\begin{equation}
\label{eq:coset_prob_surf_GKP}
    \begin{aligned}
        P([[\boldsymbol{\chi}]]) &= \sum_{\boldsymbol{u}\in2\Lambda(S^T)}\prod_{i=1}^NP_\sigma(\sqrt{\pi}(\boldsymbol{\eta}'_{\boldsymbol{s}, i} - \boldsymbol{\chi}_i - \boldsymbol{u}_i))\\
    &= \prod_{i=1}^N\sum_{\boldsymbol{u}_i\in2\Lambda(S_i^T)}P_\sigma(\sqrt{\pi}(\boldsymbol{\eta}'_{\boldsymbol{s}, i} - \boldsymbol{\chi}_i - \boldsymbol{u}_i)),
    \end{aligned}
\end{equation}
where $\boldsymbol{\eta}'_{\boldsymbol{s}, i}$, $\boldsymbol{\chi}_i$ and $\boldsymbol{u}_i$ are two-component vectors, the components of $\boldsymbol{\eta}'_{\boldsymbol{s}}$,  $\boldsymbol{\chi}$ and $\boldsymbol{u}$ in the $i$-th qubit respectively.
%
%
In Eq.~\eqref{eq:coset_prob_surf_GKP}, we have expressed the coset probability as a product of probability for each qubit. 
To proceed, we will need to truncate the summation for each qubit.
Since $\Lambda(S_i^T)$ is a two-dimensional lattice, it is possible to exhaustively search through the lattice and find $N_v$ closest points to $\boldsymbol{\eta}'_{\boldsymbol{s}, i} - \boldsymbol{\chi}_i$.
In practise, we find that $N_v\leq 4$ is sufficient for the codes explored, and Eq.~\eqref{eq:coset_prob_surf_GKP} constitutes a generalization of Eq.~\eqref{eq:def_P_chi_0} for the surface-square GKP code.

\begin{figure}[!ht]
\centering
\includegraphics[width=\linewidth]{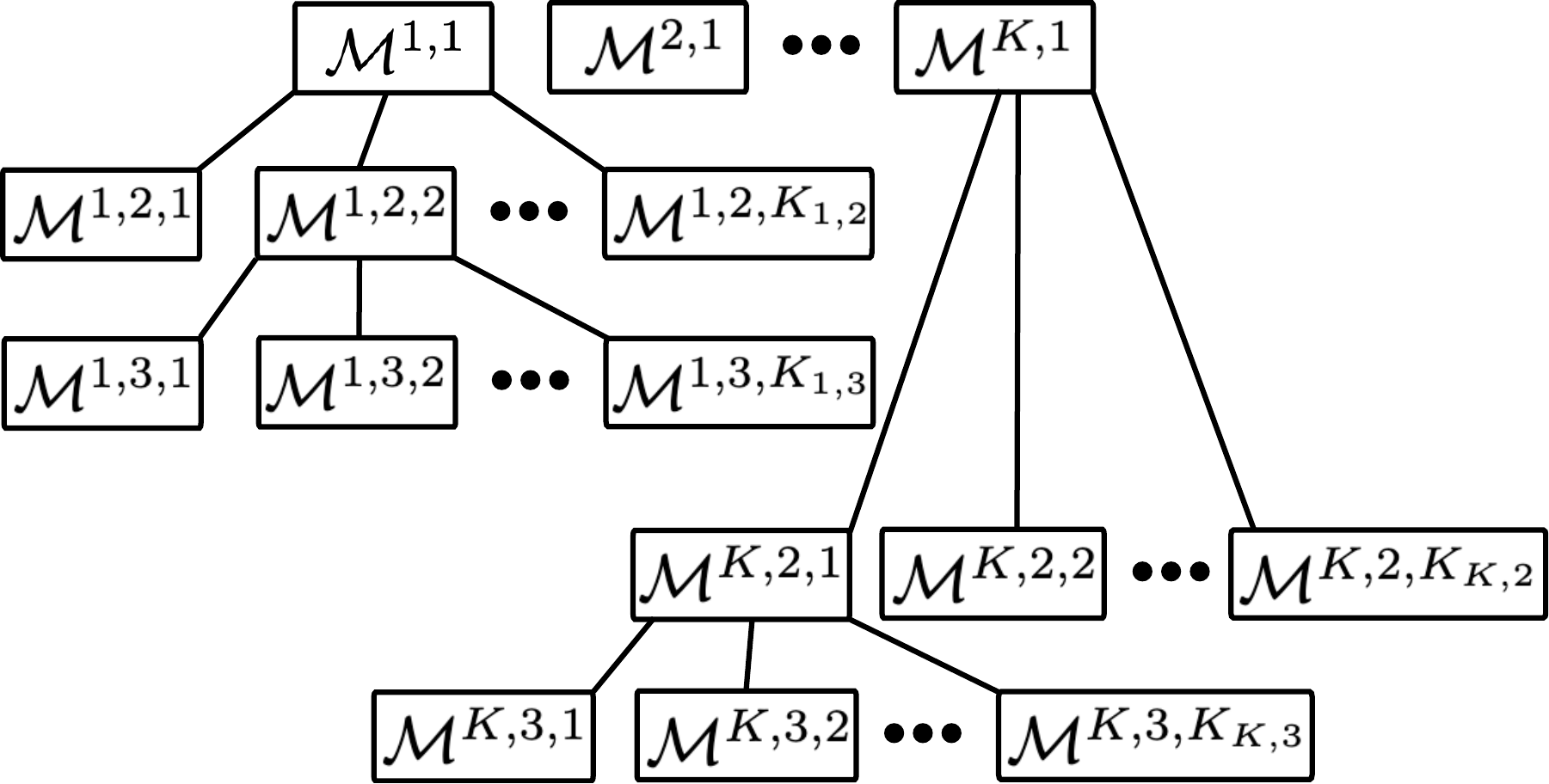}
 \caption{Decoding tree for more effectively decoding a general surface-GKP code. See Appendix \ref{sec: Details of approximate MLD for the surface-GKP code} for more description. 
}
 \label{fig:decoding_tree_v2}
\end{figure}

After implementing the above ``brute-force'' approach, we find that the algorithm is less effective than expected, primarily due to the fact that most of the identified matchings have negligible contribution to the coset probability. To more effectively identify the $K$ most important representatives, we adopt a decoding tree approach as depicted in Fig.~\ref{fig:decoding_tree_v2}. Despite the tree is similar to the one shown in Fig.~\ref{fig:decoding_tree}, we note that the tree in Fig.~\ref{fig:decoding_tree_v2} does not have a root and it starts from several nodes $\mathcal{M}^{i,j=1}$ instead. The nodes $\mathcal{M}^{i,j=1}$ corresponds to the coset representative obtained by assembling $\boldsymbol{a}^{(p),i}$ and $\boldsymbol{a}^{(q),i,j=1}$ where the latter denotes the first coset representative for the $\hat{q}$ subspace after modifying the candidate error with $\boldsymbol{a}^{(p),i}$ as defined in Eq.~\eqref{eq:modified_candidate_error_q}.
We can identify the $K$ nodes $\mathcal{M}^{i,j=1}$ by first identifying the first $K$ $\boldsymbol{a}^{(p),i}$ followed by identifying the corresponding $\boldsymbol{a}^{(q),i,j=1}$. After that, we identify the node with the largest coset probability, which is chosen to be $\mathcal{M}^{1,1}$, or $i=1$, in Fig.~\ref{fig:decoding_tree_v2} as an example. 
We may then continue to find the second coset representative $\boldsymbol{a}^{(q),i=1,j=2}$ for the same modified candidate error that is defined with respect to $\boldsymbol{a}^{(p),i=1}$. 
This leads to candidates $\mathcal{M}^{i=1,j=2,k}$ where $1\leq k\leq K_{i=1,j=2}$ and $K_{i=1,j=2}$ denotes the number of candidates for $\mathcal{M}^{i=1,j=2}$. 
The coset representative with the second largest probability 
may be found among the union of sets $\left\{\mathcal{M}^{i=1,j=2,k}, 1\leq k\leq K_{i=1,j=2}\right\}$ and $\left\{\mathcal{M}^{i, j=1}, 2\leq i\leq K\right\}$ which is chosen to be $\mathcal{M}^{i=1,j=2,k=2}$ as an example in Fig.~\ref{fig:decoding_tree_v2}. The similar idea can be repeatedly applied to find coset representatives with subsequent largest probability. Comparing to the brute-force approach, this approach avoids the need to explore coset representatives that have negligible contribution, which greatly reduces the runtime. We have adopted this approach in our numerical {simulations} in Sec.~\ref{sec: Surface-hexagonal GKP code}.

To summarize, we present an algorithm to approximate the MLD for a general surface-GKP code {that} finds $K$ coset representatives {and then calculates} the coset probability with Eq.~\eqref{eq:coset_prob_surf_GKP}. 
As already noted in Sec.~\ref{sec: Coset structures for general surface-GKP codes}, the MWMs found by the $K$-MWM decoder need not contain the closest lattice point in the symplectic dual lattice of the surface-GKP code for the candidate error. The latter corresponds to the MWM of a hypergraph, which is known to be hard to identify. However, as we show in Sec.~\ref{sec:Numerical results}, the $K$-MWM decoder constitutes a valid approximation for MLD for general surface-GKP codes.

\section{Surface code subject to correlated errors as a special case of general surface-GKP code}
\label{sec: Surface code subject to correlated errors as a special case of general surface-GKP code}

In this section, we generalize the discussion in Sec.~\ref{sec: Surface code subject to graphlike errors as a special case of surface-square GKP code} to surface code subject to correlated errors, and show that {its coset probability can be treated as a special case of that of the general surface-GKP code considered in App.~\ref{sec: Details of approximate MLD for the surface-GKP code}. }

Recall from Sec.~\ref{sec: The MLD for correlated errors} that the MLD for surface code with correlated errors aims to evaluate the following coset probability (see Eq.~\eqref{eq:P_e}-\eqref{eq:P_l_correlated})
\begin{align}
\label{eq:P_e_app}
    P([\boldsymbol{\eta}]) =\sum_{\boldsymbol{g}\in\mathcal{G}}\prod_{i=1}^Np(\boldsymbol{\eta}_i\ominus\boldsymbol{g}_i),
\end{align}
where
\begin{align}
\label{eq:P_e_2}
    p(\boldsymbol{\eta}_i\ominus\boldsymbol{g}_i) =
    \begin{cases}
  1 & \boldsymbol{\eta}_i\ominus\boldsymbol{g}_i = [0,0]\\
  \frac{{\epsilon^X}}{1-\epsilon} & \boldsymbol{\eta}_i\ominus\boldsymbol{g}_i = [1,0]\\
  \frac{{\epsilon^Z}}{1-\epsilon} & \boldsymbol{\eta}_i\ominus\boldsymbol{g}_i = [0,1]\\
  \frac{{\epsilon^Y}}{1-\epsilon} & \boldsymbol{\eta}_i\ominus\boldsymbol{g}_i = [1,1]
\end{cases}.
\end{align}
Here $\epsilon=\epsilon^X+\epsilon^Y+\epsilon^Z$, $\boldsymbol{\eta}\equiv[\boldsymbol{\eta}^X; \boldsymbol{\eta}^Z]\in\mathbb{F}_2^{2N}$ denotes a physical error and $\boldsymbol{g}\equiv[\boldsymbol{g}^X; \boldsymbol{g}^Z]\in\mathbb{F}_2^{2N}$ denotes a stabilizer. We use $\boldsymbol{\eta}_i$ and $\boldsymbol{g}_i$ to denote the components of $\boldsymbol{\eta}$ and $\boldsymbol{g}$ on the $i$-th qubit.
We have assumed the error rates of different qubits are the same to simplify the presentation below, but it can be  generalized to the case where the error rates of different qubits are different. 
In order to decode the surface code with correlated errors as a surface-GKP code, we consider concatenating the surface code to $N$ non-square one-mode GKP code. 
From Eq.~\eqref{eq:M_general}, its generator matrix is given by
\begin{align}
\label{eq:M_general_app}
    M_\text{surf} = M_\text{surf-sq}S^T \equiv M_\text{surf-sq}\begin{bmatrix}
        \gamma_1I_N & \gamma_2I_N \\ \gamma_3I_N  & \gamma_4I_N
    \end{bmatrix}^T.
\end{align}
Here the one-mode GKP codes are assumed to be identical, which is sufficient for our purpose below.
Following the same argument in Sec.~\ref{sec: Coset structures for general surface-GKP codes}, we have that the GKP lattice $\sqrt{2}\Lambda_\text{surf}$ contains the lattice $2\Lambda(S^T)$ as a sublattice. 
Hence for any $\boldsymbol{u}\in\sqrt{2\pi}\Lambda_\text{surf}$, it can be written as $\boldsymbol{u}=\sqrt{\pi}S(\boldsymbol{g}+\boldsymbol{v})$ where $\boldsymbol{g}\in\mathbb{F}_2^{2N}$ for certain stabilizer, and $\boldsymbol{v}\in2Z_{2N}$.
For the purpose of simulating the qubit code using GKP code, we will consider the discrete candidate error $\boldsymbol{\eta}_{\boldsymbol{s}}=\sqrt{\pi}S\boldsymbol{\eta}'$ for some $\boldsymbol{\eta}'\in\mathbb{F}_2^{2N}$.
Put them together, the coset probability of the surface-GKP code reads
\begin{equation}
\label{eq:MLD_5_app}
    \begin{aligned}
    &\sum_{\boldsymbol{u}\in\sqrt{2\pi}\Lambda_\text{surf}}P_\sigma(\boldsymbol{\eta}_{\boldsymbol{s}}-\boldsymbol{u})\\    
    =&\sum_{\boldsymbol{g}\in\mathcal{G}}\sum_{\boldsymbol{v}\in2Z_{2N}}P_\sigma(\sqrt{\pi}S(\boldsymbol{\eta}'-\boldsymbol{g}-\boldsymbol{v}))\\
    =&\sum_{\boldsymbol{g}\in\mathcal{G}}\sum_{\boldsymbol{v}\in2Z_{2N}}\prod_{i=1}^{N}P_\sigma(\sqrt{\pi}S_i(\boldsymbol{\eta}'-\boldsymbol{g}-\boldsymbol{v})_i)\\
    =&\sum_{\boldsymbol{g}\in\mathcal{G}}\prod_{i=1}^{N}\tau(S_i, \boldsymbol{b}_i ; \sigma),
    \end{aligned}
\end{equation}
where $\boldsymbol{b}\equiv \boldsymbol{\eta}'-\boldsymbol{g}$, and (we have again omitted the irrelevant factor from the Gaussian distribution)
\begin{equation}
\label{eq:def_tau_conc}
    \begin{aligned}
       \tau(S_i, \boldsymbol{b}_i; \sigma) &\equiv \sum_{\boldsymbol{v}_i\in2Z_2}P_\sigma(\sqrt{\pi}S_i(\boldsymbol{b}_i-\boldsymbol{v}_i)) \\
       &=\sum_{\boldsymbol{v}_i\in2Z_2}\exp\left(-\frac{\pi||S_i(\boldsymbol{b}_i-\boldsymbol{v}_i)||^2}{2\sigma^2}\right).
    \end{aligned}
\end{equation}
Here, we have
\begin{align}
\label{eq:def_S_i}
    S_i=\begin{bmatrix}
        \gamma_1 & \gamma_2\\
        \gamma_3 & \gamma_4
    \end{bmatrix},
\end{align}
which is the symplectic matrix for the one-mode GKP code.
In contrast to Eq.~\eqref{eq:MLD_5}, $\mathcal{G}$ in Eq.~\eqref{eq:MLD_5_app} indicates the full stabilizer group of the qubit code, and we have used the fact that  $\Lambda(S^T)$ is a direct sum of $N$ sublattices as explained in Appendix \ref{sec: Details of approximate MLD for the surface-GKP code}.

In order to reduce Eq.~\eqref{eq:MLD_5_app}-\eqref{eq:def_tau_conc} to Eq.~\eqref{eq:P_e_app}-\eqref{eq:P_e_2} so that the MLD for the qubit code can be approximated via that for the surface-GKP code, for each $\boldsymbol{b}_i$, we pick the appropriate $\boldsymbol{v}_i$ in the summation in Eq.~\eqref{eq:def_tau_conc} such that $\boldsymbol{\eta}_i=\boldsymbol{\eta}'_i\ominus\boldsymbol{v}_i$, i.e., 
\begin{align}
    \boldsymbol{\eta}_i\ominus\boldsymbol{g}_i = \boldsymbol{b}_i\ominus\boldsymbol{v}_i 
\end{align}
Hence, we can remove the summation in Eq.~\eqref{eq:def_tau_conc} and solve the equation
\begin{equation}
\label{eq:def_tau_conc_2}
    \begin{aligned}
    \exp\left(-\frac{\pi||S_i(\boldsymbol{\eta}_i\ominus\boldsymbol{g}_i)||^2}{2\sigma^2}\right) = p(\boldsymbol{\eta}_i\ominus\boldsymbol{g}_i)
    \end{aligned}
\end{equation}
for the four possibilities listed in Eq.~\eqref{eq:P_e_2}. 
In particular, upon plugging Eq.~\eqref{eq:def_S_i} and Eq.~\eqref{eq:P_e_2} into Eq.~\eqref{eq:def_tau_conc_2}, we have the following set of equations
\begin{equation}
\label{eq:set_of_eq}
    \begin{aligned}
        \frac{{\epsilon^X}}{1-\epsilon} &=\exp\left(-\frac{\pi(\gamma_1^2+\gamma_3^2)}{2\sigma^2}\right),\\
    \frac{{\epsilon^Z}}{1-\epsilon} &=\exp\left(-\frac{\pi(\gamma_2^2+\gamma_4^2)}{2\sigma^2}\right),\\
    \frac{{\epsilon^Y}}{1-\epsilon} &=\exp\left(-\frac{\pi[(\gamma_1+\gamma_2)^2+(\gamma_3+\gamma_4)^2]}{2\sigma^2}\right).
    \end{aligned}
\end{equation}
For example, if the qubit code is subject to {depolarizing} noise with ${\epsilon^X}={\epsilon^Y}={\epsilon^Z}=\epsilon/3$, then one of the solutions to Eq.~\eqref{eq:set_of_eq} reads
\begin{equation}
\label{eq:para_depolarizing_noise}
    \begin{aligned}
        \gamma_1 = \frac{\sqrt{2}}{3^{\frac{1}{4}}}, \quad \gamma_2 &= -\frac{1}{2}\gamma_1, \quad \gamma_3 = 0, \quad \gamma_4 = \frac{1}{\gamma_1},\\
        \sigma &= \left(\frac{\sqrt{3}}{\pi}\log\frac{3(1-\epsilon)}{\epsilon}\right)^{-1/2},
    \end{aligned}
\end{equation}
which holds iff $0\leq \epsilon\leq 3/4$.
Upon comparing to Eq.~\eqref{eq:basis_hex}, we notice that this is precisely the one-mode hexagonal GKP code. 
Hence we conclude that in order to simulate a qubit code subject to depolarizing noise, we can use the corresponding concatenated-hexagonal GKP code, with the noise variance given in Eq.~\eqref{eq:para_depolarizing_noise}.

In summary, for surface code subject to correlated noise, the corresponding surface-GKP code can be constructed with code parameters and noise variance determined by Eq.~\eqref{eq:set_of_eq}.
When {the surface-GKP code is} subject to discrete displacement errors that follow {Bernoulli} distribution, {it} behaves identically to the surface code, hence the MLD for the latter can be approximated by approximating the MLD for the former, following the approach in Appendix \ref{sec: Details of approximate MLD for the surface-GKP code}. 
{We emphasize again that such mapping should be viewed as a tool to decode concatenated GKP codes and their corresponding qubit stabilizer codes within a unified framework. It should not be used to translate the fidelities between the two families of codes, since the underlying error models are qualitatively different.}

\section{Minimum weight cycle of a graph}
\label{sec: Minimum weight cycle of a graph}

\begin{algorithm}
\caption{MWC($G$)}
\label{alg: MinWeightCycle}
{\bf Input: } The decoding graph $G=(\mathcal{V}, \mathcal{E}, \mathcal{W}, \mathcal{V}_h)$; \\ 
{\bf Output: } The minimum weight cycle $\mathcal{C}\subset\mathcal{E}$ \\ 
$\mathcal{C} \leftarrow \left\{\right\}$ //The candidate for the minimum weight cycle\\ 
$w \leftarrow \infty$ //The weight of the candidate\\ 
\For{$e\in\mathcal{E}$}{
$v_1, v_2 \leftarrow e[1], e[2]$ //Get the two vertices of $e$\\
$\mathcal{P}\leftarrow$ ShortestPath($G, v_1, v_2$) \\
$\mathcal{C}_2\leftarrow \mathcal{P} \cup e$ \\
\If{$w(\mathcal{C}_2) < w$}
    {
        $w\leftarrow w(\mathcal{C}_2)$\\
        $\mathcal{C}\leftarrow \mathcal{C}_2$
    }
 }
\end{algorithm}

In this section, we describe an algorithm to find the minimum weight cycle (MWC) for a decoding graph $G=(\mathcal{V}, \mathcal{E}, \mathcal{W}, \mathcal{V}_h)$ with $\mathcal{V}_h=\emptyset$. The algorithm is based on Dijkstra's shortest path algorithm with time complexity $O(|\mathcal{E}|^2\log|\mathcal{V}|)$ and space complexity $O(|\mathcal{V}|^2)$. The idea of finding the MWC is to remove every edge $e\equiv \left\{v_1, v_2\right\}$ from the graph one by one, and find the shortest path between $v_1$ and $v_2$, followed by adding the edge $e$ back to the shortest path to complete the cycle. 
Since the MWC is guaranteed to contain at least one edge from the graph, it can be obtained by simply finding the one with minimum weight among all the candidates. The algorithm is presented as MWC($G$) in Algorithm \ref{alg: MinWeightCycle} where we assume an algorithm ShortestPath($G, v_1, v_2$), such as the Dijkstra's shortest path algorithm, is known to find the shortest path between two vertices in the graph. 

\section{An example of decoding with $K$ MWMs}
\label{sec: An example of decoding with $K$ MWMs}

In this section, we provide an example code with matchings exhibit low enough weights and different logical classes than the first MWM. In this case, the inclusion of subsequent MWMs would change the decision of logical class.

\begin{figure}
\centering
\includegraphics[width=\linewidth]{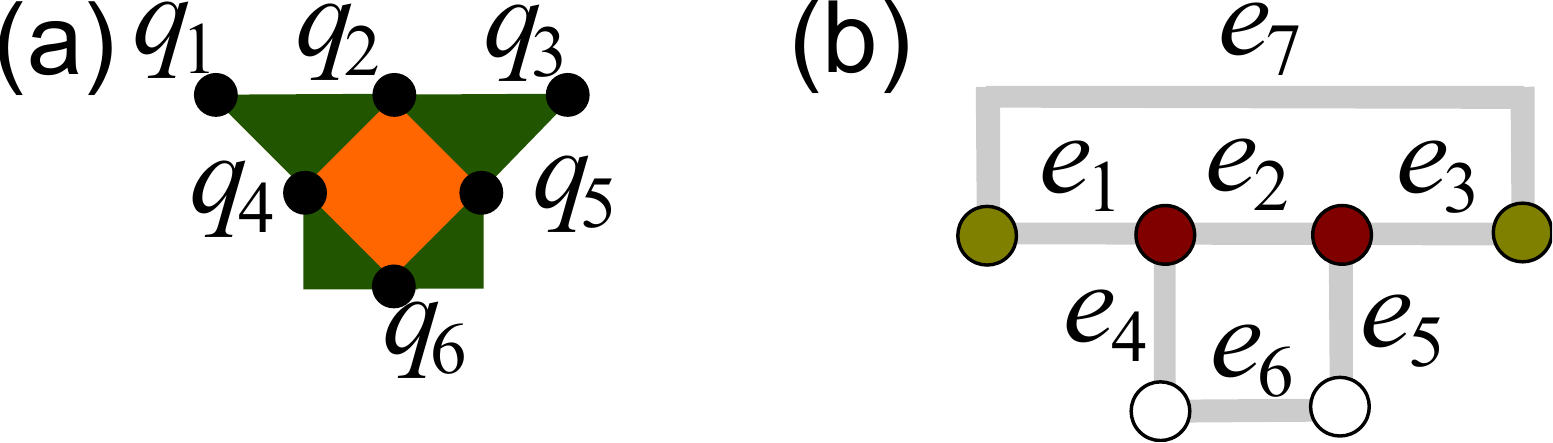}
 \caption{
 (a) A six-qubit code with the $X$-type and $Z$-type
stabilizers shown in orange and green respectively. 
 (b) The decoding graph of the code where the edge $e_i$ corresponds to the qubit $q_i$ and white vertices correspond to $Z$-type stabilizers.
 The red vertices represent nontrivial stabilizer measurement outcomes.
 The virtual edge $e_7$ between the virtual vertices is shown explicitly.
 }
 \label{fig:example_decoding}
\end{figure}

In Fig.~\ref{fig:example_decoding}(a), we illustrate the six-qubit code considered where the $X$-type and $Z$-type
stabilizers are shown in orange and green respectively. The stabilizer generators and logical operators are listed below
\begin{align}
    &\left\{Z_1Z_2Z_4,Z_2Z_3Z_5,Z_4Z_6,Z_5Z_6, X_2X_4X_5X_6\right\},\\
    &\bar{X}_L=X_1X_2X_3, \quad \bar{Z}_L=Z_1.
\end{align}
This code can be obtained by removing certain qubits and stabilizers from the distance-3 surface code. 
Assuming we are interested in correcting the $X$-type errors, the corresponding decoding graph is shown in Fig.~\ref{fig:example_decoding}(b) where, for the purpose of this demonstration, we have highlighted two vertices and assumed that the edges take the following weights
\begin{align}
    w(e_1)=w(e_3)&=w(e_4)=w(e_5)=w(e_6)=0.1,\\
    w(e_2)&=0.5, \quad w(e_7)=10^{-7}.
\end{align}
We note that $e_7$ is the edge between the two virtual vertices with an infinitesimal weight (taken to be $10^{-7}$ here).
As a result, there are in total four matchings
\begin{equation}
\begin{aligned}
\label{eq:MWMs_six_qubit_codes}
    \mathcal{M}_1 &= \left\{e_1, e_3, e_7\right\},\\
    \mathcal{M}_2 &= \left\{e_4, e_5, e_6\right\},\\
    \mathcal{M}_3 &= \left\{e_2\right\},\\
    \mathcal{M}_4 &= \left\{e_1, e_2, e_3,e_4, e_5, e_6, e_7\right\},
\end{aligned}    
\end{equation}
which have been ordered based on their weights, i.e.,
\begin{align*}
    w(\mathcal{M}_1) = 0.2+10^{-7}, w(\mathcal{M}_2) = 0.3+10^{-7}, \\
    w(\mathcal{M}_3) = 0.5+10^{-7}, w(\mathcal{M}_4) = 1.0+10^{-7}. 
\end{align*}
We can group the matchings based on their logical classes. For example, since $\mathcal{M}_1$ corresponds to the error $X_1X_3$ which anticommutes with $\bar{Z}_L$, it would yield a logical $X$ error. With that, we have
\begin{align}
    \Xi_{\bar{I}} &= \left\{\mathcal{M}_2, \mathcal{M}_3\right\},\\
    \Xi_{\bar{X}} &= \left\{\mathcal{M}_1, \mathcal{M}_4\right\},
\end{align}
where we have used the notation in Eq.~\eqref{eq:approx_P_l_MWMs}. Using the same equation, we have
\begin{align}
    P_{\bar{I}} &\approx\exp(-0.3)+\exp(-0.5) \approx 1.3473,\\
    P_{\bar{X}} &\approx\exp(-0.2)+\exp(-1.0) \approx 1.1866.
\end{align}
Since $P_{\bar{I}}>P_{\bar{X}}$, we conclude that the most probable logical class that is consistent with the syndrome reads $\bar{I}$. We note that this is in contrast to the most probable physical error $X_1X_3$ which is in the $\bar{X}$ class.

\begin{widetext}

\begin{figure}[H]
\centering
\includegraphics[width=\linewidth]{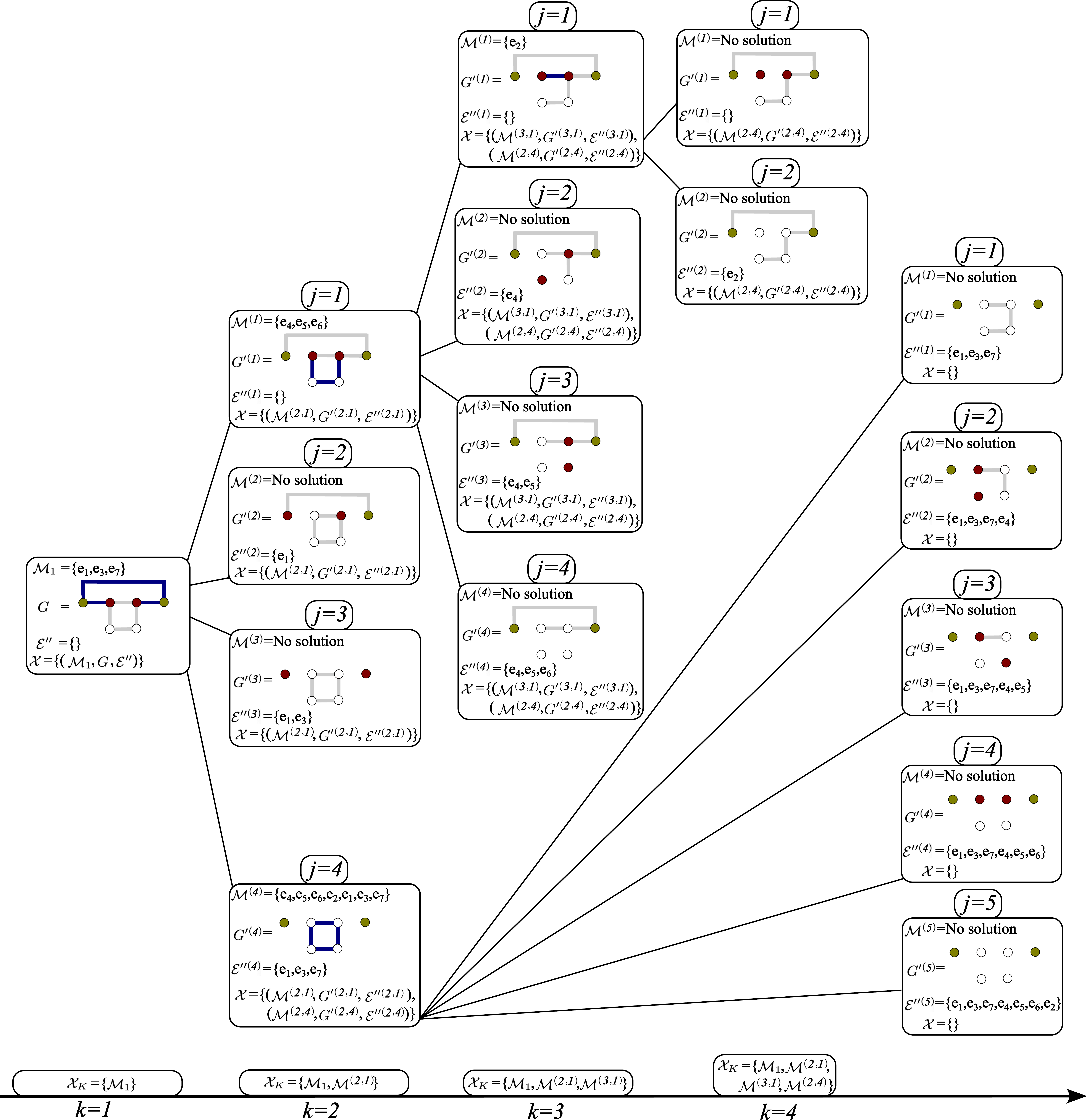}
 \caption{Execution of Algorithm \ref{alg: MWMs_v1} to find the MWMs of the decoding graph shown in Fig.~\ref{fig:example_decoding}(b). See the text for description.
}
 \label{fig:decoding_tree_example_code}
\end{figure}

\end{widetext}

We proceed to illustrate that all the MWMs shown in Eq.~\eqref{eq:MWMs_six_qubit_codes} of the decoding graph can be identified using the $K$-MWM decoder. In Fig.~\ref{fig:decoding_tree_example_code}, we show the result of applying Algorithm \ref{alg: MWMs_v1} to the decoding graph shown in Fig.~\ref{fig:example_decoding}(b). 
The looping variable $k$ in line 6 of Algorithm \ref{alg: MWMs_v1} is shown along the horizontal axis, and $k=1$ corresponds to the quantities obtained from lines 3 to 5. 
Alternatively, the box at $k=1$ can be viewed as the root of the decoding tree which grows along the horizontal axis. 
The nodes at the $k$-th level are the children of a node at the $(k-1)$-th level, as indicated by the lines connecting the boxes. 
As defined in lines 7-10 of Algorithm \ref{alg: MWMs_v1}, the graph $G'$ for the nodes at the $k$-th level is essentially the graph of their parent node.
For instance, the graph $G'$ for $k=3$ is the graph $G'^{(1)}$ at the $j=1$ step of the $k=2$ level. 
The labels of the edges in the graphs are the same as those in Fig.~\ref{fig:example_decoding}(b) which have been omitted for clarity.
With that, we can calculate $(\mathcal{M}^{(j)}, G'^{(j)}, \mathcal{E}''^{(j)})$ following lines 12-21 of Algorithm \ref{alg: MWMs_v1} for each $j$-th step in the $k$-th iteration and the results are shown in each box. 
We also need to update $\mathcal{X}$ at each $j$-th step per lines 22-24. 
The set of identified MWMs $\mathcal{X}_K$ after the $k$-th iteration, which is updated in lines 26-27, is shown on the axis. 
In order to distinguish the matchings found at the $j$-th step of the $k$-th iteration, we have used $\mathcal{M}^{(k,j)}$ in both $\mathcal{X}_K$ and $\mathcal{X}$.
We note that the matchings in $\mathcal{X}$ are ordered based on the weight $w(\mathcal{M}^{(k,j)})$ and we pick the one with the smallest weight as the parent node for the next level of the tree, per line 7. 
We remark that $\mathcal{M}^{(j)}=\text{MWM}(G'^{(j)})\cup\mathcal{E}''^{(j)}$ per line 21, and $\text{MWM}(G'^{(j)})$ is the set of edges shown in blue in each box. 
For instance, at the $j=4$ step in the $k=2$ iteration, we have $\mathcal{M}^{(j=4)}=\left\{e_2,e_4,e_5,e_6\right\}\cup\left\{e_1,e_3,e_7\right\}$ where the first term is the MWM of the subgraph in $G'^{(4)}$. This subgraph has no highlighted vertex, as discussed in Sec.~\ref{sec:Finding the second matching for the syndrome graph}, its MWC serves as its MWM in this case. 
In some cases, there is no solution for the decoding graph shown in the box. This is the case for $j=2, k=2$ because there is an odd number of highlighted vertices in a subgraph, and for $j=4, k=3$ because there is no cycle in the subgraph without a highlighted vertex.
A box with no solution has no child and the algorithm will stop when all the boxes have been visited or equivalently $\mathcal{X}=\emptyset$. 
As one can confirm, the ordered set $\mathcal{X}_K$ at $k=4$ is precisely the matchings shown in Eq.~\eqref{eq:MWMs_six_qubit_codes}.

\bibliography{GKP_facts.bib}

\end{document}